\documentclass[12pt]{article}
\usepackage{epsfig,cite}
\usepackage {amsmath}
\usepackage{amssymb}
\include{epsf}
\newcount\timecount
\newcount\hours \newcount\minutes  \newcount\temp \newcount\pmhours
\hours = \time
\divide\hours by 60
\temp = \hours
\multiply\temp by 60
\minutes = \time
\advance\minutes by -\temp
\def\hour{\the\hours}
\def\minute{\ifnum\minutes<10 0\the\minutes
            \else\the\minutes\fi}
\def\clock{
\ifnum\hours=0 12:\minute\ AM
\else\ifnum\hours<12 \hour:\minute\ AM
      \else\ifnum\hours=12 12:\minute\ PM
            \else\ifnum\hours>12
                 \pmhours=\hours
                 \advance\pmhours by -12
                 \the\pmhours:\minute\ PM
                 \fi
            \fi
      \fi
\fi
}

\def\monthname{\relax\ifcase\month 0/\or January\or February\or
   March\or April\or May\or June\or July\or August\or September\or
   October\or November\or December\else\number\month/\fi}

\def\bold#1{\setbox0=\hbox{$#1$}%
     \kern-.025em\copy0\kern-\wd0
     \kern.05em\copy0\kern-\wd0
     \kern-.025em\raise.0433em\box0 }

\newcommand{\br}{{\rm BR}}

\newcommand{\ETslash}{/ \hspace{-.7em} E_T}

\def\beq{\begin{equation}}
\def\eeq{\end{equation}}

\def\ga{\mathrel{\raise.3ex\hbox{$>$\kern-.75em\lower1ex\hbox{$\sim$}}}}
\def\la{\mathrel{\raise.3ex\hbox{$<$\kern-.75em\lower1ex\hbox{$\sim$}}}}
\def\gev{{\rm \, Ge\kern-0.125em V}}
\def\tev{{\rm \, Te\kern-0.125em V}}
\def\gyr{{\rm \, G\kern-0.125em yr}}

\def\tb{\tan \beta}

\def\gappeq{\mathrel{\rlap {\raise.5ex\hbox{$>$}}
{\lower.5ex\hbox{$\sim$}}}}
\def\lappeq{\mathrel{\rlap{\raise.5ex\hbox{$<$}}
{\lower.5ex\hbox{$\sim$}}}}
\def\Toprel#1\over#2{\mathrel{\mathop{#2}\limits^{#1}}}

\def\stau{\widetilde \tau}

\def\m12{m_{1\!/2}}

\def\stau{\tilde{\tau}}

\def\bea{\begin{eqnarray}}
\def\eea{\end{eqnarray}}

\newcommand{\bmm}{\ensuremath{\br(B_s \to \mu^+\mu^-)}}

\begin{document}
\begin{titlepage}
\pagestyle{empty}
\baselineskip=21pt
\rightline{KCL-PH-TH/2012-46, LCTS/2012-32, CERN-PH-TH/2012-331}
\rightline{UMN--TH--3127/12, FTPI--MINN--12/38}
\vskip 0.2in
\begin{center}
{\large{\bf The End of the CMSSM Coannihilation Strip is Nigh}}
\end{center}
\begin{center}
\vskip 0.3in
{\bf Matthew Citron}$^{1}$,
{\bf John~Ellis}$^{2,3}$,
{\bf Feng Luo}$^{2}$,
{\bf Jad~Marrouche}$^1$,\\
{\bf Keith~A.~Olive}$^{4,5}$
and {\bf Kees~J.~de~Vries}$^1$
\vskip 0.3in
{\small {\it
$^1${High Energy Physics Group, Blackett Laboratory, Imperial College, \\ Prince Consort Road, London SW7 2AZ, UK}\\
$^2${Theoretical Particle Physics and Cosmology Group, Department of
  Physics, \\ King's College London, London WC2R 2LS, United Kingdom}\\
$^3${Theory Division, CERN, CH-1211 Geneva 23, Switzerland}\\
$^4${School of Physics and Astronomy, University of Minnesota, Minneapolis, MN 55455, USA}\\
$^5${William I. Fine Theoretical Physics Institute, School of Physics and Astronomy,\\
University of Minnesota, Minneapolis, MN 55455, USA}\\
}}

\vskip 0.25in
{\bf Abstract}
\end{center}
\baselineskip=18pt \noindent

A recent global fit to the CMSSM incorporating current constraints on supersymmetry, 
including missing transverse energy searches at the LHC, BR$(B_s \to \mu^+ \mu^-)$ and the direct XENON100
search for dark matter, favours points towards the end of the stau-neutralino (${\tilde \tau_1}$-$\chi$)
coannihilation strip with relatively large $m_{1/2}$ and $10 \lappeq \tb \lappeq 40$
and points in the $H/A$ rapid-annihilation funnel with $\tb \sim 50$. The coannihilation
points typically have $m_{\tilde \tau_1}-m_\chi \lappeq 5$~GeV, and a significant fraction,
including the most-favoured point, has $m_{\tilde \tau_1}-m_\chi < m_\tau$. In such a
case, the $\tilde \tau_1$ lifetime would be so long that the $\tilde \tau_1$ would be detectable
as a long-lived massive charged particle that may decay inside or outside the apparatus.
We show that CMSSM scenarios close to the tip of the coannihilation strip for $\tan \beta \lappeq 40$
are already excluded by LHC searches for massive charged particles,
and discuss the prospects for their detection in the CMS and ATLAS detectors via
time-of-flight measurements, anomalous heavy ionization or decays into one or more soft
charged particles.

\vfill
\leftline{December 2012}
\end{titlepage}
\baselineskip=18pt

\section{Introduction}

In the constrained Minimal Supersymmetric extension of the Standard Model (CMSSM)~\cite{funnel,cmssm},
in which the soft supersymmetry-breaking parameters are assumed to be universal at
some input GUT scale, and $R$-parity is assumed to be conserved so that the lightest
supersymmetric particle (LSP, presumed here to be the lightest neutralino, $\chi$) is stable
\cite{ehnos,cmssmwmap},
there are several distinct regions in parameter space where the relic $\chi$ density
falls within the range allowed by the WMAP experiment and other astrophysical and
cosmological observations \cite{wmap}. These include the $\chi-$stau coannihilation strip
\cite{efo},
the funnel region where there are rapid annihilations through s-channel $H/A$ poles \cite{funnel,efgosi},
and the focus-point strip \cite{fp}.

Recent LHC searches for missing-energy events~\cite{ATLAS7,CMS7,ATLAS8,CMS8}, 
combined with the measurements of $M_h$~\cite{ATLASH,CMSH}
and $B_s \to \mu^+ \mu^-$ decay~\cite{bmm,LHCbnew} as well as the direct XENON100 search
for dark matter~\cite{XE100}, constrain severely the CMSSM parameter space. The LHC has
excluded much of the coannihilation strip, whilst XENON100 has put another
nail in the coffin of the focus-point strip~\cite{MC8}. A recent global analysis finds almost
equally good fits in the rapid-annihilation funnel and towards the end of the
coannihilation strip, both being compatible with the LHC measurement of $M_h$ and other constraints~\cite{MC8}.
The rapid-annihilation funnel is favoured by LHC missing-energy ($\ETslash$)
searches and $M_W$, in particular, whereas the coannihilation region is favoured by $g_\mu - 2$ and
$B_s \to \mu^+ \mu^-$. 

It is well known that the  ${\tilde \tau_1} - \chi$ mass difference decreases monotonically
along the coannihilation strip towards its tip at large $m_\chi$.
Some distance before the tip of the strip, the mass difference falls below
$m_\tau$, so that the two-body decay ${\tilde \tau_1} \to \chi + \tau$
becomes kinematically forbidden, three- and four-body decays
dominate, and the ${\tilde \tau_1}$ lifetime exceeds
$10^{-9}$~s~\cite{Jittoh, Kaneko}. In this case, any ${\tilde \tau_1}$ produced inside an LHC detector
would appear as a long-lived charged particle that may or may not decay before escaping.
The sensitivities of the conventional LHC searches for missing-energy events
to this portion of the CMSSM parameter space should be examined in both cases.
The LHC experiments ATLAS and CMS are capable of detecting
long-lived charged particles, and have published limits on their possible
production in various scenarios in which they decay outside the apparatus, 
including some supersymmetric frameworks such as gauge-mediated 
supersymmetry breaking (GMSB), but not the CMSSM~\cite{CMSmcp,ATLASmcp}.

In this paper we first show that the ${\tilde \tau_1}$ is long-lived in
a favoured fraction of the remaining CMSSM parameter space, and then argue
that searches for long-lived charged particles should be important features
of a comprehensive strategy for supersymmetry searches and future global fits. 
We then compare the available limits on long-lived
charged particles with the other LHC searches included
in previous global fits, and argue that the end of the CMSSM coannihilation strip is
nigh. The small portion that has not already been excluded could be
explored by a combination of LHC searches for $\ETslash$ events and massive charged particles 
with the 8-TeV data taken during 2012.

\section{Summary of Results from a Global CMSSM Fit}

The starting point of our study is a recent global analysis~\cite{MC8} of
the CMSSM including the ATLAS search for jets + $\ETslash$ events with
5/fb of data at 7~TeV~\cite{ATLAS7}, 
the (presumed) Higgs mass measurement $M_h \sim 125$~GeV~\cite{ATLASH,CMSH},
a combination of the available constraints on \bmm~\cite{bmm}~\footnote{The results in~\cite{MC8} are very little
affected by incorporation of the recent measurement of \bmm\ by the LHCb Collaboration~\cite{LHCbnew},
whose central value is quite close to the combination of previous data, and very consistent with the
predictions of the CMSSM and NUHM1 fits in~\cite{MC8}. See {\tt http://mastercode.web.cern.ch/mastercode/} for current Mastercode results.},
the latest XENON100 constraint on direct dark matter scattering~\cite{XE100},
and low-energy constraints including $g_\mu - 2$ and precision electroweak data.
As already mentioned, the XENON100 results disfavour the focus-point region
of the CMSSM parameter space, 
the ATLAS $\ETslash$ results disfavour points
on the coannihilation strip with small $m_{1/2}$, and (relatively) large values of
$m_{1/2}$ are also favoured by the Higgs mass.

The global analysis in~\cite{MC8} found two `islands' of CMSSM parameter
space with comparable values of the global likelihood: a low-mass region
with $m_{1/2} \sim 910$~GeV, $m_0 \sim 300$~GeV,  and $\tb \sim 16$,
and a high-mass region with $m_{1/2} \sim 1890$~GeV, $m_0 \sim 1070$~GeV 
and $\tb \sim 45$. The former lies in the coannihilation region and the
latter is in the rapid-annihilation funnel region. As shown in Table 2 of~\cite{MC8},
the low-mass CMSSM region is favoured by $g_\mu - 2$, $A_{\rm fb}(b)$ and \bmm,
whereas the high-mass region is favoured by $M_W$, $A_\ell({\rm SLD})$ and
the ATLAS 5/fb $\ETslash$ analysis.

\begin{figure}
\vskip 0.5in
\begin{minipage}{8in}
\epsfig{file=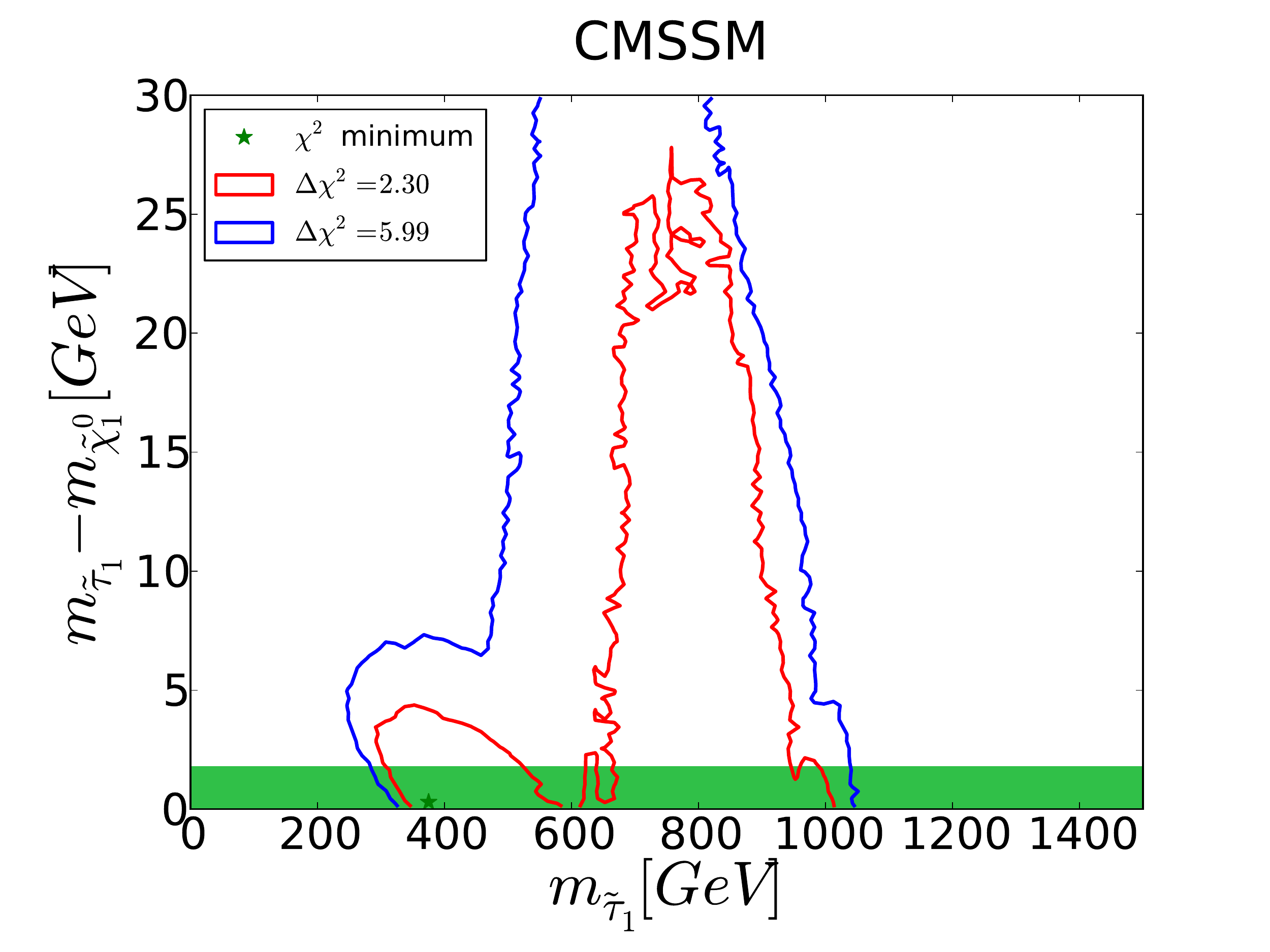,height=2.5in}
\epsfig{file=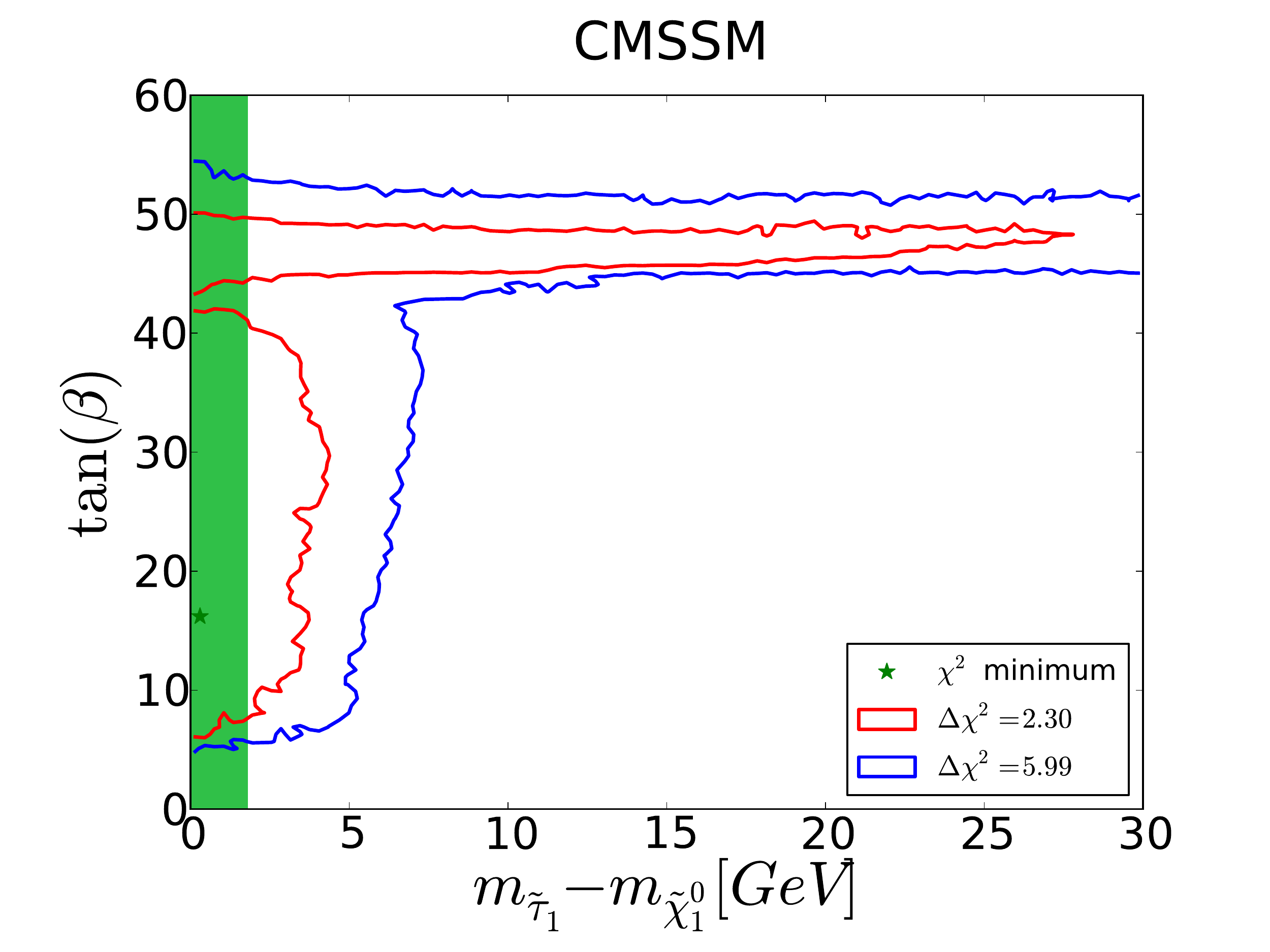,height=2.5in}
\hfill
\end{minipage}
\caption{\it
Portions of the $(m_{\tilde \tau_1}, m_{\tilde \tau_1}-m_\chi)$ plane (left panel) and
the $(m_{\tilde \tau_1}-m_\chi, \tb)$ plane (right panel), displaying the best-fit point (green star) and
the 68\% and 95\% CL contours (red and blue lines, respectively) found in a recent global analysis~\protect\cite{MC8} 
of the CMSSM. The bands with $m_{\tilde \tau_1}-m_\chi < m_\tau$ are shaded green.
}
\label{fig:MC8}
\end{figure}

The left panel of Fig.~\ref{fig:MC8} shows how the different `islands' found in~\cite{MC8}
populate different ranges of $m_{\tilde \tau_1}$ (horizontal axis) and
$m_{\tilde \tau_1}-m_\chi$ (vertical axis). Here and in the right panel the red (blue) lines are the
$\Delta \chi^2 = 2.30 (5.99)$ contours corresponding approximately to 68 (95)\% CL
contours, and the green stars mark the best-fit point found in~\cite{MC8}.
To guide the eye, a green band shows where the mass difference between the $\tilde \tau_1$ and $\chi$ 
is smaller than $m_\tau$. We note that this mass difference is $0.42$ GeV for the best-fit point and that,
more generally, points in the 
lower-mass region with $m_{\tilde \tau_1} \sim 400$~GeV, corresponding to the coannihilation strip, tend to have small mass 
differences, generally $\lappeq 5$~GeV at the 68\% CL and often $< m_\tau$.
On the other hand, points in the higher-mass region, corresponding to the rapid-annihilation funnel,
have mass differences that may extend above $25$~GeV at the 68\% CL. 
It is the small mass difference that makes the coannihilation mechanism
sufficiently efficient to bring the relic $\chi$ density down into the
WMAP range, whereas larger mass differences are possible in the region where rapid
$\chi \chi$ annihilation occurs via the direct s-channel $H/A$ poles. We note in the
left panel of Fig.~\ref{fig:MC8} a tendency for smaller values of $m_{\tilde \tau_1}-m_\chi$
to be favoured as $m_{\tilde \tau_1}$ increases from $\sim 300$~GeV to $\sim 600$~GeV,
as expected since $m_{\tilde \tau_1}-m_\chi \to 0$ as one approaches the tip of the coannihilation strip.
The right panel of Fig.~\ref{fig:MC8} shows how the different `islands' found in~\cite{MC8} populate
different regions of the $(m_{\tilde \tau_1}-m_\chi, \tb)$ plane. We see that the points
with $\tb < 43$ have $m_{\tilde \tau_1}-m_\chi < 5$~GeV, whereas those with
$\tb > 43$ may have values of $m_{\tilde \tau_1}-m_\chi$ extending above 25~GeV
at the 68\% CL. The band with $m_{\tilde \tau_1}-m_\chi < m_\tau$ is again shaded green.

\begin{figure}
\vskip 0.5in
\vspace*{-0.75in}
\begin{minipage}{8in}
\epsfig{file=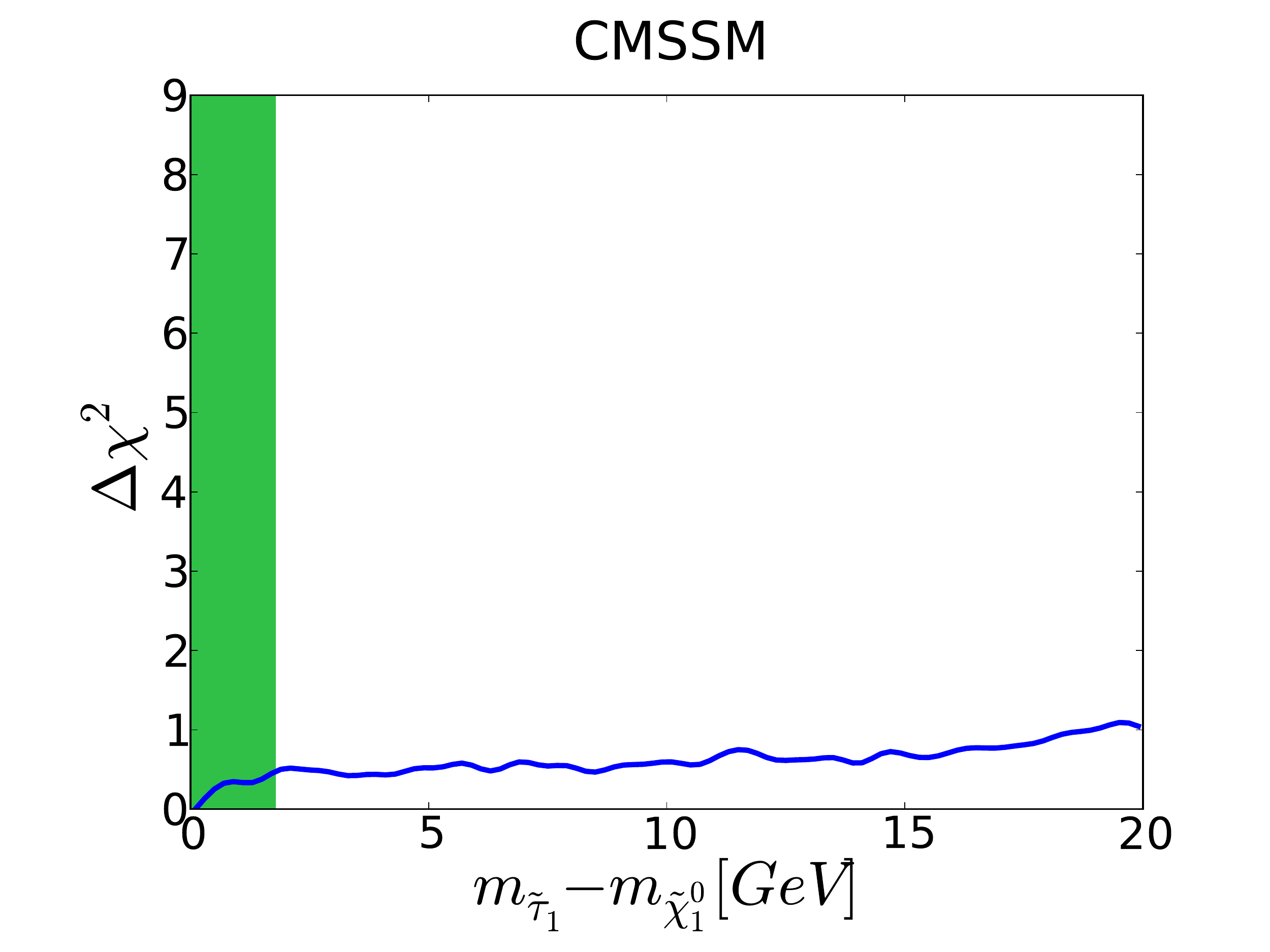,height=2.5in}
\epsfig{file=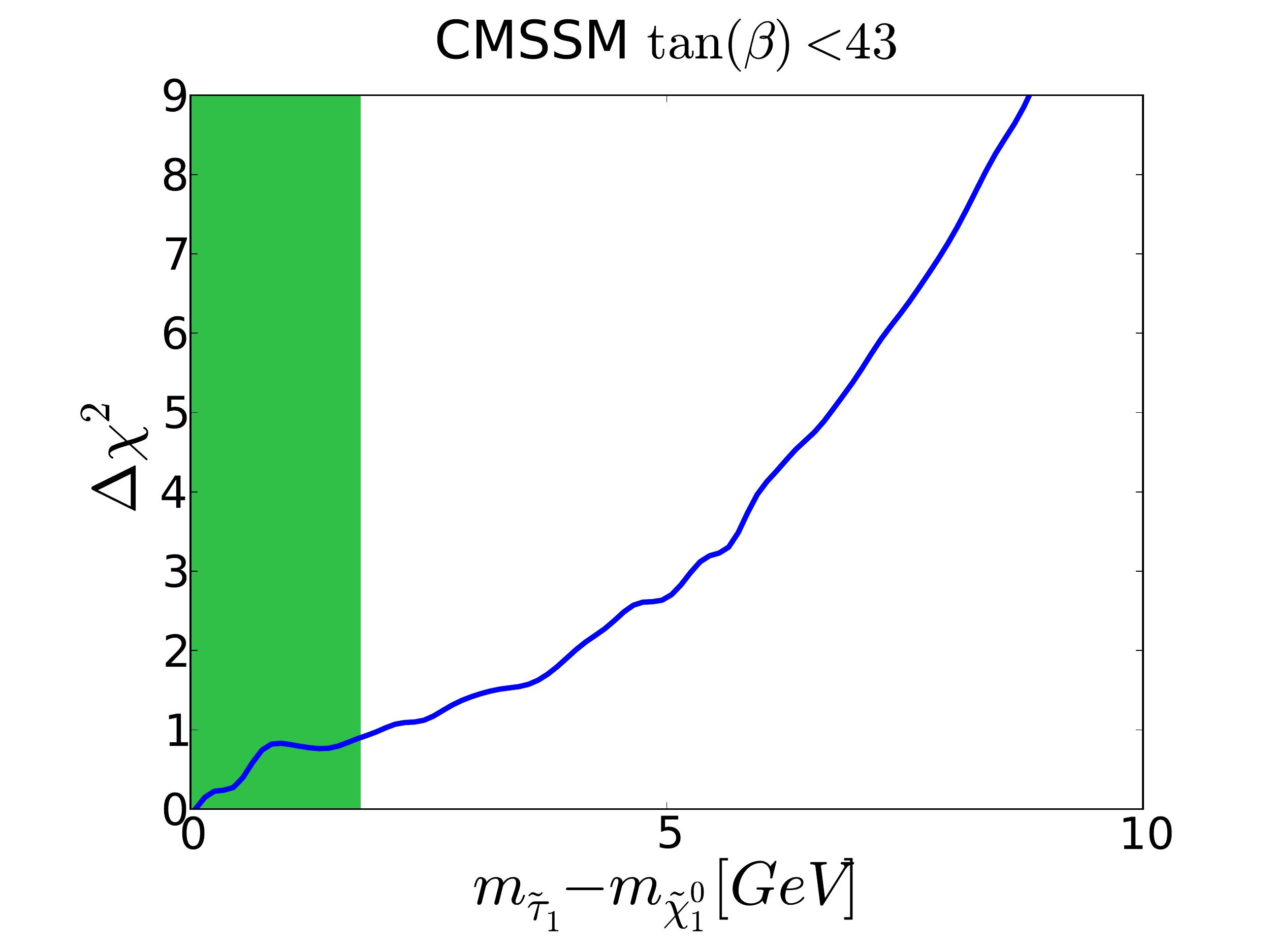,height=2.5in}
\hfill
\end{minipage}
\caption{
{\it The difference in $\chi^2$ compared with the best-fit point
as a function of $m_{\tilde \tau_1}-m_\chi$ for the full CMSSM
sample (left panel) and the coannihilation subsample with $\tb < 43$ (right panel), as
found in a recent global analysis~\protect\cite{MC8}.
The ranges $m_{\tilde \tau_1}-m_\chi < m_\tau$ are shaded green.
}} 
\label{fig:MC8L}
\end{figure}

In the left panel of Fig.~\ref{fig:MC8L} we display the global $\Delta \chi^2$ function as a function of
$m_{\tilde \tau_1}-m_\chi$ in the full CMSSM sample analyzed in~\cite{MC8}.
We see that small values of $m_{\tilde \tau_1}-m_\chi$ are slightly favoured, though values
 up to $\sim 20$~GeV are allowed at the $\Delta \chi^2 \sim 1$ level. However,
 if we restrict to the $\tb < 43$ region, as shown in the right panel of Fig.~\ref{fig:MC8L},
 we see that the preference for small $m_{\tilde \tau_1}-m_\chi$ is much more marked.
 Specifically, we see again that the best-fit point has $m_{\tilde \tau_1}-m_\chi < m_\tau$,
 inside the range shaded green.

The CMSSM results from~\cite{MC8} shown above include the constraints
provided by results from the ATLAS search for jets + $\ETslash$ events with
$\sim 5$/fb of data at 7 TeV~\cite{ATLAS7}~\footnote{A recent ATLAS analysis of  jets + $\ETslash$ events with
$\sim 5$/fb of data at 8 TeV~\cite{ATLAS8} quotes a lower limit $m_{1/2} \gappeq 700$~GeV in the range of $m_0$ close to the
coannihilation region for $\tan \beta = 10$, corresponding to $m_{\tilde \tau_1} \gappeq 300$~GeV,
which does not exclude the best-fit CMSSM point found in~\cite{MC8}.}, but not the constraints from searches for
long-lived charged particles. However, the latter are the relevant
constraints when $m_{\tilde \tau_1}-m_\chi < m_\tau$ and will become even more relevant for
future CMSSM searches at the LHC, as we discuss below in more detail.

\begin{figure}
\vskip 0.5in
\vspace*{-0.75in}
\begin{center}
\begin{minipage}{8in}
\epsfig{file=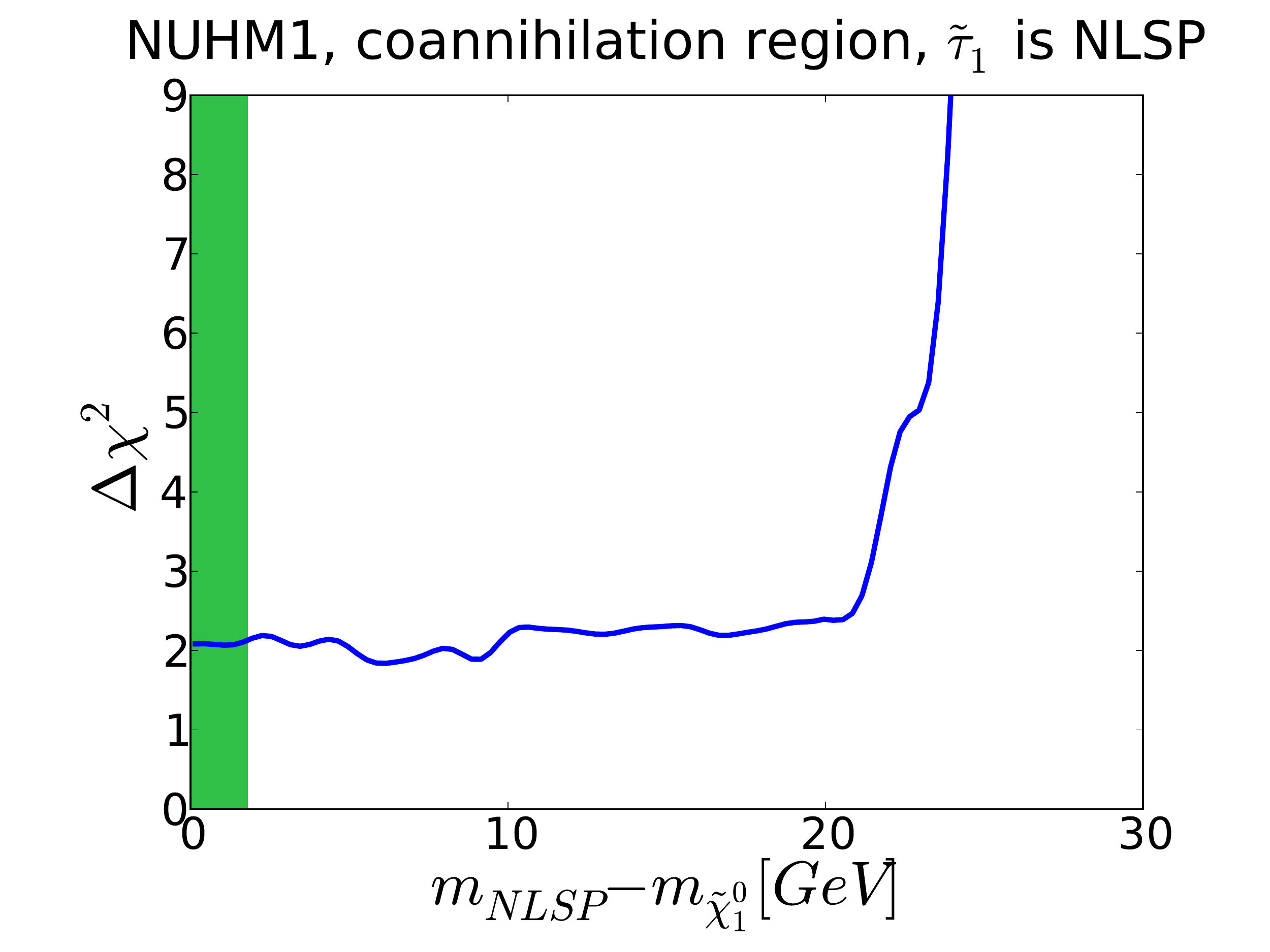,height=2.5in}
\epsfig{file=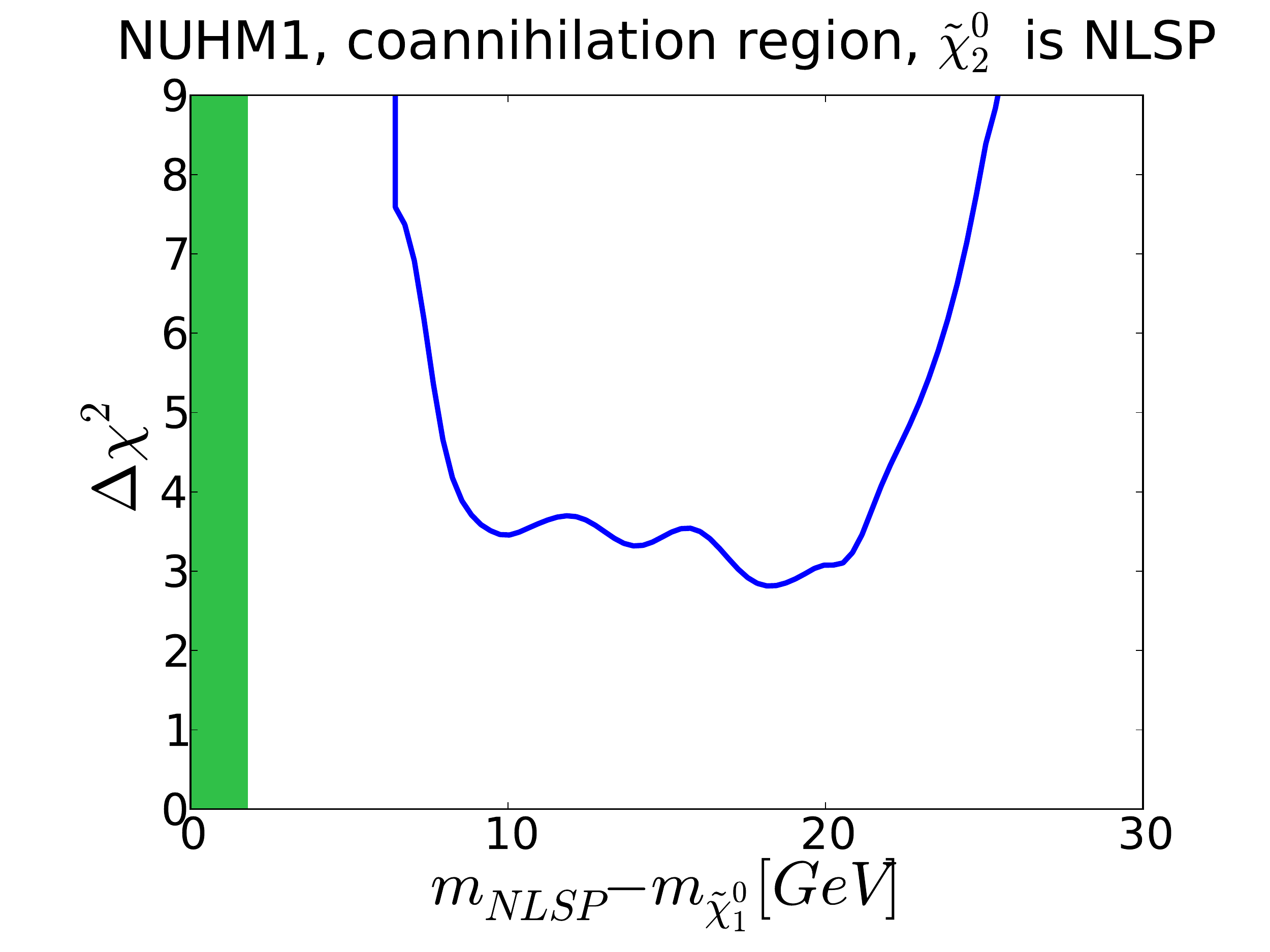,height=2.5in}
\hfill
\end{minipage}
\end{center}
\caption{
{\it The difference in $\chi^2$ for points in the NUHM1 coannihilation region discussed 
in a recent global analysis~\protect\cite{MC8}, compared with the best-fit point,
as a function of $m_{NLSP}-m_\chi$.
The left (right) panel shows NUHM1 points with a ${\tilde \tau_1}$ ($\tilde \chi_2^0$) NLSP,
and the ranges where $m_{NLSP}-m_\chi < m_\tau$ are shaded green.
}} 
\label{fig:MC8NUHM1}
\end{figure}

We have made a similar analysis of the region of the NUHM1 parameter space
at large $m_{1/2}$ and small $m_0$ where chargino-neutralino-stau coannihilations
bring the relic $\chi$ density into the WMAP range, as commented in~\cite{MC8}.
In most cases in this region, the NLSP is the ${\tilde \tau_1}$, but in
some cases it is the next heavier neutralino, $\tilde \chi_2^0$, or another sparticle. The left (right) panel of
Fig.~\ref{fig:MC8NUHM1} displays the $\chi^2$ function as a function of  $m_{NLSP}-m_\chi$
at values $\lappeq 20$~GeV for points with a ${\tilde \tau_1}$ ($\tilde \chi_2^0$) NLSP in this portion of the NUHM1 parameter space.
We see that in the ${\tilde \tau_1}$ case (left panel of Fig.~\ref{fig:MC8NUHM1})
the $\chi^2$ function is relatively flat. 
This is because the appearance of coannihilations with the
charginos and heavier neutralinos and/or other sparticles enables $m_{\tilde \tau_1}-m_\chi$
to be somewhat larger than along the CMSSM coannihilation strip.
Note that at the best fit point of the full NUHM1, we have $m_{\tilde \tau_1}-m_\chi = 2.98$ GeV
and while the relic density is still determined by stau coannihilations, the
mass difference is large enough so that the stau is not long-lived in this case.
On the other hand, the $\chi^2$ function is a distorted parabola in the $\tilde \chi_2^0$ NLSP case
(right panel of Fig.~\ref{fig:MC8NUHM1}): at large $m_{\tilde \chi_2^0}-m_\chi$ the relic LSP
density lies above the WMAP range, and at small $m_{\tilde \chi_2^0}-m_\chi$ it lies below the WMAP range.
We also note that the minimum value of $\chi^2$ in
this coannihilation region, which occurs in the ${\tilde \tau_1}$ NLSP
case, is $\sim 1.8$ higher than at the absolute minimum in the
NUHM1. Because of the larger mass difference and the higher $\chi^2$,
we conclude that there is less reason than in the CMSSM
to study the low-mass-difference region in the NUHM1.
Accordingly, we do not discuss it further in this paper,
though some of our later considerations within the CMSSM context may also apply to the NUHM1
and other models.

\section{Towards the Tip of the Coannihilation Strip}

\begin{figure}[h!]
\vskip 0.5in
\vspace*{-0.75in}
\begin{minipage}{8in}
\epsfig{file=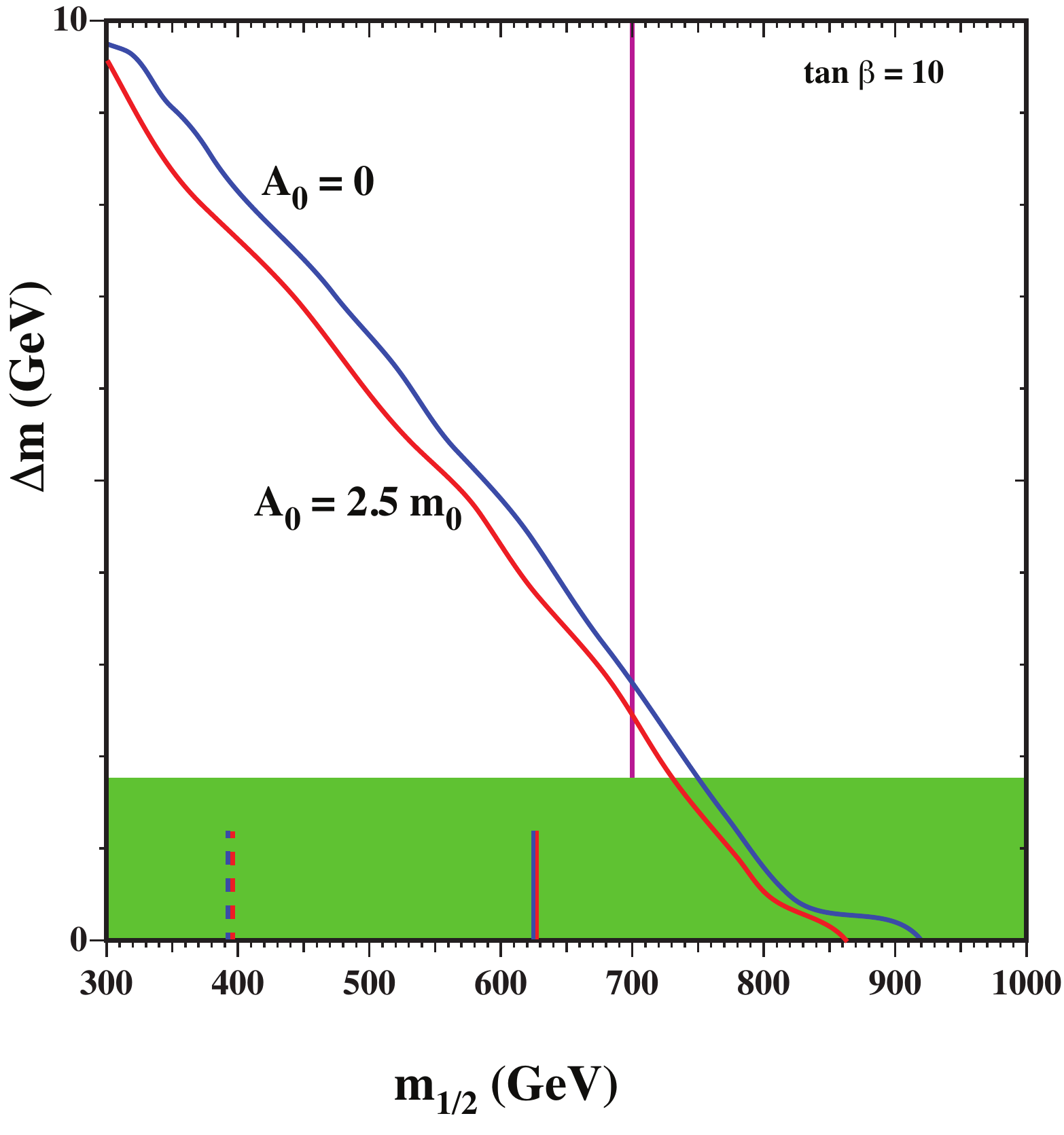,height=3.3in}
\epsfig{file=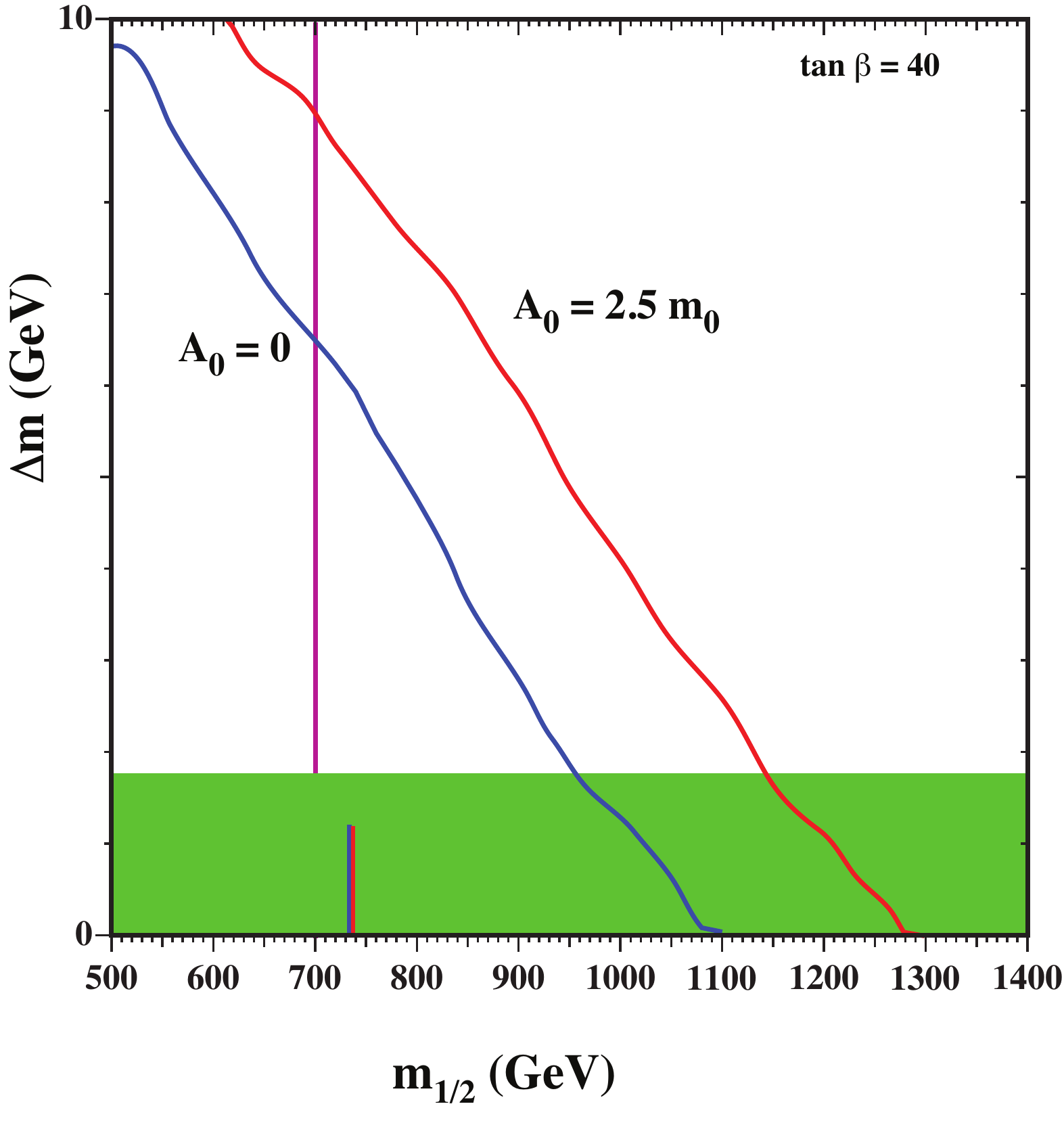,height=3.3in}
\end{minipage}
\caption{
{\it The mass difference $\Delta m \equiv m_{\tilde \tau_1}-m_\chi$ 
as a function of $m_{1/2}$ along the CMSSM
coannihilation strips calculated using {\tt SSARD}~\protect\cite{SSARD} 
for $\tb = 10$ (left panel) and $\tb = 40$ (right panel), 
and for $A_0 = 0$ (blue lines) and $2.5 \, m_0$ (red lines).
The bands with $m_{\tilde \tau_1}-m_\chi < m_\tau$ are shaded green.
The lower limit on $m_{1/2}$ from the 8-TeV ATLAS 5/fb $\ETslash$ search at the LHC~\protect\cite{ATLAS8}
is represented by maroon lines, and the lower limits from searches for the direct and total
production of metastable charged particles~\protect\cite{ATLASmcp} are shown as dashed and solid
lines, respectively, inside the green bands (see Section 6 for details).
}} 
\label{fig:DeltaM}
\end{figure}

As remarked above, the mass difference $m_{\tilde \tau_1}-m_\chi$
required to bring the relic $\chi$ density within the WMAP range decreases
monotonically as $m_{1/2}$ increases along a coannihilation strip, vanishing
at its tip. Examples of this are displayed in Fig.~\ref{fig:DeltaM}, for $\tb = 10$
in the left panel and for $\tb = 40$  in the right panel, in both cases for the choices $A_0 = 0$ and $2.5 \, m_0$
(solid blue and red lines, respectively). Along these strips, $m_0$ has been chosen
so the neutralino relic density lies within the WMAP range, using calculations with the {\tt SSARD} code~\cite{SSARD}.
The locations of the tips of the strips
vary with both $\tb$ and $A_0$, but
the qualitative behaviour of $m_{\tilde \tau_1}-m_\chi$ as a function of the
distance from the tip is similar for all of the coannihilation strips.

Fig.~\ref{fig:DeltaM} also displays the lower limit $m_{1/2} > 700$~GeV
quoted by the ATLAS Collaboration for the choices $\tan \beta = 10$ 
and $A_0 = 0$ in a recent $\ETslash$ analysis using 5/fb of data from the LHC 
running at 8 TeV~\cite{ATLAS8}. It was shown in~\cite{MC8} that a previous
lower limit from ATLAS using 5/fb of 7-TeV LHC data~\cite{ATLAS7} was largely independent
of $\tan \beta$ and $A_0$. Combining the 7-TeV ATLAS data with other
constraints on the CMSSM, that analysis also found $m_{1/2} \gappeq 700$~GeV
at the $\Delta \chi^2 = 2.3$ level, corresponding to the 68\% CL in the
$(m_0, m_{1/2})$ plane, considering all values of $\tan \beta$ and $A_0$~\cite{MC8}. 

We see that the ATLAS limit intercepts the WMAP strips for $\tb = 10, A_0 = 0 \, (2.5 \, m_0)$ where
$m_{\tilde \tau_1}-m_\chi \sim 2.8 \, (2.5)$~GeV, and the WMAP strips for
$\tb = 40, A_0 = 0 \, (2.5 \, m_0)$ where $m_{\tilde \tau_1}-m_\chi \sim 6.5 \, (9)$~GeV.
We also show, again in green shading, the band where $m_{\tilde \tau_1}-m_\chi < m_\tau$.
The efficiencies and sensitivities of
the current LHC $\ETslash$ search strategies need careful study in the limit
of small $m_{\tilde \tau_1}-m_\chi$, where in the CMSSM also the masses of the
other sleptons are closer to $m_\chi$ than in generic regions of the $(m_0, m_{1/2})$ plane. 
We also note that the sensitivity of $\ETslash$ searches by ATLAS and CMS to points within
the green band may be affected by the fact that events with a massive long-lived ${\tilde \tau_1}$ in the final state
have a different experimental signature. One should, in particular, consider the possibilities
that the ${\tilde \tau_1}$ may decay inside the detector into either one or three light charged
particles, as well as the possibility that it escapes from the detector before decaying.

Clearly, the ATLAS $\ETslash$ search has not yet reached the band where
$m_{\tilde \tau_1}-m_\chi < m_\tau$ along any of these WMAP strips, although it is
very close for $\tan \beta = 10$. Equally clearly, future ATLAS and CMS searches using the full data set at 8~TeV
and higher energies will reach this band sooner or later, depending on the
values of $\tb$ and $A_0$. 

How sacrosanct are the WMAP strips? CMSSM parameter sets lying {\it above} these
strips would give {\it too much} cold dark matter, if one assumes that $R$ parity
is conserved (as in the CMSSM), that the lightest neutralino $\chi$ is the LSP, 
and that the Universe has expanded adiabatically 
since the epoch of thermal dark matter decoupling, with only dark matter and Standard Model 
particles contributing to its expansion (as in conventional Big Bang cosmology). 
Thus, discovery of the $R$-conserving CMSSM with parameters measured to be above the WMAP strip 
would imply that either conventional Big Bang cosmology should be modified or some
other sparticle is the LSP, e.g., the gravitino. Conversely,
CMSSM parameter sets lying {\it below} these strips would give {\it too little} LSP dark matter.
However, in this case one could postulate the presence of some other contribution to the
cold dark matter density, e.g., axions, without having to rethink either Big Bang cosmology
or the nature of the LSP. Therefore, parameter sets below the WMAP strips, in particular
inside the green bands at lower masses, cannot be excluded absolutely. Hence it is relevant to consider the 
sensitivity of current constraints within the green band also at smaller $m_{1/2}$ below the coannihilation strips.

In the following sections, we first study the sensitivity of the ATLAS $\ETslash$ search for
small $m_{\tilde \tau_1}-m_\chi > m_\tau$
in the neighbourhoods of the coannihilation strips, and then the sensitivities of the
LHC $\ETslash$ searches inside the green shaded bands where $m_{\tilde \tau_1}-m_\chi < m_\tau$.

\section{Sensitivity of the ATLAS ~$\ETslash$ Search in the Coannihilation Strip Region}

The stau-neutralino coannihilation strips, in addition to featuring a small mass difference
between the ${\tilde \tau_1}$ and the $\chi$, as seen in Fig.~\ref{fig:DeltaM}, may also
feature relatively small mass differences between the $\chi$ and the other sleptons,
${\tilde e_{R,L}}, {\tilde \mu_{R,L}}$ and ${\tilde \tau_2}$, depending on the values
of $\tan \beta$ and $A_0$ as well as $m_{1/2}$. Since the amount of $\ETslash$
and the transverse momenta of any final-state leptons in events in which the
${\tilde e_{R,L}}, {\tilde \mu_{R,L}}$ and ${\tilde \tau_{1,2}}$ are produced depend on these 
mass differences, it is natural to ask whether there is any impact on the sensitivity of LHC
searches for supersymmetry in the neighbourhood of the coannihilation strips. 

In this connection, we
note that the granularity of the published results of LHC $\ETslash$ searches in an
$(m_0, m_{1/2})$ plane is typically ${\cal O}(50)$~GeV~\cite{ATLAS7,CMS7,ATLAS8,CMS8}, so any fine structure close
to the ${\tilde \tau_1}- \chi$ LSP boundary would not have been apparent. We also note
that the ATLAS 8-TeV $\ETslash$ search has a somewhat reduced reach in $m_{1/2}$ for $m_0 \lappeq 500$~GeV~\cite{ATLAS8}.
For these reasons, we have extended validations~\cite{MC8} of the previous ATLAS $\ETslash$ 
search in simulations based on {\tt PYTHIA} and {\tt Delphes} 
to study in more detail possible effects on the $m_{1/2}$ reach in the coannihilation strip 
region where $m_{\tilde \tau_1}-m_\chi = {\cal O}(5)$~GeV. We concentrate on the
search for events with $\ETslash$ but no identified accompanying leptons, since this
has previously been found to be the most sensitive search at the large values of the ratio $m_{1/2}/m_0$ near the
coannihilation strips.

We have checked the evolution of the sparticle spectrum for CMSSM
scenarios with $m_{1/2} = 700$~GeV (similar to the 95\% CL lower limit given by
ATLAS for small $m_0$~\cite{ATLAS8}), $\tan \beta = 10, 40$ and $A_0 = 0, 2.5 \, m_0$,
varying $m_0$ so that $\Delta m \equiv m_{\tilde \tau_1}-m_\chi$ increases 
from being adjacent to the green band where the ${\tilde \tau_1}$ is long-lived to $\simeq 10$~GeV.
We do not see any notable variations in the sparticle spectrum,
specifically none in the hierarchy of sparticle masses. On the other hand, the hierarchies of
sparticle masses for $\tan \beta = 40$ are rather different from those for $\tan \beta = 10$,
particularly in the $A_0 = 2.5 \, m_0$ case.

\begin{figure}[h!]
\begin{minipage}{8in}
\hspace{3.7cm}
\epsfig{file=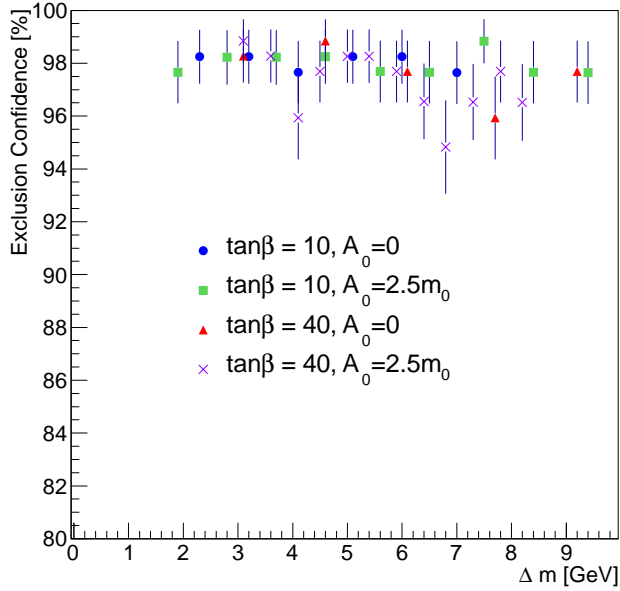,height=5in}
\hfill
\end{minipage}
\caption{
{\it Estimates using {\tt PYTHIA} and {\tt Delphes} simulations
of the exclusion sensitivity of the ATLAS 5/fb $\ETslash$ search along the
line in the $(m_0, m_{1/2})$ found in~\cite{ATLAS8} to yield 95\% CL exclusion for $\tan \beta = 10$ and $A_0 = 0$.
Our simulations are for $\tan \beta = 10$ and $40$ and the choices $A_0 = 0$ and $2.5 \, m_0$, 
and are plotted as functions of
$\Delta m \equiv m_{\tilde \tau_1}-m_\chi$. We recall that for $\tan \beta = 10$
the WMAP-compatible coannihilation strip has $\Delta m \sim 3$~GeV,
whereas for $\tan \beta = 40$ it has $\Delta m \sim 6.5$ to $9$~GeV. 
}} 
\label{fig:sensitivity}
\end{figure}

Fig.~\ref{fig:sensitivity} displays the exclusion confidence levels we estimate
using {\tt PYTHIA}~\cite{PYTHIA} and {\tt Delphes}~\cite{Delphes} to simulate the 8-TeV ATLAS 5/fb $\ETslash$ search
for a selection of scenarios with $m_{1/2} = 700$~GeV, the same values $\tan \beta = 10, 40$ and
$A_0 = 0, 2.5 \, m_0$ and varying $\Delta m \lappeq 10$~GeV.
In these cases, the WMAP-compatible coannihilation strip has $m_{\tilde \tau_1}-m_\chi \sim 2.5$~GeV
to $9$~GeV, as seen in Fig.~\ref{fig:DeltaM}. We see no significant differences, outside the statistical 
uncertainties in our analysis of the exclusion confidence levels, between simulations for different values 
of $\tan \beta$ and $A_0$. This is in agreement with the
previous analysis in~\cite{MC8} but now with finer granularity in the relevant region of small 
$m_{\tilde \tau_1}-m_\chi$~\footnote{Since this analysis is similar to
the {\tt PYTHIA} and {\tt Delphes} validation described in~\cite{MC8}, we do not discuss details here.}.

On the basis of this study, we assume for our purposes here
that the 95\% CL limit in the $(m_0, m_{1/2})$ plane
established by ATLAS~\cite{ATLAS8} on the basis of an $\ETslash$ 
analysis with 5/fb of 8-TeV data may be extended through the WMAP coannihilation
strip range of $m_{\tilde \tau_1} - m_\chi$ down to the band close to the LSP boundary
where the ${\tilde \tau_1}$ becomes long-lived, simply by assuming that
the results are independent of $m_{1/2}$ in this region.
The vertical maroon lines in Fig.~\ref{fig:DeltaM} demonstrate
the corresponding impact of the $\ETslash$ constraint on the CMSSM
coannihilation strips discussed earlier. 

We see that a significant fraction of the remaining portion of the coannihilation 
strip not excluded at the 95\% CL by the above analysis has $\Delta m < m_\tau$,
in which case the ${\tilde \tau_1}$ is long-lived. This conclusion
is particularly strong for $\tan \beta = 10$. The experimental sensitivity to
$m_{1/2}$ requires a separate, dedicated analysis in the low-$\Delta m$
regions, as we now discuss.

\begin{figure}[h!]
\vskip -2.5in
\hspace*{-.70in}
\begin{minipage}{8in}
\epsfig{file=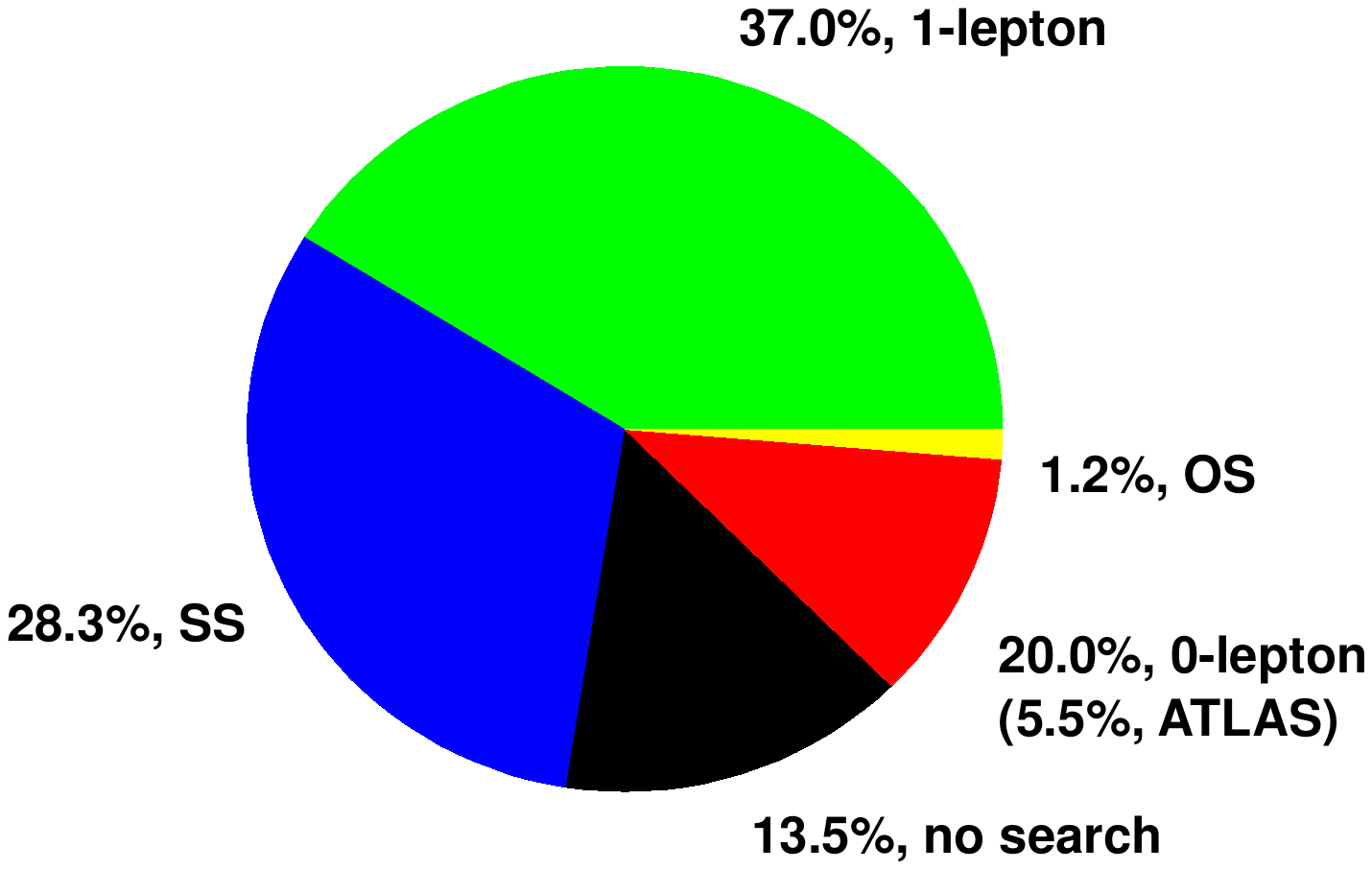,height=5.5in}
\hspace{-2cm}
\epsfig{file=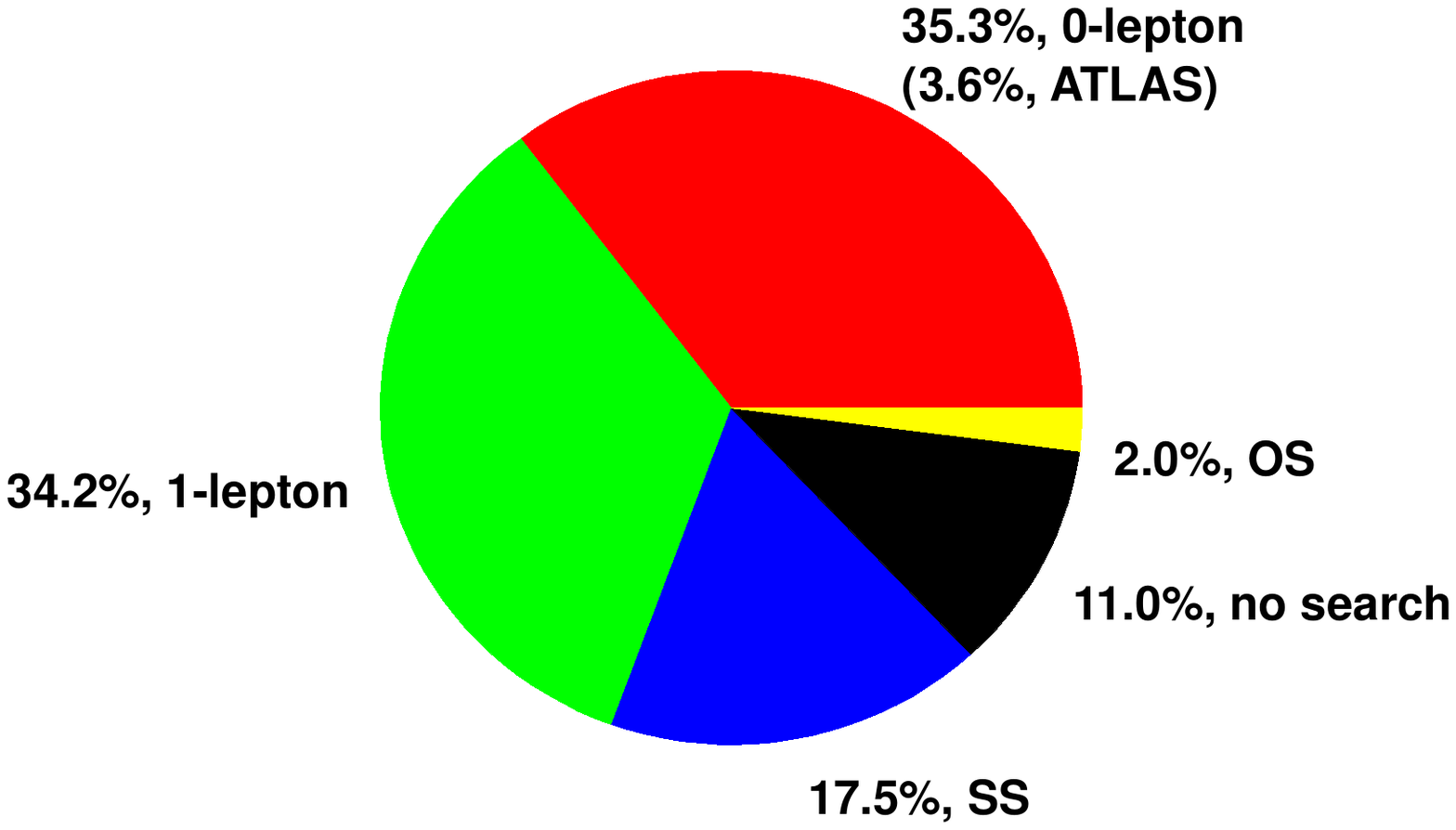,height=5.5in}
\end{minipage}
\vspace*{-0.7in}
\caption{
{\it Pie charts displaying the fractions of 100,000 events generated using {\tt PYTHIA}
with parameters corresponding to the CMSSM best-fit point with $\Delta m < m_\tau$ 
(left) and an analogous point with $m_{\tilde \tau_1}$ increased so that
$\Delta m > m_\tau$. Indicated in parentheses are the fractions of these events that
survive the ATLAS triggers and selections for zero-lepton + $\ETslash$ events
according to simulations using {\tt Delphes}.
}} 
\label{fig:piecharts}
\end{figure}

When $\Delta m > m_\tau$, supersymmetric cascades are likely to include ${\tilde \tau_1} \to
\tau \chi$ decays followed by $\tau$ decays that are mainly hadronic.
On the other hand, if $\Delta m < m_\tau$, the ${\tilde \tau_1}$ decays would not
occur promptly, and the ${\tilde \tau_1}$ would not be registered as a conventional physics object. 
This change in the pattern of cascade decays is reflected in the populations of different experimental 
signatures, as shown in Fig.~\ref{fig:piecharts}. The pie charts show the fractions of a pair of 100,000
supersymmetric events generated in {\tt PYTHIA} simulations that yield $\ETslash$ events 
with no leptons (red segments) and with single leptons (green segments), as well
as same- and opposite-sign dilepton events (SS, blue segments; OS, yellow segments),
and events with no distinctive experimental signature (black segments). Comparing
the simulation of the best-fit point (left) that has $\Delta m < m_\tau$ 
with that of a point that has $m_{\tilde \tau_1}$ increased so as to
lie just outside the green band of small $\Delta m$ (right), we see that the
best-fit point has a larger fraction of same-sign dilepton events (28.3\% vs 17.5\%) and a smaller fraction
of zero-lepton $\ETslash$ events (20.0\% vs 35.3\%)~\footnote{For completeness,
we note that the fractions of one-lepton events are similar in the two cases (37.0\% vs
34.2\%).}. These differences are due essentially to the disappearance of
hadronic $\tau$ decays from the supersymmetric cascades mentioned above.
{\it A priori}, this raises the prospect that the efficiency and hence the sensitivity
of the ATLAS $\ETslash$ searches discussed above would be reduced inside the green
band where $\Delta m < m_\tau$.

However, we then passed the event samples simulated using {\tt PYTHIA} through {\tt Delphes}
to estimate which fractions of the generated events would pass the ATLAS trigger and event
selection for the zero-lepton + $\ETslash$ category, assuming that the long-lived ${\tilde \tau_1}$
would escape the detector as a massive charged particle before decaying. The results are shown in parentheses in
Fig.~\ref{fig:piecharts}: 5.5\% of the simulated best-fit events survived the simulated ATLAS
analysis chain, compared with 3.6\% of the events with larger $\Delta m$.
This result indicates that the sensitivity of the ATLAS zero-lepton + $\ETslash$ search
for a low-$\Delta m$ point inside the green band is at least as high as for a point just
outside the green band. This indicates that the (maroon) LHC limit in Fig.~\ref{fig:DeltaM}
may (conservatively) be extrapolated into the green band at low $\Delta m$.

However, a more careful consideration of prospective experimental signatures, and hence sensitivity, in the
region where $\Delta m < m_\tau$ requires a discussion of the ${\tilde \tau_1}$ decay
lifetime and branching ratios, to which we now turn our attention.

\section{Lifetime and Branching Ratios for ${\tilde \tau_1}$ Decay}

As was discussed
in~\cite{Jittoh}, if $\Delta m \equiv m_{\tilde \tau_1} - m_\chi > m_\tau$ the
dominant ${\tilde \tau_1}$ decay is two-body, namely ${\tilde \tau_1} \to \tau \chi$,
which occurs promptly with such a short lifetime that no ${\tilde \tau_1}$ track
is detectable, as assumed above. However, if $\Delta m < m_\tau$
the dominant decays are three- and four-body,
so the ${\tilde \tau_1}$ lifetime is much longer, and it may decay either
inside or outside the detector.

\begin{figure}
\vskip 0.5in
\vspace*{-0.75in}
\begin{minipage}{8in}
\epsfig{file=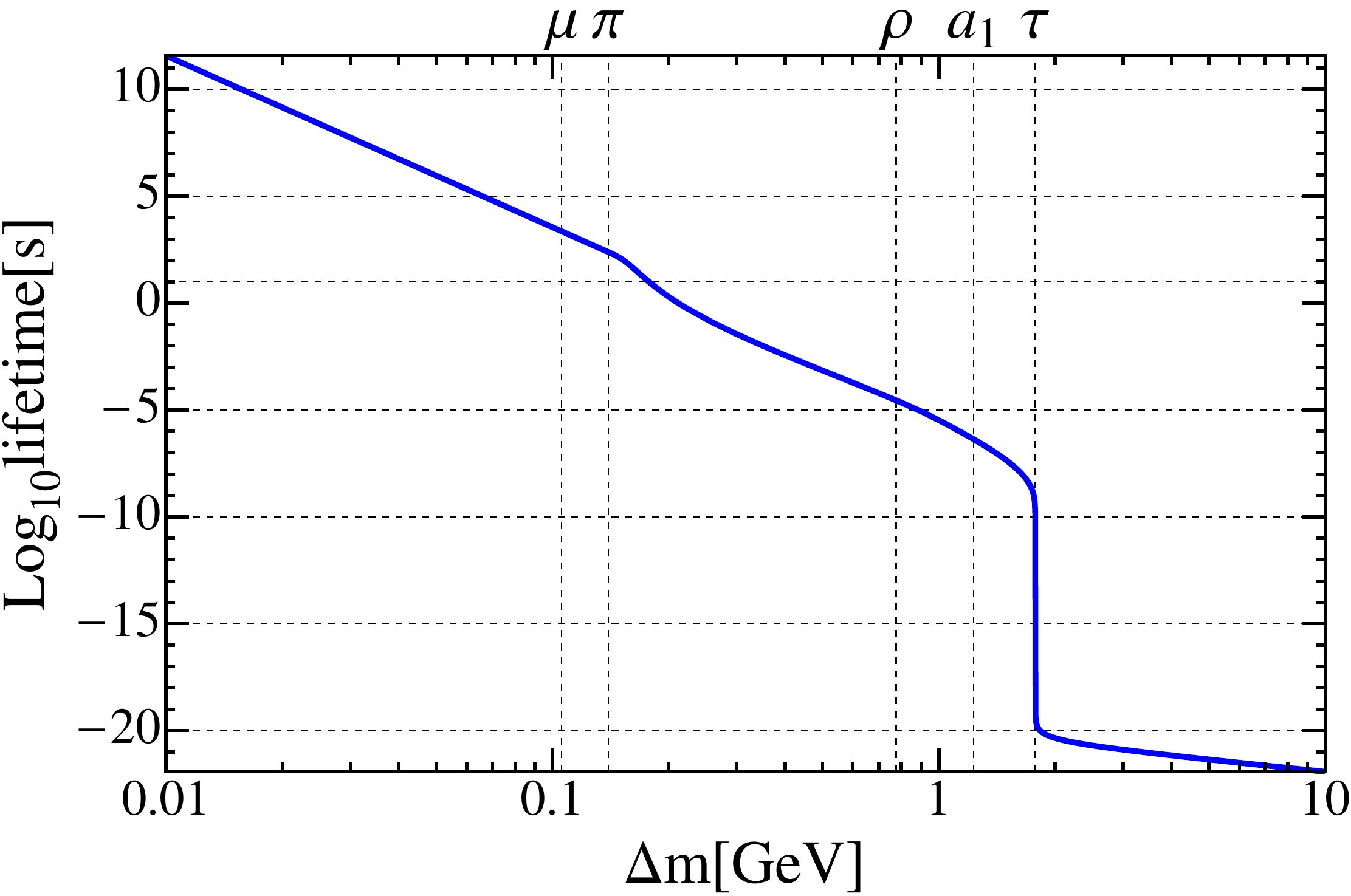,height=2.1in}
\epsfig{file=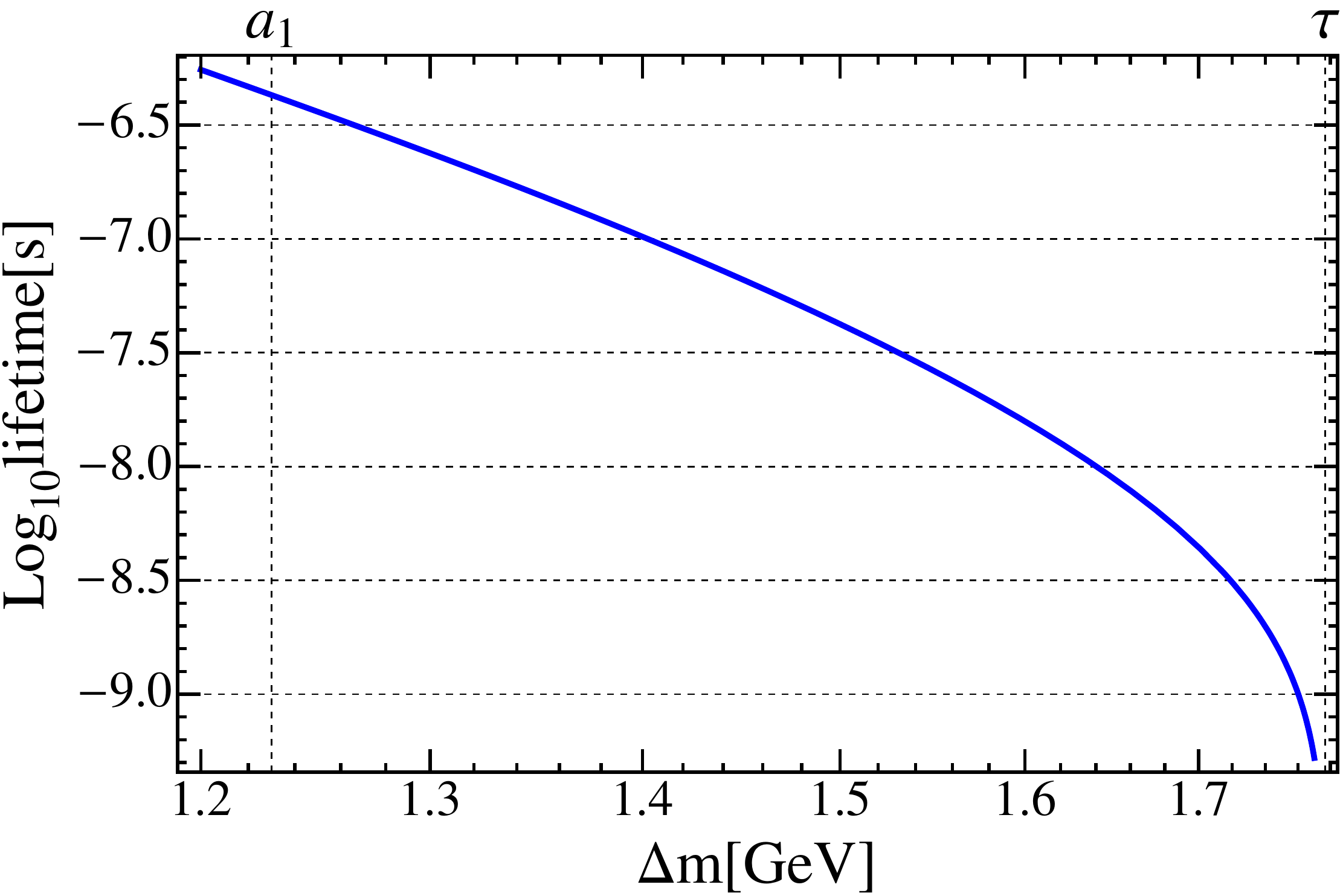,height=2.1in}
\hfill
\end{minipage}
\caption{
{\it The ${\tilde \tau_1}$ lifetime calculated for $m_{\tilde \tau_1} = 300$~GeV and a
${\tilde \tau_L} - {\tilde \tau_R}$ mixing angle $\theta_\tau = \pi/3$, as a function of
$\Delta m \equiv m_{\tilde \tau_1} - m_\chi$. The left panel covers the range
$10~{\rm MeV} < \Delta m < 10~{\rm GeV}$ where the lifetime is between $\sim 10^{12}$ and $\sim 10^{-22}$~s, 
and the right panel shows in more detail the restricted
range $1.2~{\rm GeV} < \Delta m < m_\tau$ where the lifetime is between $\sim 1$ and $\sim 400$~ns.
The vertical dashed lines correspond to the $\tau$, $a_1$, $\rho$, $\pi$ and $\mu$ masses, 
indicated by the labels on the top of the figures.
}} 
\label{fig:ourlifetime}
\end{figure}

We have recalculated the ${\tilde \tau_1}$ lifetime for the same supersymmetric model
parameters as assumed in~\cite{Jittoh}, namely $m_{\tilde \tau_1} = 300$~GeV and a
${\tilde \tau_L} - {\tilde \tau_R}$ mixing angle $\theta_\tau = \pi/3$~\footnote{Details of our
calculation are given in the Appendix, where we also
discuss the aspects of our calculation that differ from that of~\cite{Jittoh}.}.
We display our result in Fig.~\ref{fig:ourlifetime} as a function of $\Delta m \equiv
m_{\tilde \tau_1} - m_\chi$. As one would expect, the ${\tilde \tau_1}$ decays promptly
with a lifetime $\lappeq 10^{-20}$~s if $\Delta m > m_\tau$. On the other hand, as seen
in more detail in the right panel of Fig.~\ref{fig:ourlifetime}, when $m_\tau > \Delta m > 1.2$~GeV the
${\tilde \tau_1}$ lifetime is between 1 and 400~ns, corresponding to a significant likelihood
of observing the ${\tilde \tau_1}$ decay inside an LHC detector, as we discuss below.
We note in passing that, whereas the total ${\tilde \tau_1}$ decay rate is very sensitive to
$\Delta m$ (typically $\sim \Delta m^5$ or more), it is much less sensitive to $m_{\tilde \tau_1}$
($\propto 1/m_{\tilde \tau_1}$), and hence the results in Fig.~\ref{fig:ourlifetime} are typical of
the range of $m_{1/2}$ likely to be of interest to the LHC experiments in the near future.

\begin{figure}
\vskip 0.5in
\begin{minipage}{8in}
\epsfig{file=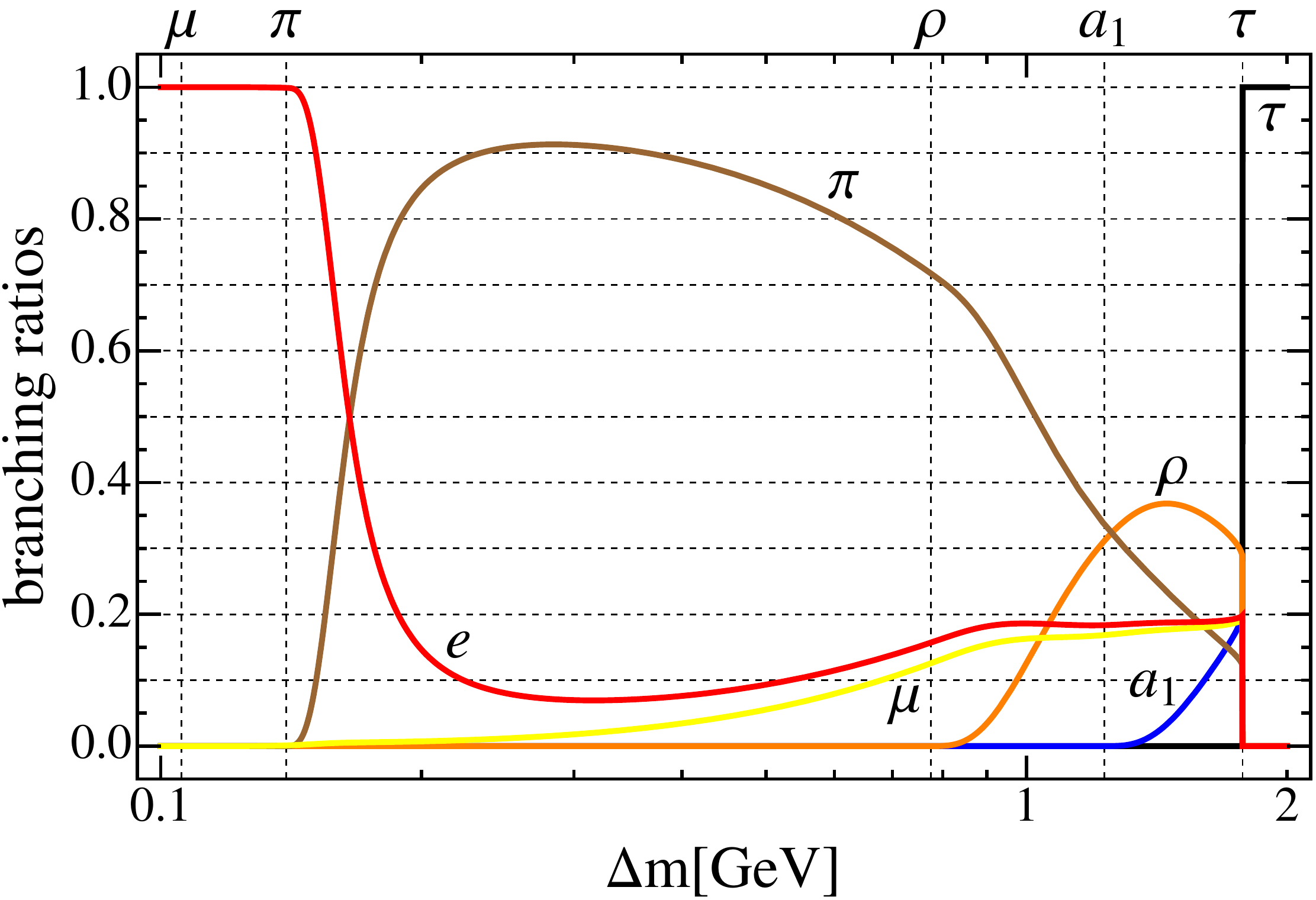,height=2.1in}
\epsfig{file=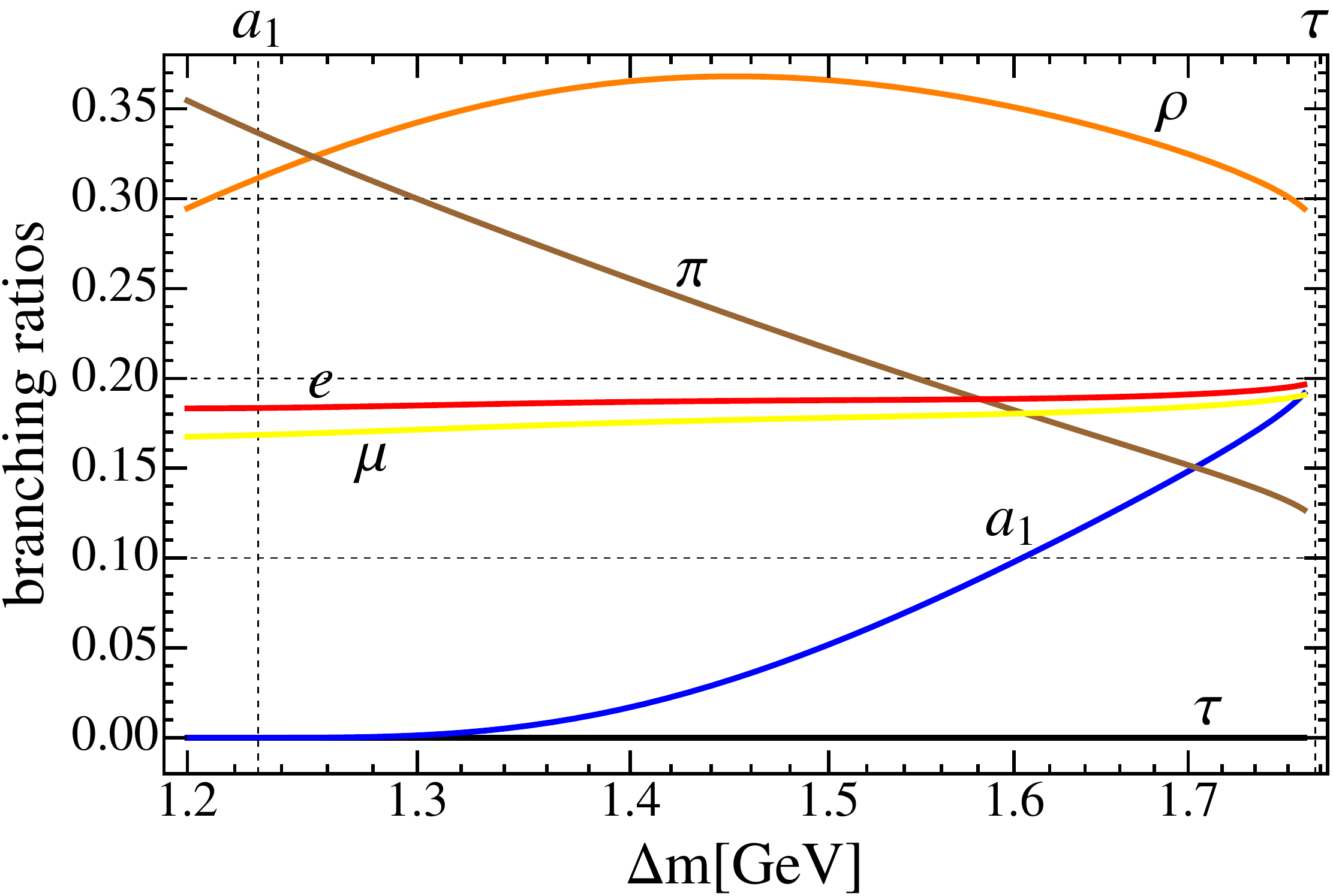,height=2.1in}
\hfill
\end{minipage}
\caption{
{\it The principal ${\tilde \tau_1}$ branching ratios calculated for $m_{\tilde \tau_1} = 300$~GeV and a
${\tilde \tau_L} - {\tilde \tau_R}$ mixing angle $\theta_\tau = \pi/3$, as a function of
$\Delta m \equiv m_{\tilde \tau_1} - m_\chi$. The left panel covers the range
$100~{\rm MeV} < \Delta m < 2~{\rm GeV}$, and the right panel shows in more detail the restricted
range $1.2~{\rm GeV} < \Delta m < m_\tau$. The black, blue, orange, brown, yellow, and red lines 
are for the final states with $\tau$, $a_1(1260)$, $\rho(770)$, $\pi$, $\mu$, and $e$, respectively,
indicated by the labels adjacent to the corresponding curves. 
}} 
\label{fig:BRs}
\end{figure}

What would be the experimental  signature of ${\tilde \tau_1}$ decay inside an LHC
detector? In Fig.~\ref{fig:BRs} we show results of our calculations of the dominant ${\tilde \tau_1}$
decay branching ratios. As expected, the dominant branching ratio for $\Delta m > m_\tau$
is the two-body decay ${\tilde \tau_1} ^-\to \tau^- \chi$. In the range $m_\tau > \Delta m \gappeq 0.8$~GeV,
there is competition among the three-body decays ${\tilde \tau_1}^- \to \pi^- \nu_\tau \chi$,
${\tilde \tau_1}^- \to \rho^- \nu_\tau \chi$ and ${\tilde \tau_1}^- \to a_1^- \nu_\tau \chi$ (which were not
considered in~\cite{Jittoh}), and the four-body decays ${\tilde \tau_1}^- \to e^- {\bar \nu_e} \nu_\tau \chi$
and ${\tilde \tau_1}^- \to \mu^- {\bar \nu_\mu} \nu_\tau \chi$. At lower $\Delta m$, the decay
${\tilde \tau_1}^- \to \pi^- \nu_\tau \chi$ is dominant for $\Delta m \gappeq 0.16$~GeV, and then
${\tilde \tau_1}^- \to e^- {\bar \nu_e} \nu_\tau \chi$ at $\Delta m \lappeq 0.16$~ GeV. A general conclusion, then,
is that four potential signatures may be of interest to the LHC experiments, namely decays
producing $e^-$, $\mu^-$, $\pi^-$ (perhaps accompanied by one or more $\pi^0$ mesons from $\rho^-$ or
$a_1^-$ decays), and $\pi^- \pi^+ \pi^-$ from $a_1^-$ decays. 

The right panel of Fig.~\ref{fig:BRs} displays in more detail the dominant branching ratios
in the mass range $m_\tau > \Delta m > 1.2$~GeV where, according to our previous
discussion, the lifetime $\lappeq 400$~ns, and the decay 
signatures are most likely to be of interest to the LHC experiments.
Over most of this range of $\Delta m$, the decay ${\tilde \tau_1}^- \to \rho^- \nu_\tau \chi$
is expected to dominate, with the next most important 
decay modes being ${\tilde \tau_1}^- \to \pi^- \nu_\tau \chi$,
${\tilde \tau_1}^- \to e^- {\bar \nu_e} \nu_\tau \chi$ and ${\tilde \tau_1}^- \to \mu^- {\bar \nu_\mu} \nu_\tau \chi$.
The branching ratio for the mode ${\tilde \tau_1}^- \to a_1^- \nu_\tau \chi$ exceeds 5\% for
$\Delta m \gappeq 1.5$~GeV, indicating that the $\pi^- \pi^+ \pi^-$ decay signature would have
a branching ratio of a few \% in this mass range.

In the following, we discuss first and primarily the case where the ${\tilde \tau_1}$
is sufficiently long-lived to escape from the detector, returning subsequently and
more briefly to the case of ${\tilde \tau_1}$ decay inside the detector.

\section{LHC Limits on Long-Lived Massive Charged Particles}

So far, the most stringent LHC limits on the production of long-lived massive charged particles
have been published by the ATLAS Collaboration, based on 4.7/fb of data at 7 TeV in
the centre of mass~\cite{ATLASmcp}. Upper limits are given on both the total production of
metastable charged particles and on their direct production. In the former case
comparison is made with the total cross section expected in a gauge-mediated
supersymmetry-breaking (GMSB) scenario, and in the latter case with the cross section
for producing directly three light slepton flavours, calculated in both cases using {\tt PROSPINO}~\cite{PROSPINO}.

\begin{figure}[h!]
\vspace*{-1.25in}
\begin{minipage}{8in}
\epsfig{file=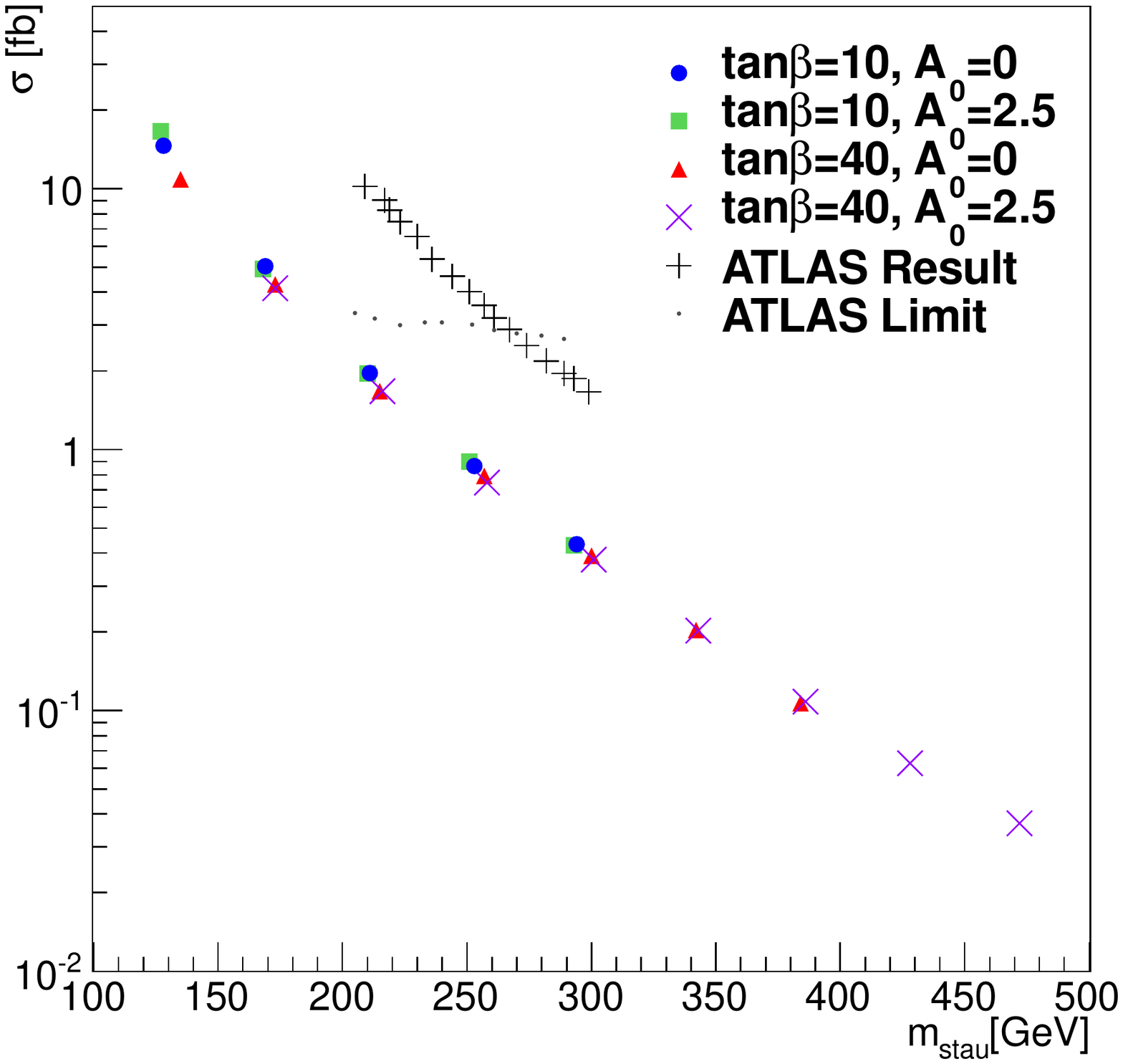,height=4.3in}
\epsfig{file=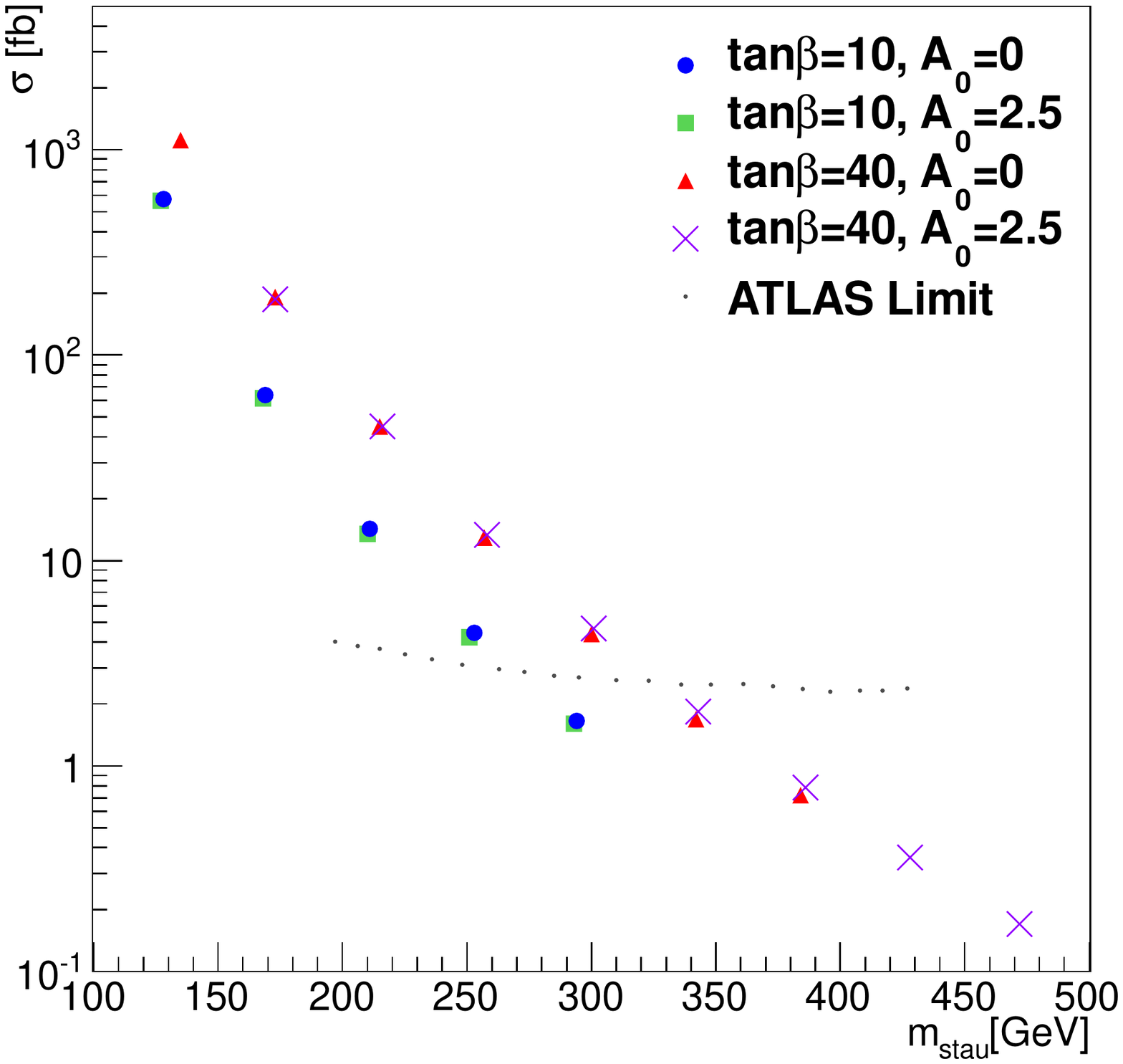,height=4.3in}
\hfill
\end{minipage}
\caption{
{\it The direct (left panel) and total (right panel) ${\tilde \tau_1}$ production cross sections as
functions of $m_{\tilde \tau_1}$ along the boundaries between the $\chi$ and ${\tilde \tau_1}$ LSP regions in the CMSSM
 for $\tb = 10, A_0 = 0$ (blue circles) and $2.5 \, m_0$ (green squares), and 
$\tb = 40, A_0 = 0$ (red triangles) and $2.5 \, m_0$ (mauve crosses). Also shown as black dots are the upper limits on the
corresponding cross sections from the ATLAS Collaboration~\protect\cite{ATLASmcp} and (in the left panel only)
the result of the ATLAS calculation of the cross section for three light slepton flavours.
}} 
\label{fig:ATLASMCP}
\end{figure}

We use here ${\tilde \tau_1}$ production cross section calculations obtained from {\tt PYTHIA}. We see
in the left panel of Fig.~\ref{fig:ATLASMCP} that, as expected,
the {\it direct} production cross section is essentially independent of both $\tan \beta$ and $A_0$ 
within the range explored ($\tan \beta = 10, 40, A_0 = 0, 2.5 \, m_0$). 
We find that {\tt PYTHIA} agrees to within $\sim 20$\% with the 
{\tt PROSPINO} calculation for three light slepton flavours used by ATLAS
(shown as black crosses), with a difference due
to the additional QCD corrections incorporated in {\tt PROSPINO}. However, in the type of CMSSM
scenario we discuss, most of the ${\tilde e}_R, {\tilde \mu}_R$ decays do not lead to
a ${\tilde \tau_1}$, and the same is true for the decays of the heavier ${\tilde e}_L, {\tilde \mu}_L$
and ${\tilde \tau_2}$. Therefore, in our case effectively only one light slepton flavour contributes
to the direct production cross section, namely the ${\tilde \tau_1}$ itself,
rather than the three flavours considered by ATLAS. Thus, in our CMSSM scenario the effective cross section for
the direct production of metastable charged particles shown in the left panel of Fig.~\ref{fig:ATLASMCP}
is a factor $\sim$ 3 smaller than assumed by ATLAS.
For this reason, the lower limit on $m_{\tilde \tau_1}$ from the absence of direct production
is reduced from the value of 278~GeV quoted by ATLAS to $\sim 170$~GeV in the CMSSM,
corresponding to $m_{1/2} \gappeq 400$~GeV. This lower limit is displayed as dashed
lines in the green bands in Fig.~\ref{fig:DeltaM}.

On the other hand, as seen in the right panel of Fig.~\ref{fig:ATLASMCP},
the {\tt PYTHIA} calculation of the {\it total} cross section for ${\tilde \tau_1}$
production in the CMSSM yields values that depend significantly on $\tan \beta$ (and slightly on $A_0$).
The larger production rate for $\tan \beta = 40$
is because the ${\tilde \tau_1}$ is significantly lighter than the ${\tilde e_R}, {\tilde \mu_R}$ 
at large $\tan \beta$, leading to increases in the cascade decay branching ratios
producing ${\tilde \tau_1}$ relative to those producing ${\tilde e_R}, {\tilde \mu_R}$.
The difference in the total cross section implies a difference in the lower limit
on $m_{\tilde \tau_1}$ that can be obtained from the ATLAS search: $m_{\tilde \tau_1} \gappeq 270$~GeV
for $\tan \beta = 10$, corresponding to $m_{1/2} \gappeq 630$~GeV, and 
$m_{\tilde \tau_1} \gappeq 330$~GeV for $\tan \beta = 40$, corresponding to $m_{1/2} \gappeq 730$~GeV.

These lower limits on $m_{1/2}$ from the upper limits on the direct and total cross sections, which are applicable in our CMSSM
scenario for $m_{\tilde \tau_1}-m_\chi < m_\tau$, are indicated as dashed and solid lines, respectively, within the green bands in
Fig.~\ref{fig:DeltaM}. We note that they complement the lower limits from $\ETslash$
searches. In particular, though the direct limit is weaker than the $\ETslash$ limit on $m_{1/2}$,
the total production limit has comparable reach.
We emphasize that the metastable particle searches exclude points in the green band that yield an LSP density
below the estimate of the cold dark matter density based on the data from  WMAP {\it et al}.,
which would have been allowed in the presence of some other contribution to the cold dark
matter density. In the cases studied, the excluded region of the green band does not quite
extend to the strip where the neutralino LSP provides all the cold dark matter density indicated
by WMAP {\it et al.}. 

The vertical lines in the green bands in Fig.~\ref{fig:DeltaM} extend
from $\Delta m = 0$ only to 1.2~GeV, since our calculation shown in
the left panel of Fig.~\ref{fig:ourlifetime} suggests that the ${\tilde \tau_1}$ has a significant
probability of decaying inside the ATLAS or CMS detector if $1.2~{\rm GeV} < \Delta m < m_\tau$,
as we discuss below in more detail.

\section{Prospects for Searches for Long-Lived Staus}

We now discuss the possible experimental signatures of long-lived staus in the green band of interest
described above, with a view to the perspectives for future searches. To this end, we
focus on two sample scenarios: one with $\tan \beta = 10, A_0 = 0$ and the other with
$\tan \beta = 40, A_0 = 2.5 \, m_0$, both with $m_{\tilde \tau_1} \simeq 270$~GeV and
$\Delta m = m_{\tilde \tau_1}-m_\chi < m_\tau$~\footnote{The production kinematics are essentially
independent of $\Delta m$ over this range.}.

\begin{figure}[h!]
\vspace*{-2in}
\begin{minipage}{8in}
\epsfig{file=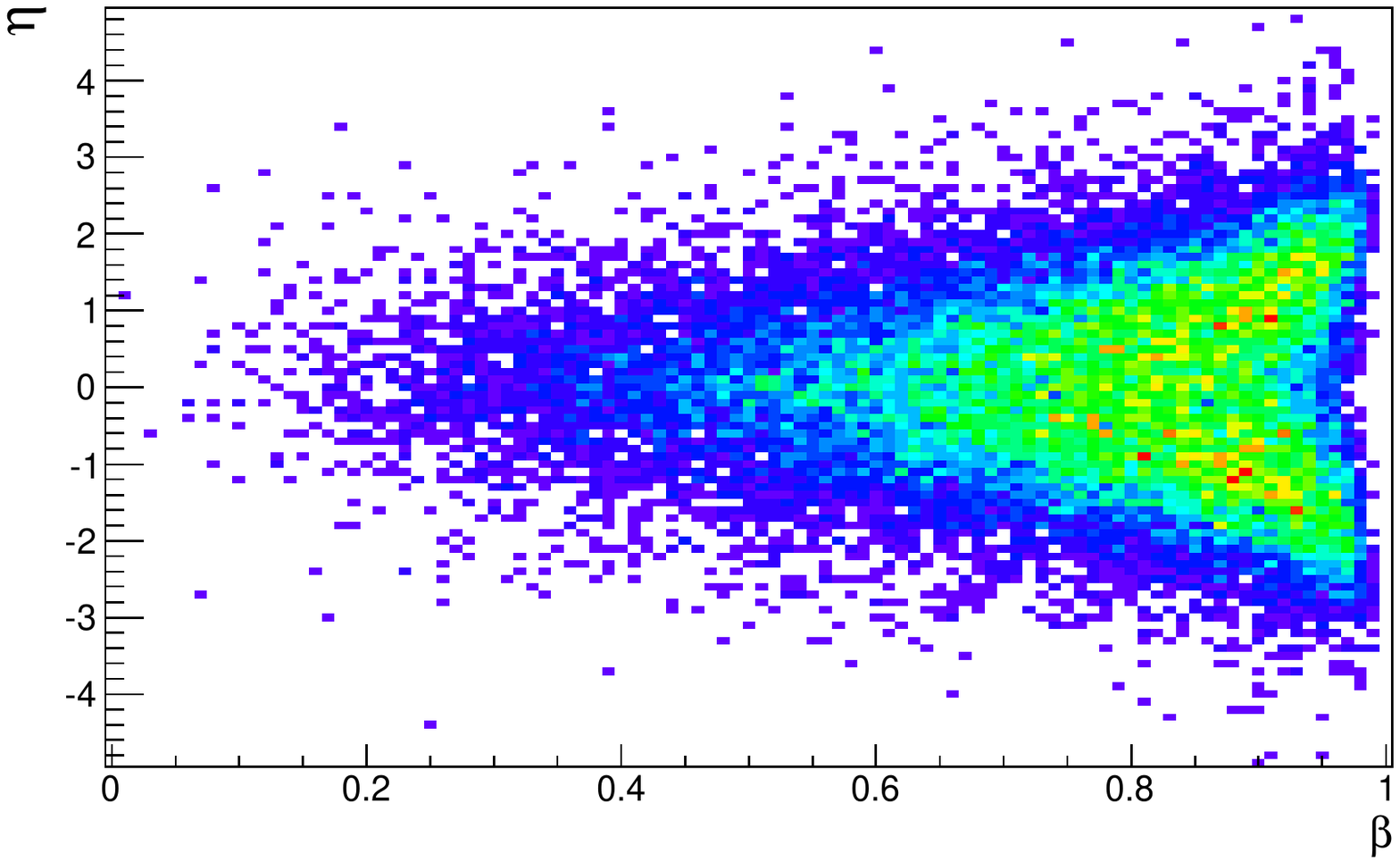,height=4.3in}
\epsfig{file=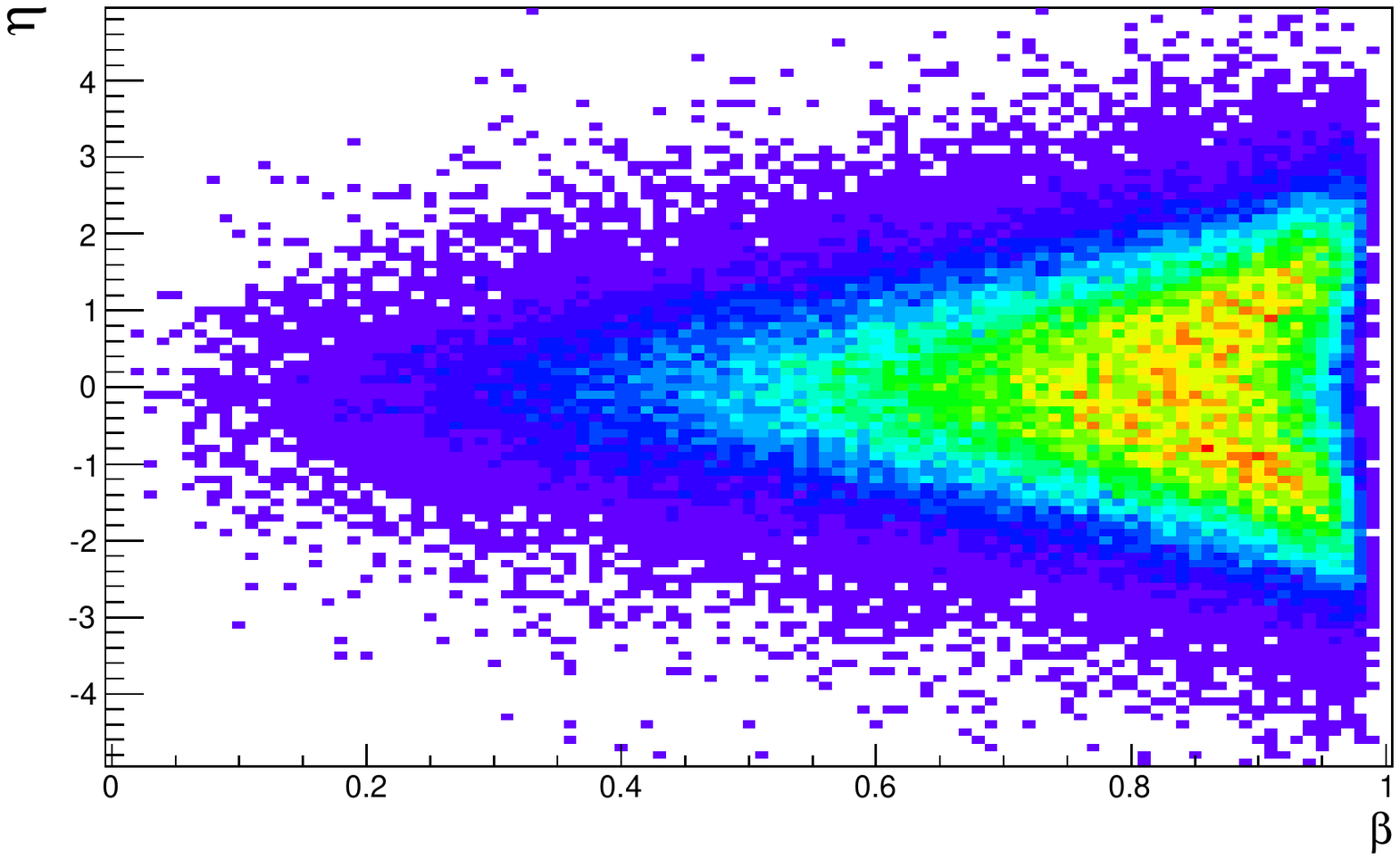,height=4.3in}
\hfill
\end{minipage}
\caption{
{\it Scatter plots of stau production events
in the $(\beta, \eta)$ plane, where $\beta$ is the ${\tilde \tau_1}$ production velocity
divided by $c$, and $\eta$ is its pseudo-rapidity at production. The left and right panels are for
$\tb = 10, A_0 = 0$ and $\tb = 40, A_0 = 2.5 \, m_0$, respectively. In both cases, we
use {\tt PYTHIA} to generate event samples with $m_{\tilde \tau_1} \simeq 270$~GeV and
$m_{\tilde \tau_1}-m_\chi < m_\tau$.
}} 
\label{fig:betaeta}
\end{figure}

Fig.~\ref{fig:betaeta} displays scatter plots of stau production events generated 
using {\tt PYTHIA} in the $(\beta, \eta)$ plane, where $\beta$ is the ${\tilde \tau_1}$ production velocity
divided by $c$, and $\eta$ is its pseudo-rapidity at production. 
These plots are for the $\tan \beta = 10, A_0 = 0$ case (left panel) and the 
$\tan \beta = 40, A_0 = 2.5 \, m_0$ case (right panel), and each contains $\simeq 100,000$ events. We recall that
a typical experimental range is $|\eta| < 2.5$, which includes the great majority of the ${\tilde \tau_1}$ events generated.

\begin{figure}[h!]
\vspace*{-2in}
\begin{minipage}{8in}
\epsfig{file=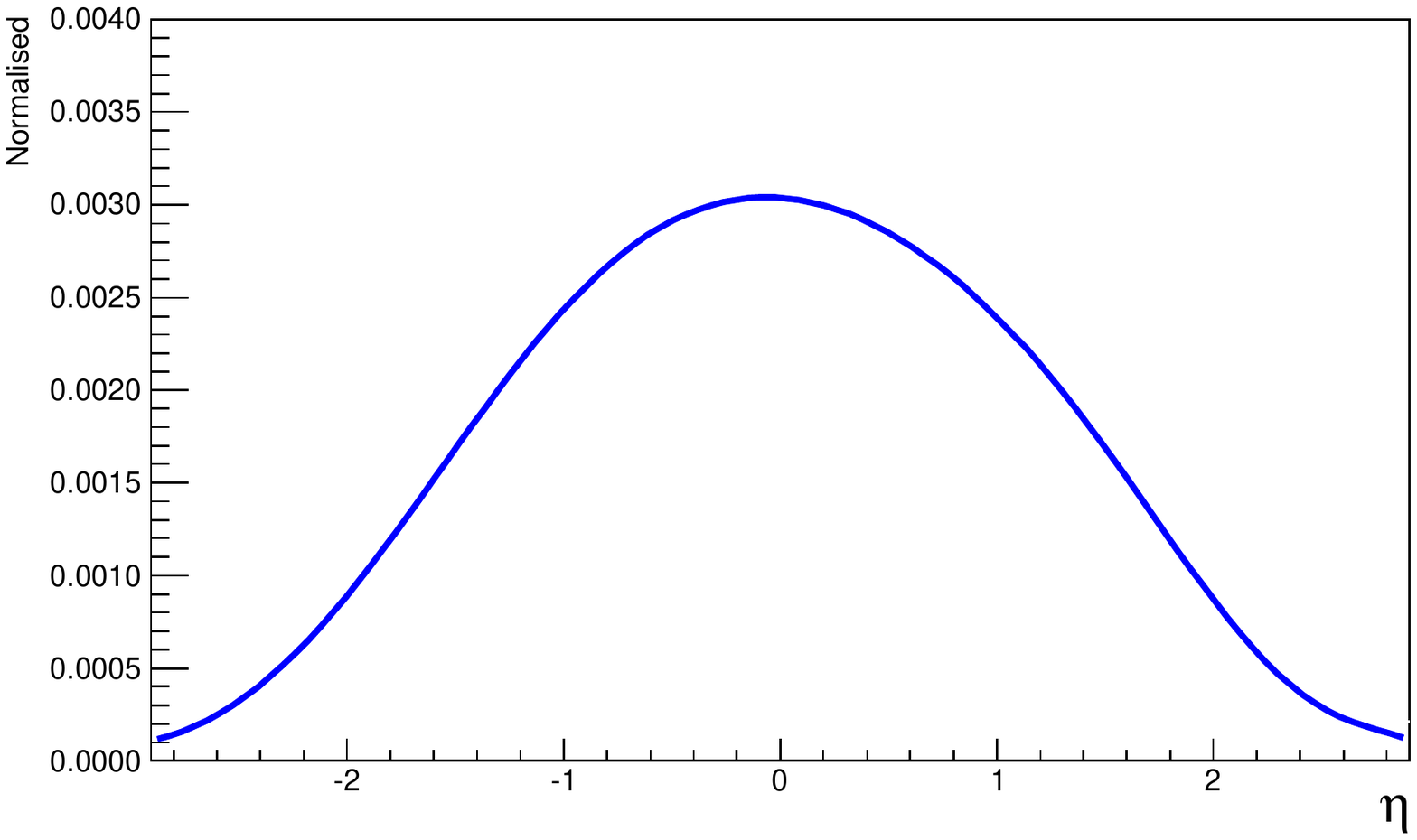,height=4.3in}
\epsfig{file=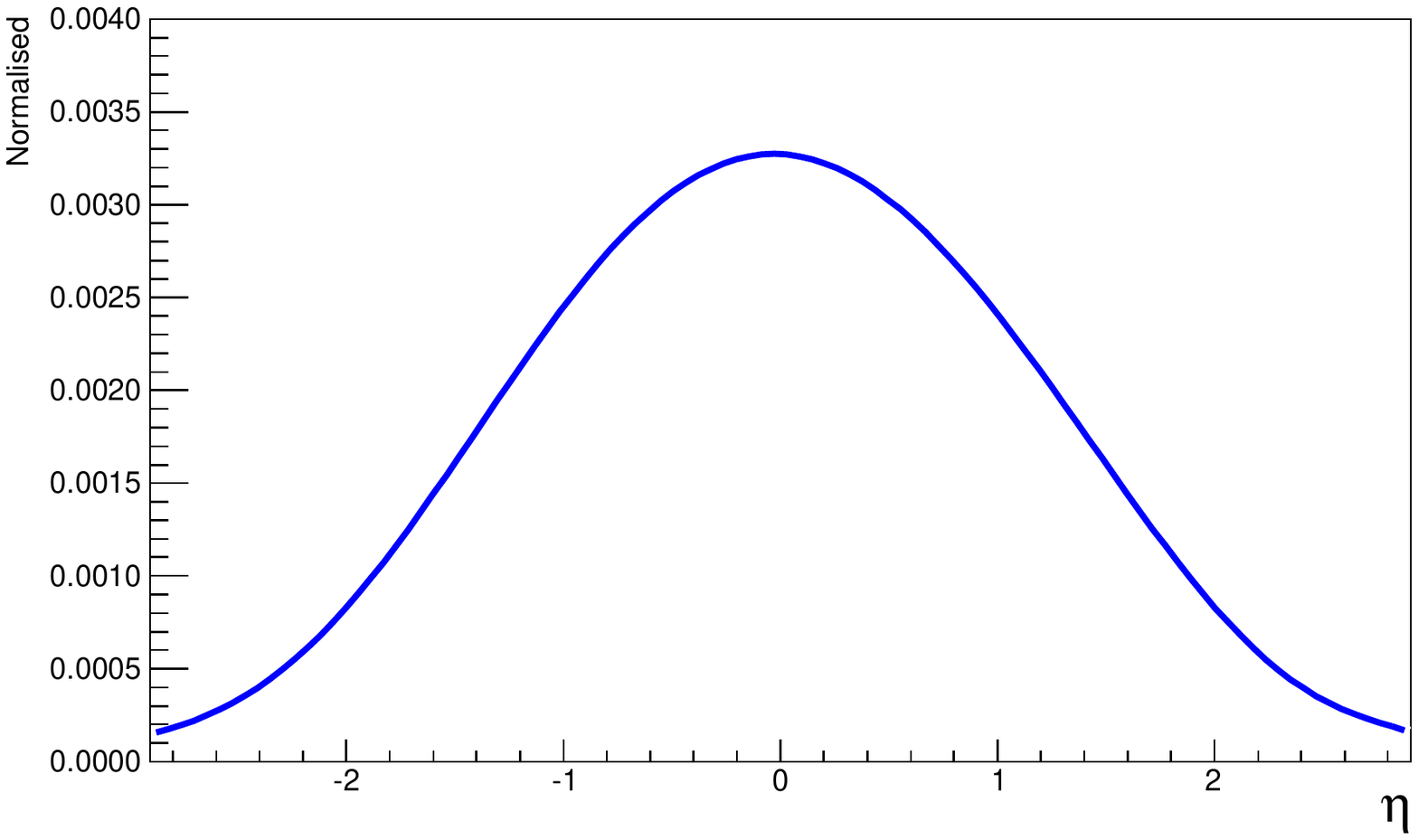,height=4.3in}
\hfill
\end{minipage}
\caption{
{\it The distributions in $\eta$ of the same ${\tilde \tau_1}$
samples as in Fig.~\ref{fig:betaeta} in the left and right panels, respectively.
}} 
\label{fig:etabeta}
\end{figure}

Fig.~\ref{fig:etabeta} displays the integrated $\eta$ distributions
for the ${\tilde \tau_1}$ in these two sample scenarios,
showing how they are peaked around $\eta = 0$ with only small tails beyond $|\eta| = 2.5$.
There is a very slight tendency for the $\eta$ distribution of the ${\tilde \tau_1}$ to be more 
sharply peaked around zero in the $\tan \beta = 40, A_0 = 2.5 \, m_0$ 
case (right panel), with an RMS spread $\Delta \eta = 1.1$ compared with 1.2
for the $\tan \beta = 10, A_0 = 0$ case (left panel),
reflecting a greater preponderance of ${\tilde \tau_1}$ production via cascade decays.

\begin{figure}
\vspace*{-1.5in}
\begin{minipage}{8in}
\epsfig{file=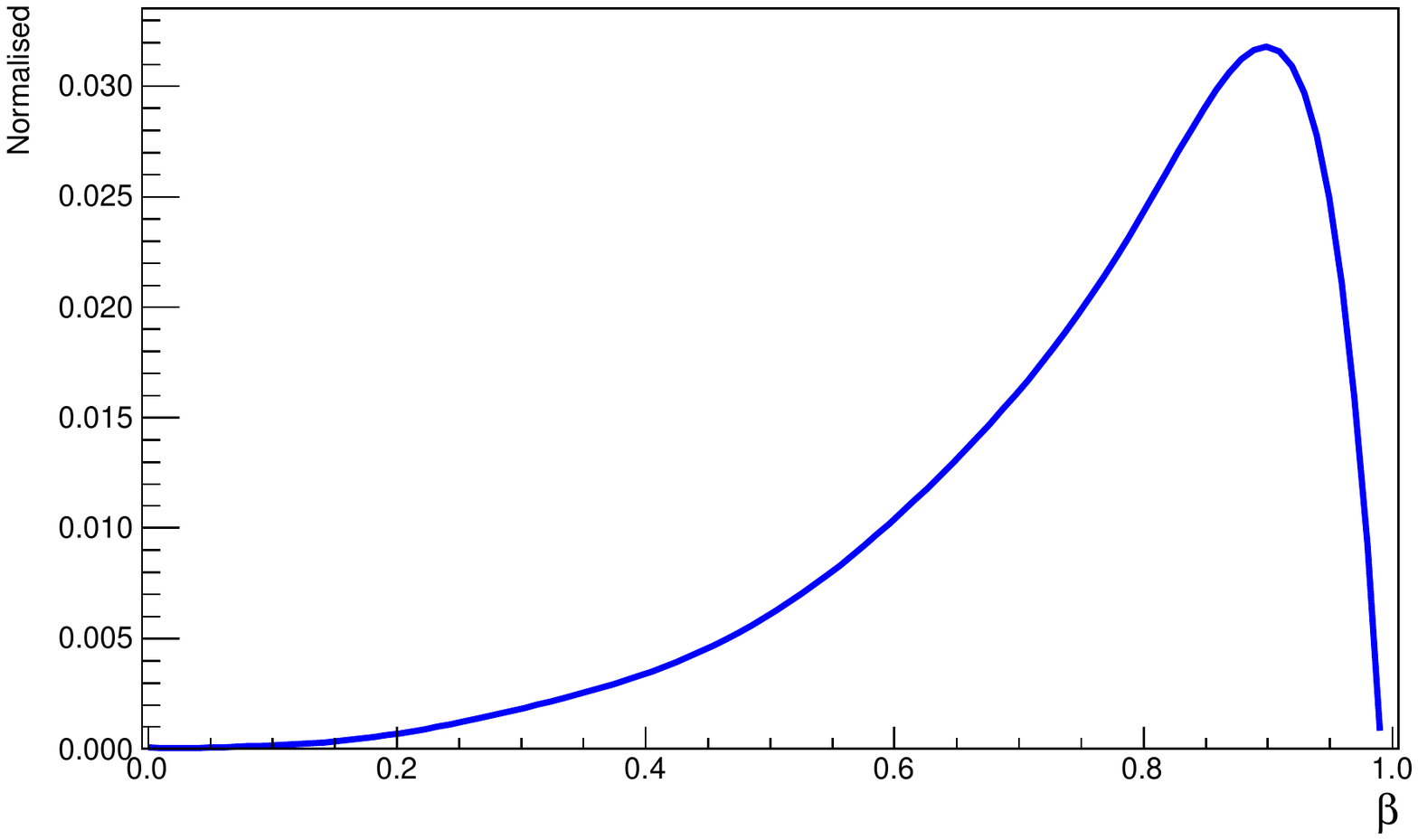,height=4.3in}
\epsfig{file=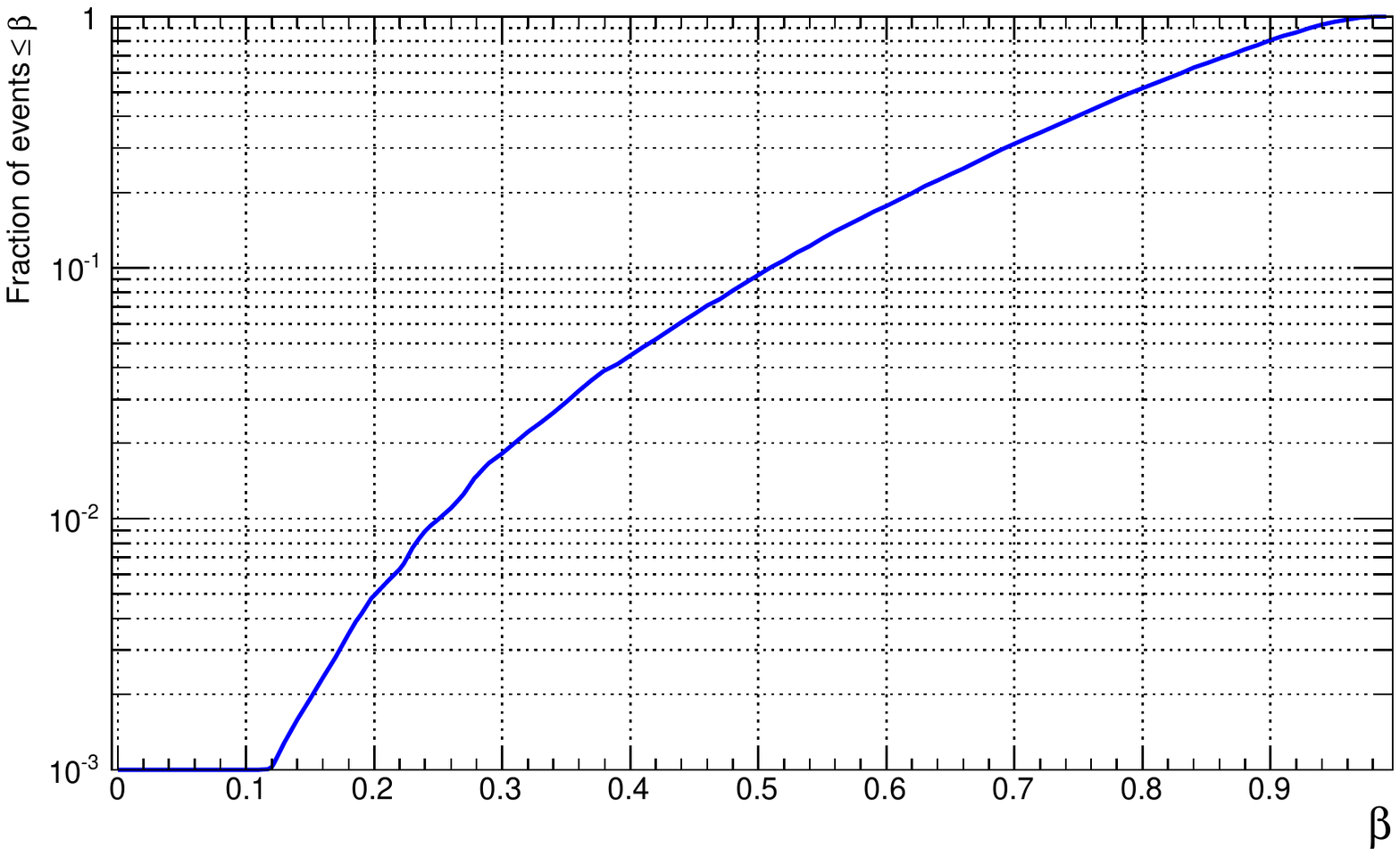,height=4.3in}
\hfill
\end{minipage}
\caption{
{\it The distribution in $\beta$ (left panel) of the same $\tan \beta = 10, A_0 = 0$ ${\tilde \tau_1}$
sample as in Fig.~\ref{fig:betaeta}, and the cumulative distribution in $\beta$
on a logarithmic scale (right panel).
}} 
\label{fig:betalog}
\end{figure}

Since the $\beta$ distributions for the ${\tilde \tau_1}$ are very similar  in the scenarios
we study, when integrated over $\eta$, we just display the example
$\tan \beta = 10, A_0 = 0$ in the left panel of Fig.~\ref{fig:betalog}.
It has $\langle \beta \rangle = 0.76$ and an RMS spread $\Delta \beta = 0.17$. The right panel of
Fig.~\ref{fig:betalog} displays on a logarithmic scale the cumulative $\beta$ distribution. We recall that
many of the slower-moving produced ${\tilde \tau_1}$s would exhibit somewhat enhanced ionization. In
particular, we see in the right panel of Fig.~\ref{fig:betalog}
that a fraction $\simeq 1$\% of the ${\tilde \tau_1}$s would have $\beta < 0.25$ and hence be likely to pass the
MoEDAL~\cite{MoEDAL} threshold for detecting heavily-ionizing charged particles.

\begin{figure}[h!]
\vspace*{-1.5in}
\begin{minipage}{8in}
\epsfig{file=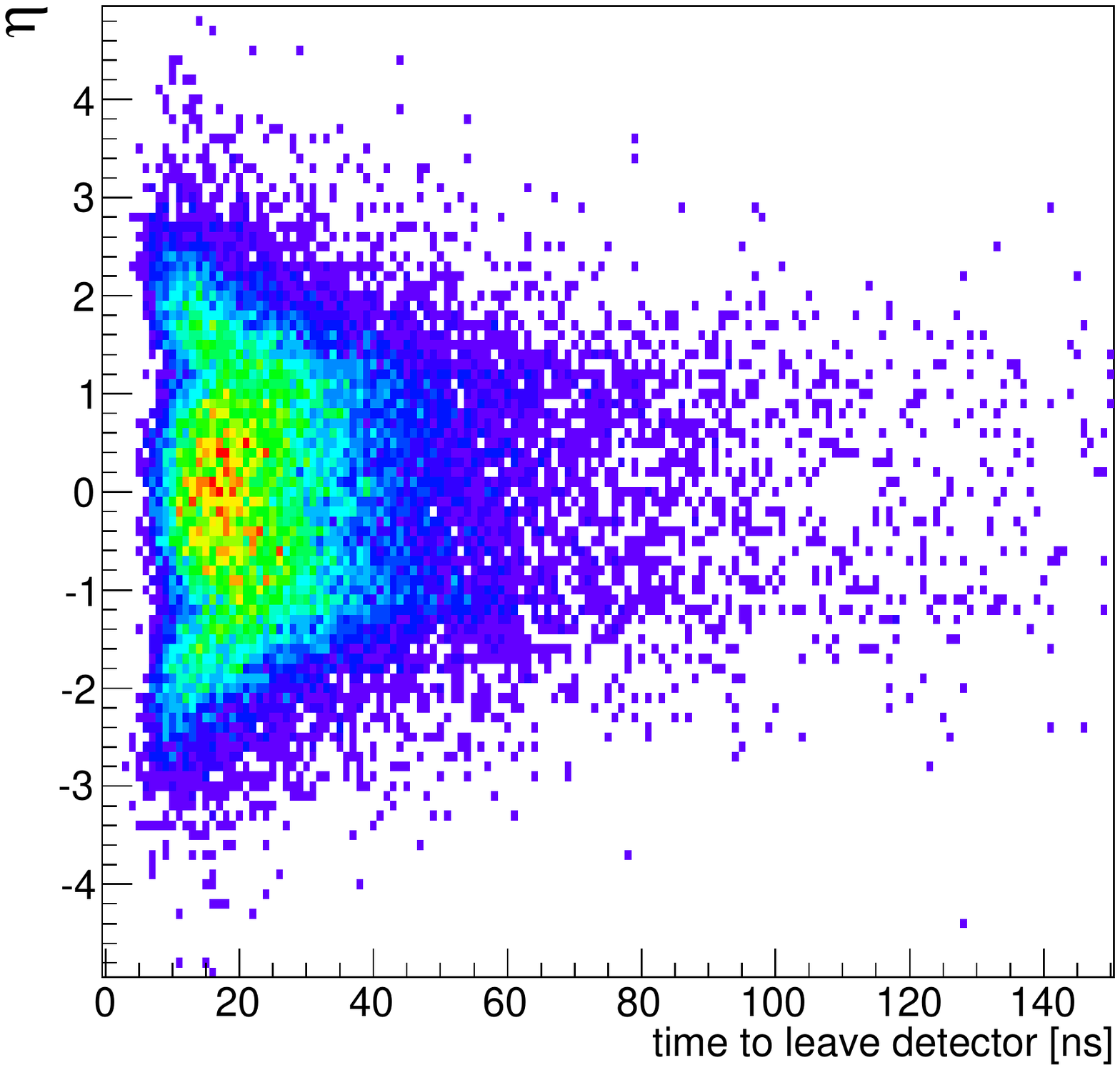,height=4.1in}
\epsfig{file=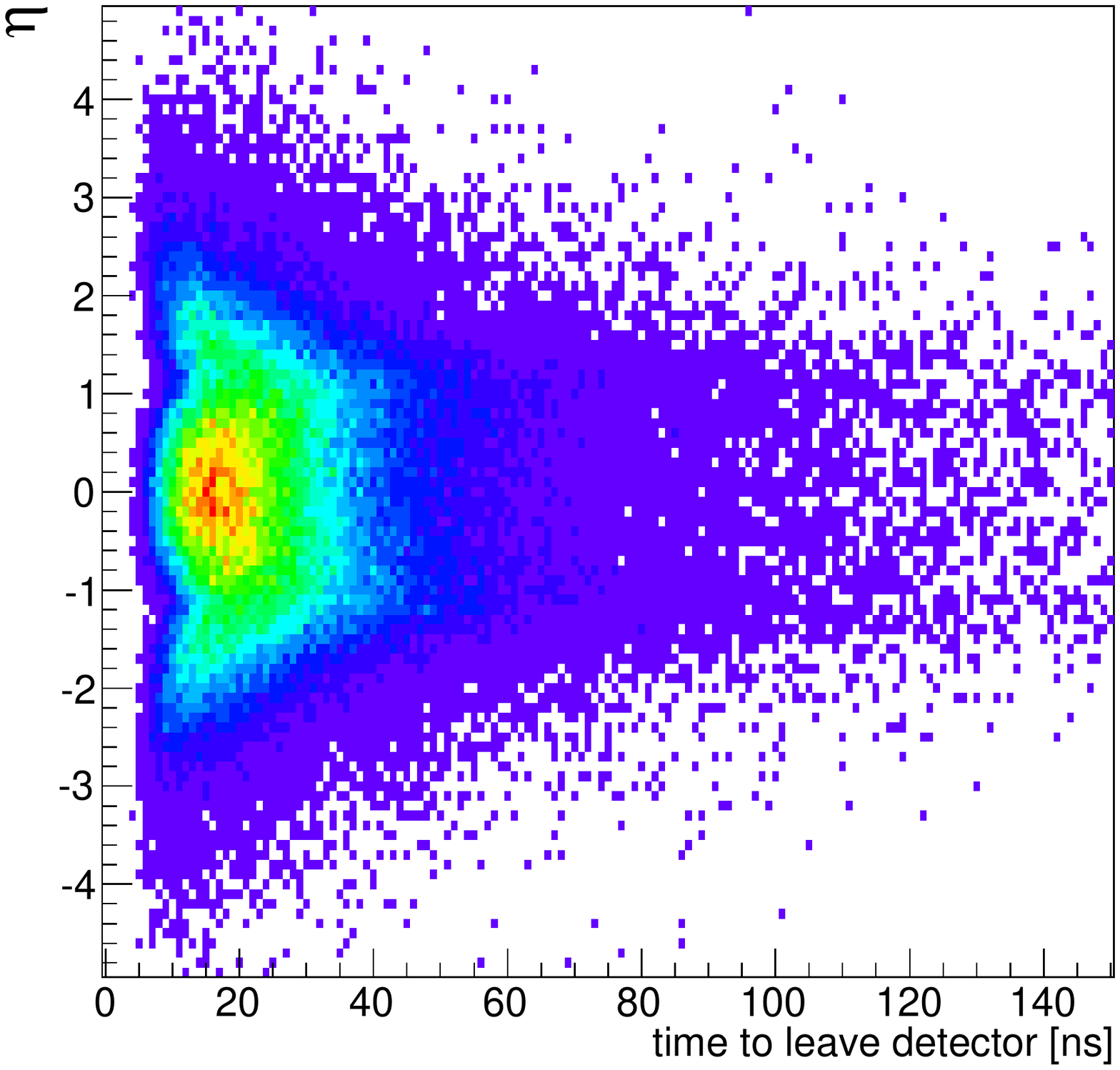,height=4.1in}
\end{minipage}
\begin{minipage}{8in}
\vspace*{-1.5in}
\epsfig{file=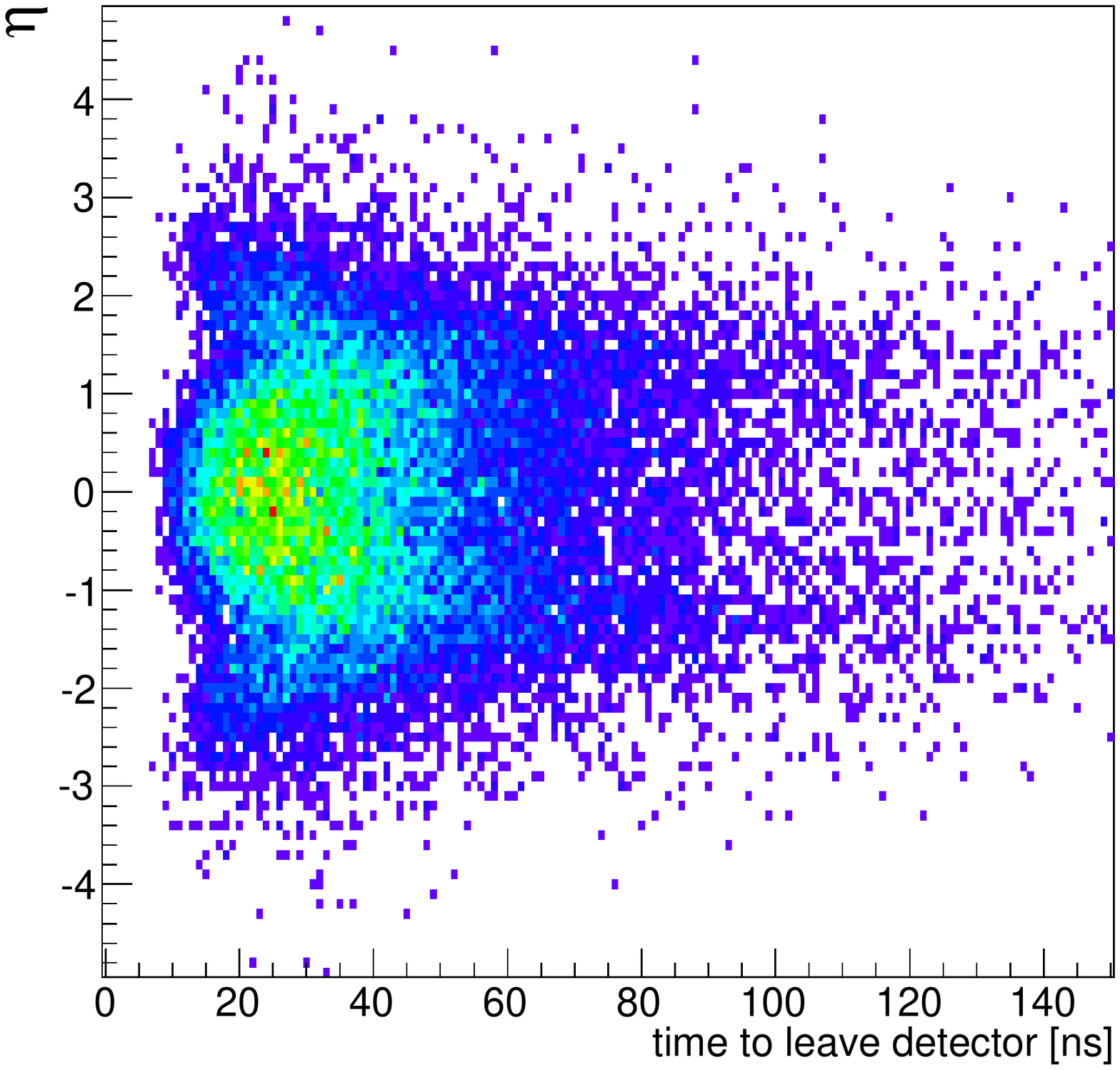,height=4.1in}
\epsfig{file=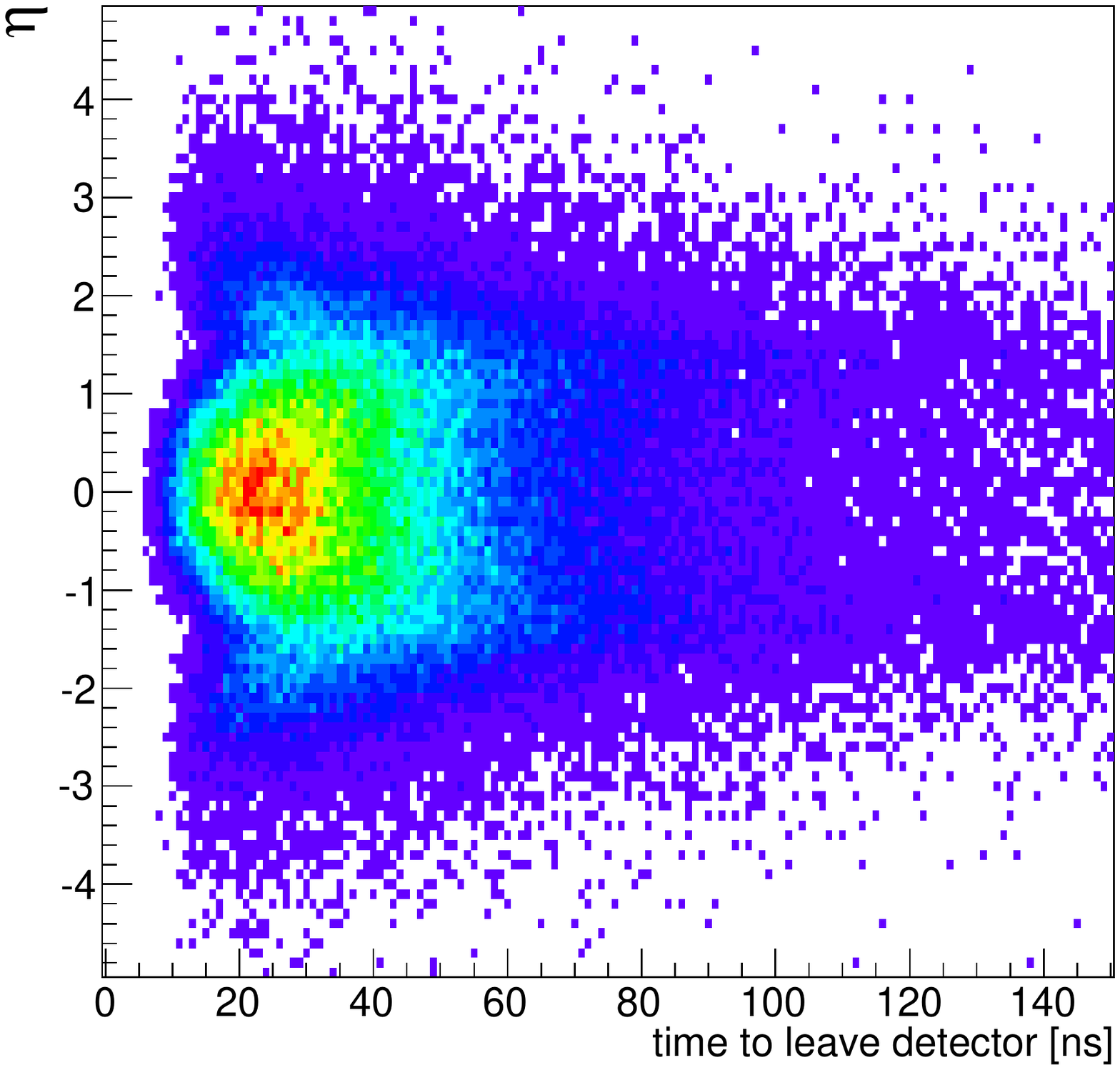,height=4.1in}
\hfill
\end{minipage}
\caption{
{\it Scatter plots in the $({\cal T}, \eta)$ plane, where ${\cal T}$ is the ${\tilde \tau_1}$ time-of-flight
through the CMS detector (upper panels) and the ATLAS detector (lower panels), using the same ${\tilde \tau_1}$
samples as in Fig.~\ref{fig:betaeta} in the left and right panels, respectively.
}} 
\label{fig:timeouteta}
\end{figure}

The $\eta$ and $\beta$ distributions are promising for time-of-flight searches for slow-moving massive particles
in the large LHC detectors, so we examine in more detail the potential time-of-flight signatures of the
produced ${\tilde \tau_1}$s, analyzing separately the cases of the CMS
and ATLAS experiments treating their respective detector geometries using {\tt Delphes}. 
Fig.~\ref{fig:timeouteta} displays scatter plots of
the time ${\cal T}$ that the ${\tilde \tau_1}$s produced inside the CMS
(upper panels) and ATLAS (lower panels) detectors would take to pass
through their respective time-of-flight systems, as functions of $\eta$. 
As previously, the left panels are for the case $\tan \beta = 10, A_0 = 0$ 
and the right panels for $\tan \beta = 40, A_0 = 2.5 \, m_0$.
As could be expected because
of the larger size of the ATLAS detector, the spread of ${\cal T}$ values
is significantly greater than for the CMS detector.

\begin{figure}[h!]
\vspace*{-1.5in}
\begin{minipage}{8in}
\epsfig{file=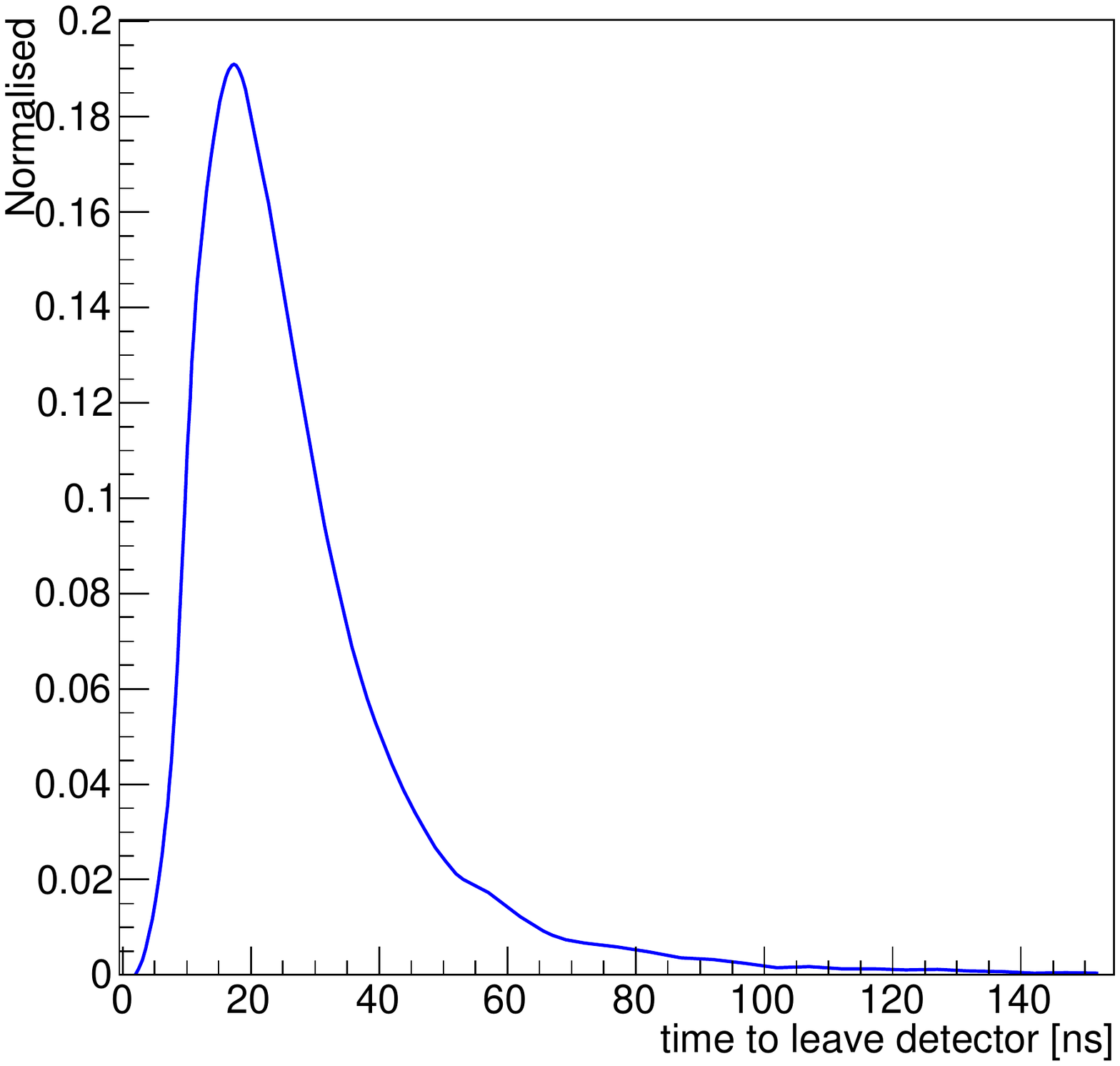,height=4.1in}
\epsfig{file=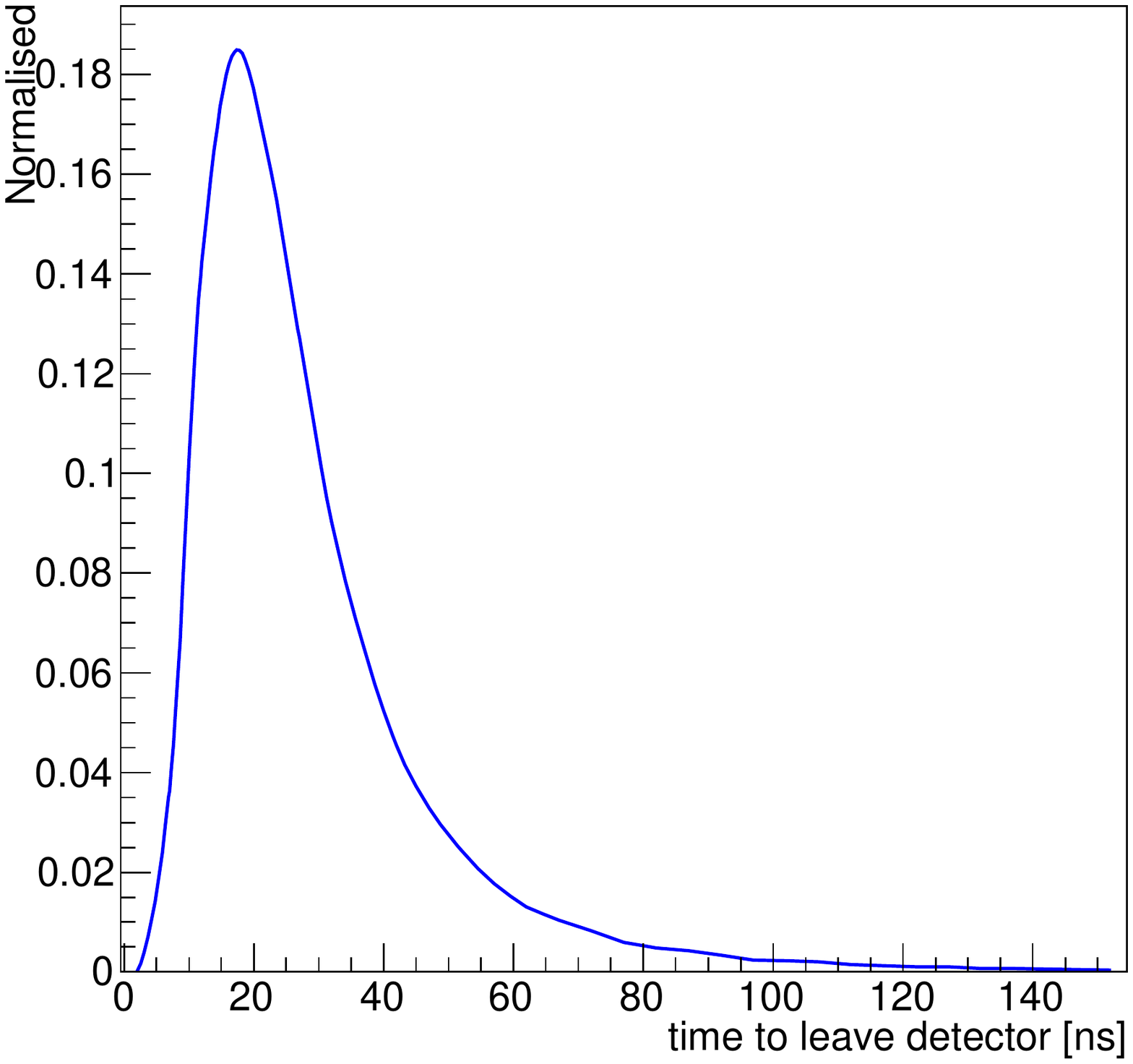,height=4.1in}
\hfill
\end{minipage}
\begin{minipage}{8in}
\vspace*{-1.5in}
\epsfig{file=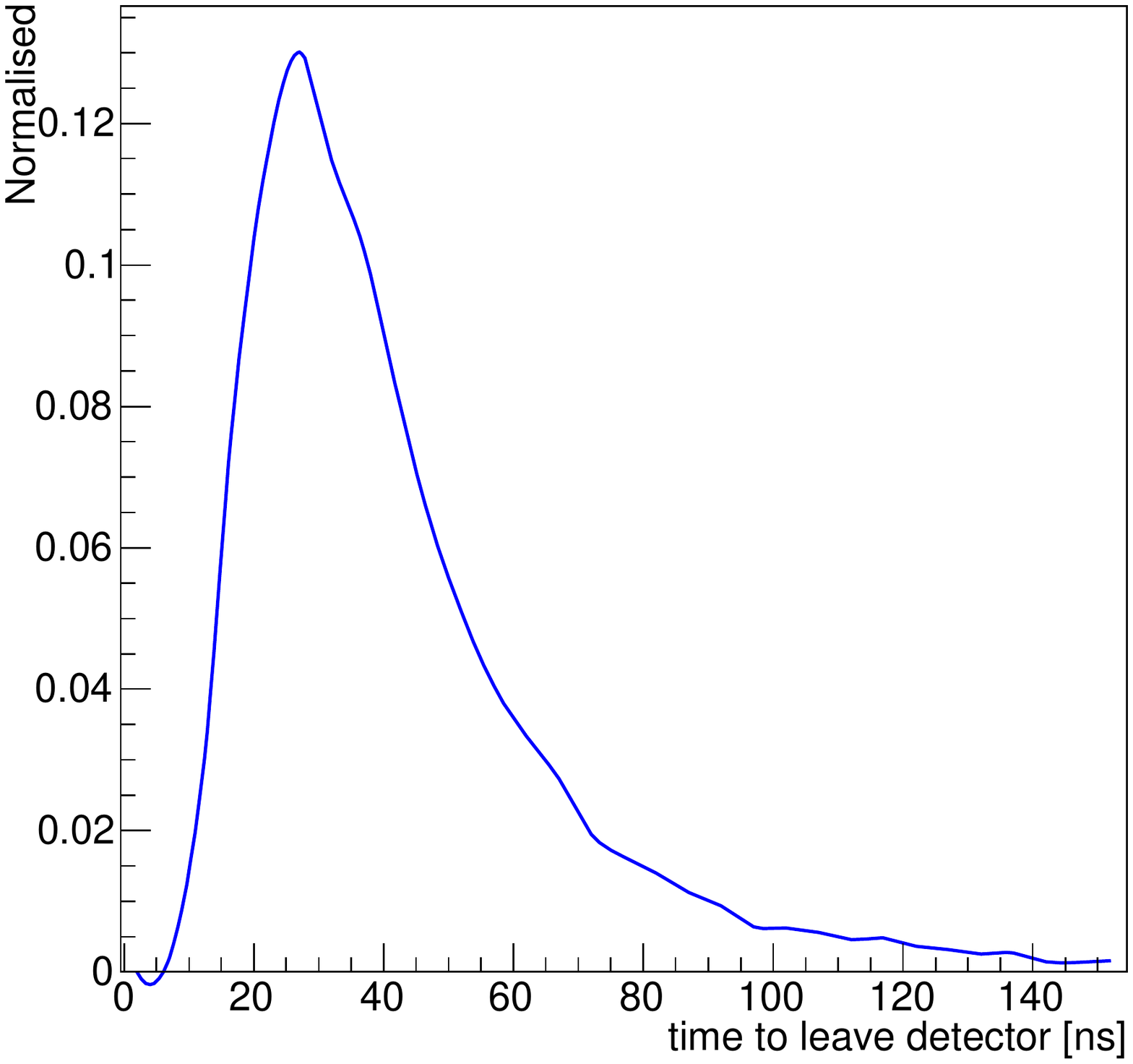,height=4.1in}
\epsfig{file=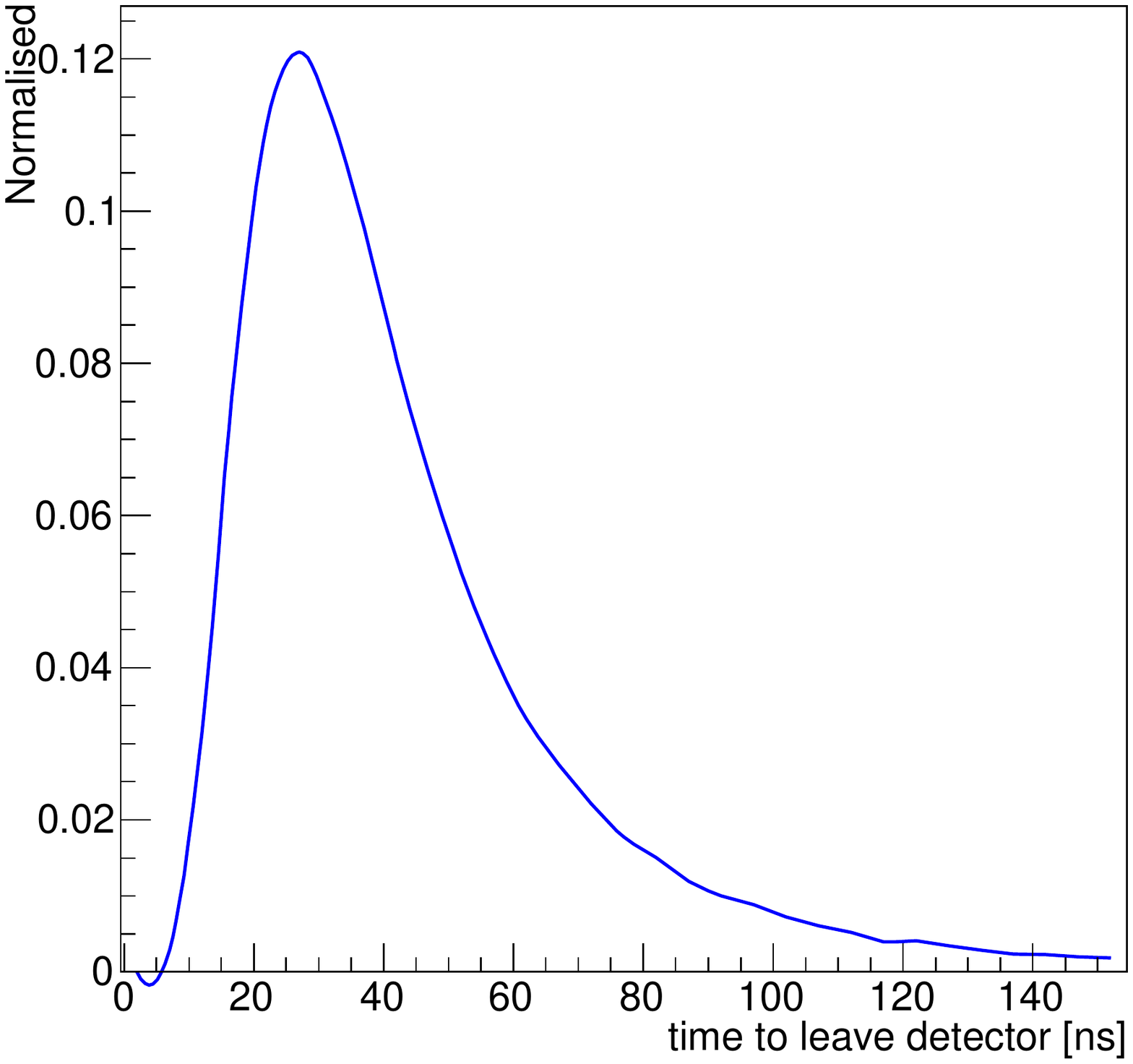,height=4.1in}
\hfill
\end{minipage}
\caption{
{\it The distributions in the time-of-flight ${\cal T}$ through the CMS detector (upper panels) and
the ATLAS detector (lower panels) for the same ${\tilde \tau_1}$
samples as in Fig.~\ref{fig:betaeta} in the left and right panels, respectively.
}} 
\label{fig:timeout}
\end{figure}

Fig.~\ref{fig:timeout} displays the ${\cal T}$ distributions for the same examples
(constrained to the ranges $|\eta| < 2.5$ corresponding to the detector acceptances):
CMS (upper panels) and ATLAS (lower panels), $\tb = 10, A_0 = 0$ (left panels) and
$\tb = 40, A_0 = 2.5 \, m_0$ (right panels). We see that the mean values of ${\cal T}$ are
quite similar in the two models studied: in the case of CMS
$\langle {\cal T} \rangle = 28$ to 29~ns, whereas in the ATLAS case
$\langle {\cal T} \rangle = 44$ to 45~ns, and the RMS spreads are also similar:
in the case of CMS $\Delta {\cal T} \simeq 20$~ns, whereas in the ATLAS case
$\Delta {\cal T} = 29$ to 31~ns. We recall that the LHC has been running with a
bunch spacing of 50~ns, and that there is discussion of reducing this to 25~ns
in the future. Thus there would appear to be a possibility of confusion between the particles produced
during different bunch crossings, which may increase in the future.

\begin{figure}[h!]
\vspace*{-1.5in}
\begin{minipage}{8in}
\epsfig{file=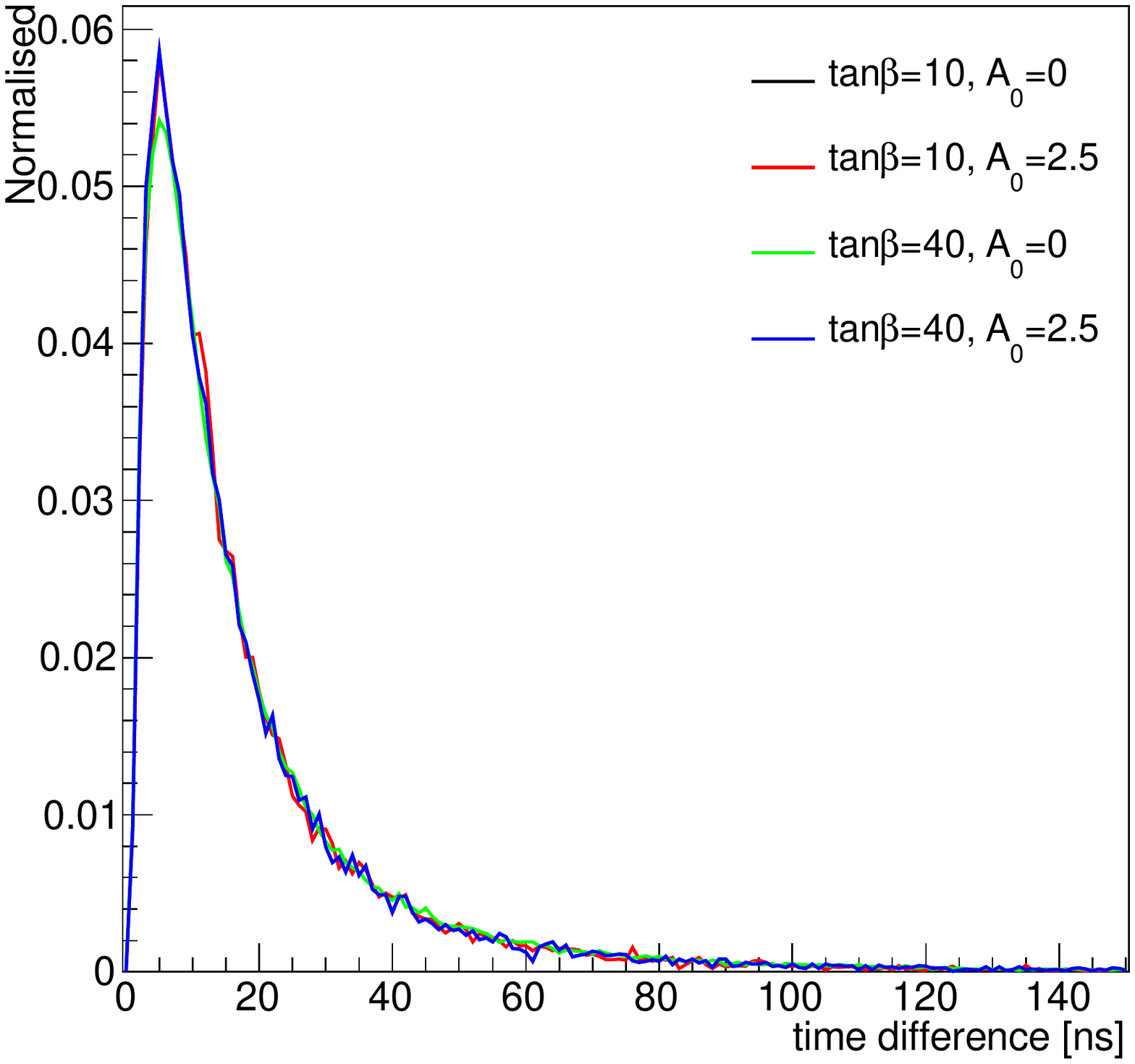,height=4.1in}
\epsfig{file=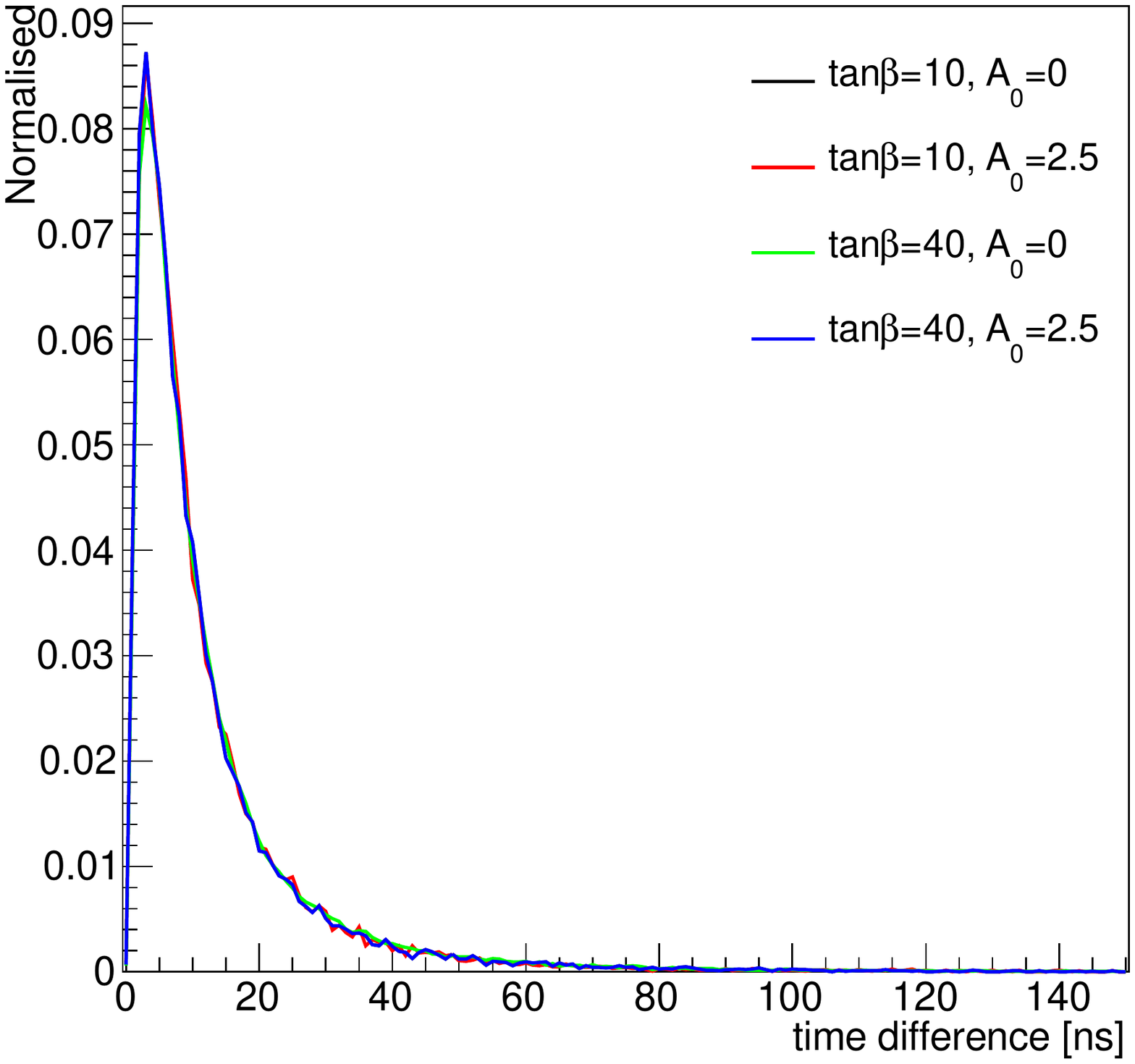,height=4.1in}
\hfill
\end{minipage}
\caption{
{\it The distribution in the time-of-flight difference between ${\tilde \tau_1}$s and muons
produced with the same values of $\eta$ and momenta in ATLAS (left panel) and CMS
(right panel).
}} 
\label{fig:deltatimeout}
\end{figure}

In order to study this possibility, in Fig.~\ref{fig:deltatimeout} we plot
distributions in a different quantity, namely the difference in the
time-of-flight between a ${\tilde \tau_1}$ and a muon with the same
initial momentum and pseudo-rapidity $\eta$. Fig.~\ref{fig:deltatimeout} displays the differences in time-of-flight
for the ATLAS detector (left panel) and the CMS detector (right panel). 
The distributions are very similar for the different models studied. As could be expected from the larger
geometrical size of the ATLAS detector, the  time-of-flight differences in CMS are generally smaller.
About 20 (12)\% of the ${\tilde \tau_1}$s produced in ATLAS (CMS) would have a time-of-flight
difference exceeding 25~ns, the minimum LHC bunch spacing currently foreseen, whereas only very
small fractions have time-of-flight differences exceeding 50~ns, the current LHC bunch spacing.

\section{Stau Decays inside the LHC Detectors}

\begin{figure}[h!]
\vspace*{-1.5in}
\begin{minipage}{8in}
\epsfig{file=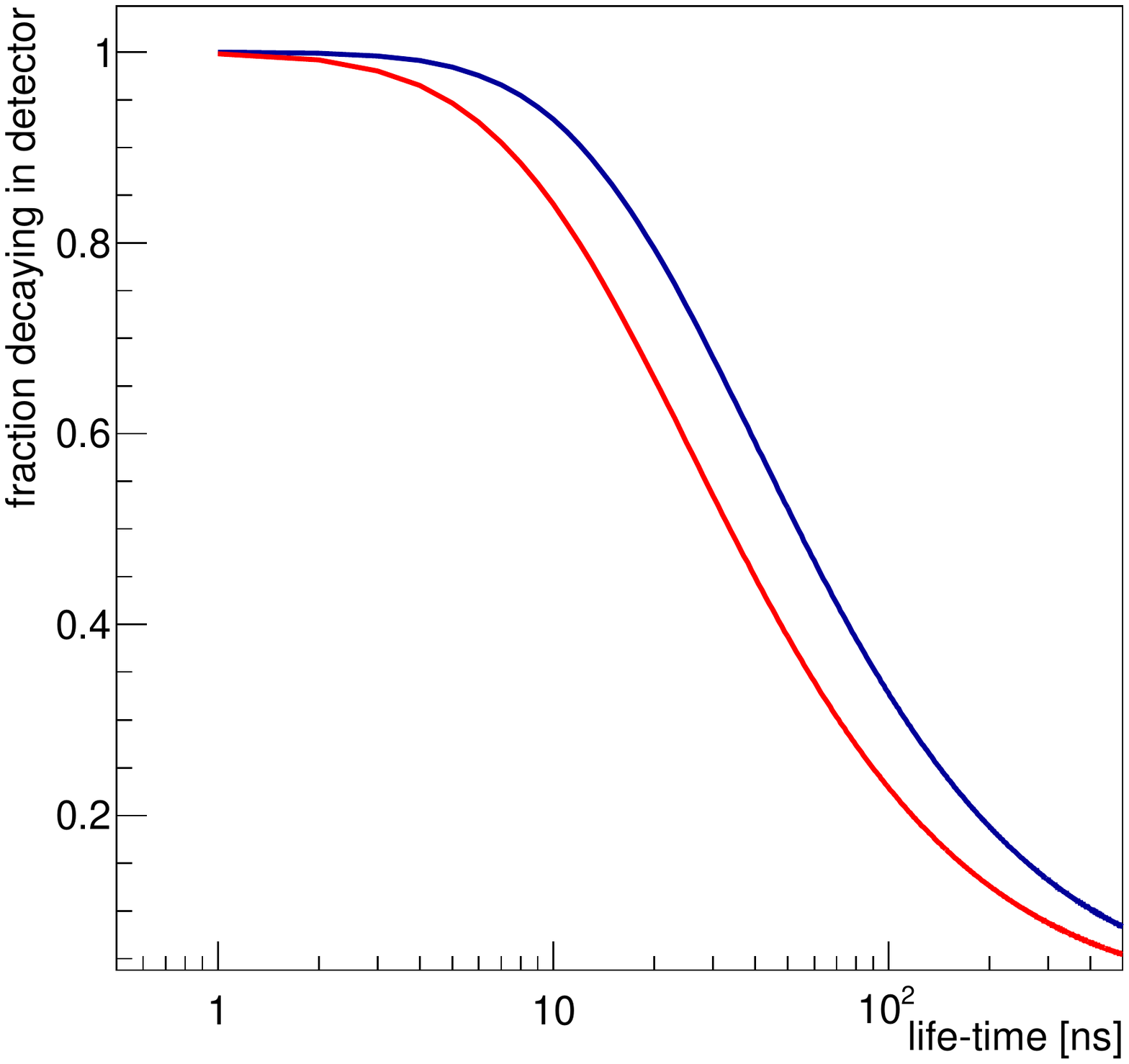,height=4.1in}
\epsfig{file=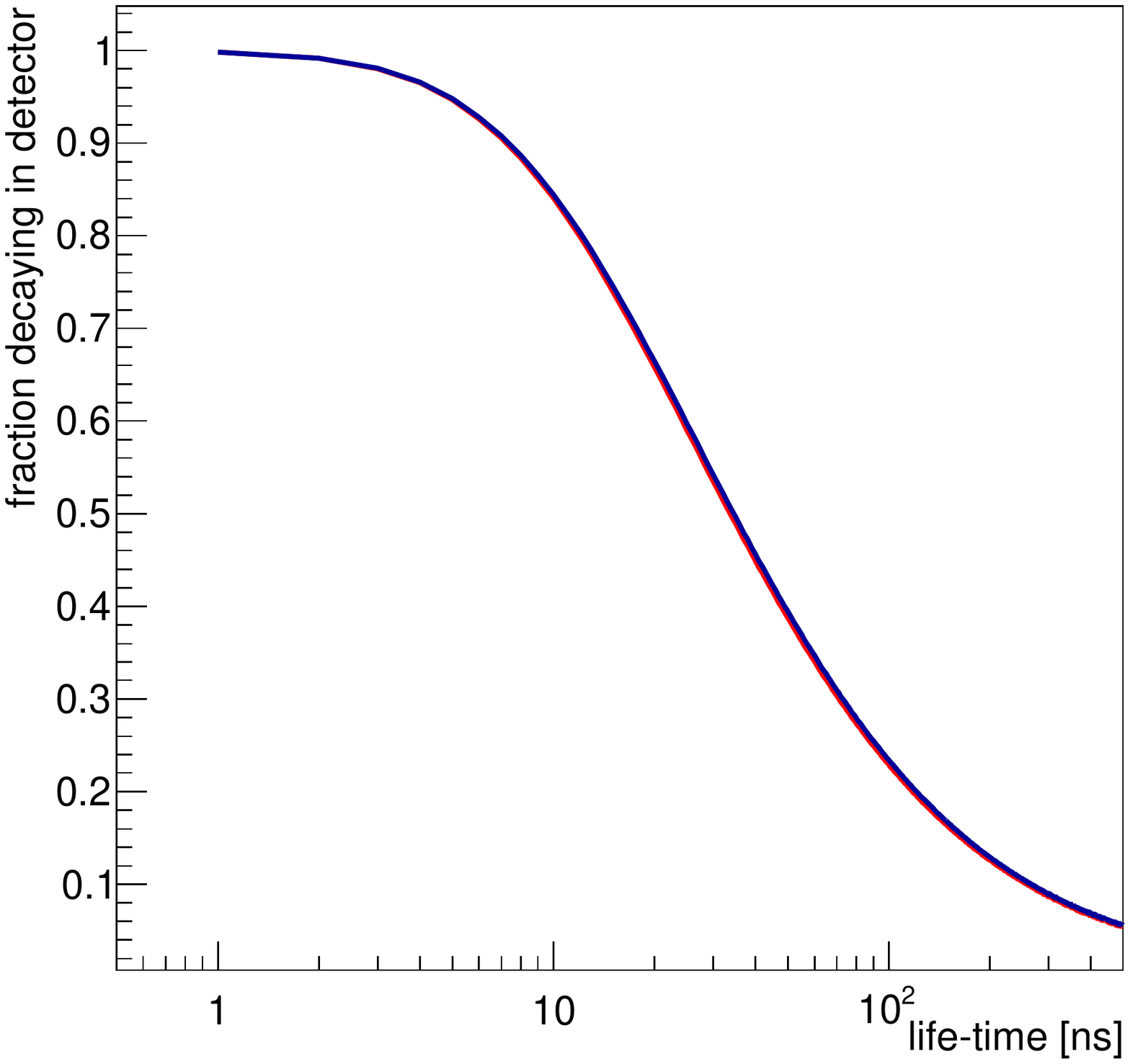,height=4.1in}
\hfill
\end{minipage}
\caption{
{\it The fractions of produced ${\tilde \tau_1}$s that would decay inside the ATLAS (left panel)
and CMS (right panel) detectors, as functions of the ${\tilde \tau_1}$ lifetime. The red (blue)
curves are for the case $\tan \beta = 10, A_0=0$ ($\tan \beta = 40, A_0=2.5\, m_0$),
respectively.
}} 
\label{fig:decayinside}
\end{figure}

Finally, we explore the possibility that the ${\tilde \tau_1}$ may decay
inside the CMS or ATLAS detector. As can be seen in Fig.~\ref{fig:decayinside},
substantial fractions $\gappeq 10$\% of the produced ${\tilde \tau_1}$s decay inside the ATLAS
detector (left panel) or the CMS detector (right panel) if the lifetime $\lappeq 250$ or 400~ns,
respectively. Comparing the calculated ${\tilde \tau_1}$ lifetime shown in Fig.~\ref{fig:ourlifetime}
with the probability distributions shown in Fig.~\ref{fig:decayinside},
we expect a non-negligible likelihood that a ${\tilde \tau_1}$ with $1.2~{\rm GeV} < \Delta m < m_\tau$ might
decay inside the detector, particularly in the case of the ATLAS detector for which $\langle {\cal T} \rangle$
is larger than for CMS. As seen in the right panel of Fig.~\ref{fig:BRs}, if $1.2~{\rm GeV} < \Delta m < m_\tau$
a significant fraction of ${\tilde \tau_1}$
decays are expected to be via the mode ${\tilde \tau_1}^- \to a_1^- \nu_\tau \chi$,
in which case the final state may contain three charged particles: $\pi^- \pi^+ \pi^-$
as well as the invisible neutrals $\nu_\tau$ and  $\chi$.

Since $\Delta m < m_\tau$, and the typical ${\tilde \tau_1}$ velocity is not large,
as seen in Fig.~\ref{fig:betalog}, the charged ${\tilde \tau_1}$ decay products
would typically have energies ${\cal O}(m_\tau)$ and their detection would pose
challenges for the experiments. In the cases of decays into single charged particles,
there would be a large mismatch between the momenta measured before and
after the decay. In the cases of decays into three charged particles, they would 
each be very soft, accentuating the challenge of detection. In either case, the
${\tilde \tau_1}$ would not be detectable as a conventional physics object, and
events producing ${\tilde \tau_1}$s would probably be classified as
missing-energy events. On the other hand, within this general class of events,
they would have the additional signature of an anomalous charged track that
changes its nature within the detector. This property might enable such events
to be distinguished from background $\ETslash$ events. We do not attempt here to 
simulate the cases where metastable particles decay inside the detector
because both the ATLAS and CMS searches for metastable particles employ 
modified tracking and object reconstruction algorithms whose accurate 
modelling lies beyond the scope of {\tt Delphes}, and hence this work.

\begin{figure}[h!]
\hspace*{1.25in}
\begin{minipage}{8in}
\begin{centering}
\epsfig{file=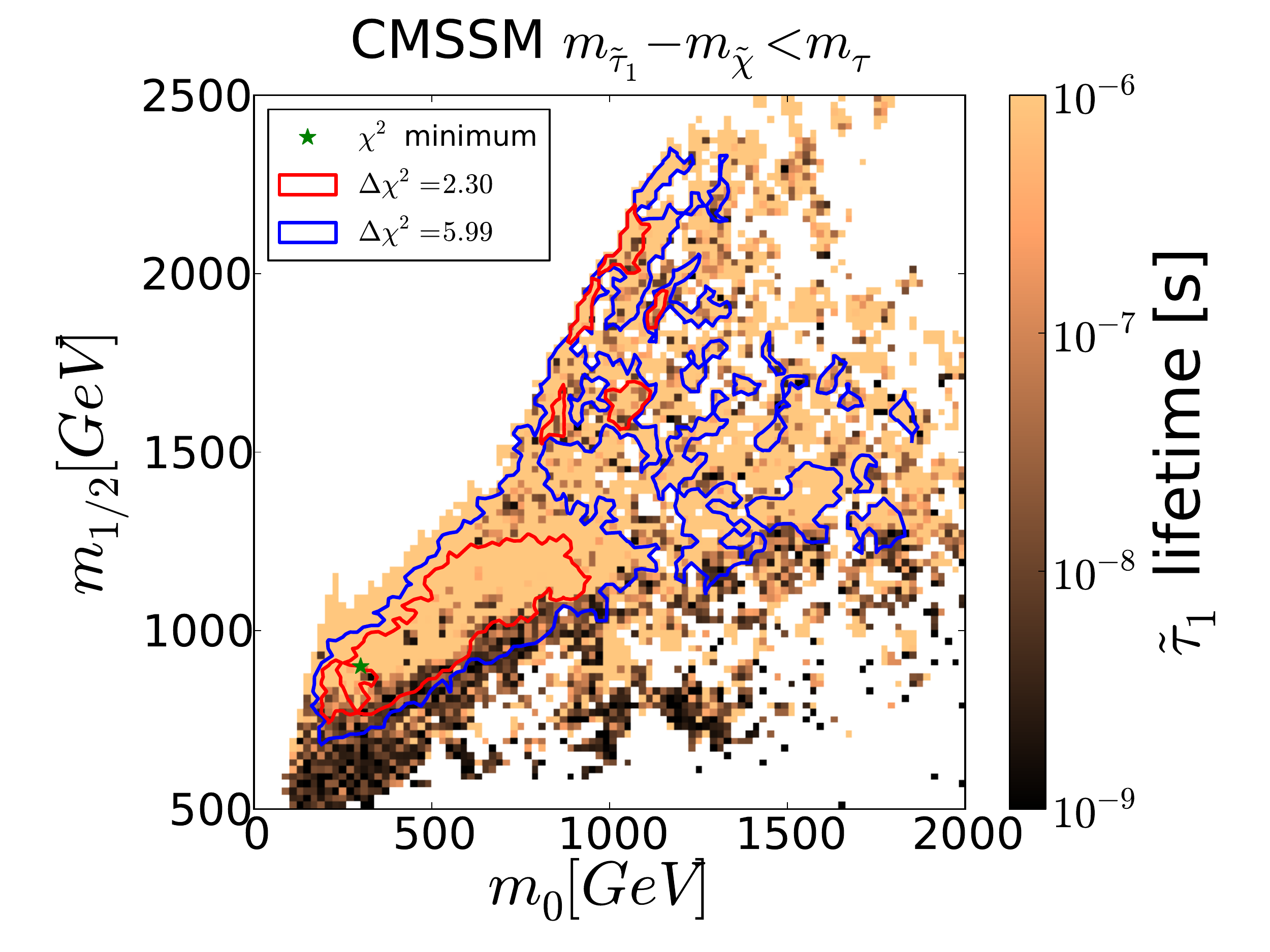,height=3.1in}
\hfill
\end{centering}
\end{minipage}
\caption{
{\it Scatter plot of 10,000 CMSSM points with $\Delta m < m_\tau$ populating bins in the $(m_0, m_{1/2})$ plane
in the ranges $0 < m_0 < 2000$~GeV, $500~{\rm GeV} < m_{1/2} < 2500$~GeV.
They were chosen from among the points analyzed in~\cite{MC8} so as to
minimize $\chi^2$ in each bin. These points
are colour-coded according to the corresponding ${\tilde \tau_1}$ lifetimes, with
shorter (longer) lifetimes displayed in darker (lighter) brown. Also shown are $\Delta \chi^2 = 2.30 (5.99)$
contours, corresponding approximately to the 68 and 95\%
CLs (red and blue, respectively).
}} 
\label{fig:lifetimeprediction}
\end{figure}

As discussed earlier, the best fit found in a global analysis of the CMSSM in~\cite{MC8}
has $\Delta m < m_\tau$ and hence a ${\tilde \tau_1}$ that decays late.
Fig.~\ref{fig:lifetimeprediction} shows a scatter plot of 10,000
CMSSM points with $\Delta m < m_\tau$ populating bins in the $(m_0, m_{1/2})$ plane
in the ranges $0 < m_0 < 2000$~GeV, $500~{\rm GeV} < m_{1/2} < 2500$~GeV.
They were chosen from among the points analyzed in~\cite{MC8} so as to
minimize $\chi^2$ in each bin~\footnote{The
speckle and irregularities in this figure are due to the limited sample size explored in~\cite{MC8}.}.
The points are colour-coded according to the corresponding ${\tilde \tau_1}$ lifetimes, with
shorter (longer) lifetimes displayed in darker (lighter) shades. 
The selected points have ${\tilde \tau_1}$ lifetimes in the range $\sim (1, 1000)$~ns and hence, 
according to the analysis above, are expected either to decay with visible tracks or to pass 
through the LHC detectors without decaying. We note, in particular, that the global best-fit point 
has a ${\tilde \tau_1}$ lifetime of 0.0026~s. We also note that larger values of
$m_{1/2}$ at fixed $m_0$ correspond, in general, to longer ${\tilde \tau_1}$ lifetimes, reflecting the fact
that these points lie closer to the tip of the coannihilation strip, where $\Delta m$
is generally reduced. We also note that the ${\tilde \tau_1}$ lifetime tends to
become shorter as $m_0$ is increased at fixed $m_{1/2}$. This reflects the fact that
these points tend to have larger values of $\tan \beta$, and the coannihilation
strip moves away from the ${\tilde \tau_1}$ LSP boundary as $\tan \beta$
increases. Fig.~\ref{fig:lifetimeprediction} also displays $\Delta \chi^2 = 2.30$ and 5.99 contours
(red and blue, respectively), corresponding approximately to the 68 and 95\% CLs. 
We see that the points within the 68\% CL
region include both some with lower masses and shorter lifetimes and some with higher masses and
longer lifetimes. Thus, this figure reinforces the message that a comprehensive strategy
for exploring the tip of the coannihilation strip will need to combine searches
for late decays inside the detector with searches for massive charged particles
that leave the detector without decaying.

\section{Summary}

We have shown in this paper that (i) one of the favoured regions
of the CMSSM parameter space remaining after LHC searches for $\ETslash$
events is towards the tip of the $\chi - {\tilde \tau_1}$ coannhiliation strip,
where the mass difference $\Delta m = m_{\tilde \tau_1} - m_\chi = {\cal O}(m_\tau)$
(indeed, the best fit found in~\cite{MC8} has $\Delta m < 1$~GeV), (ii) the ${\tilde \tau_1}$
lifetime typically exceeds 1~ns if $\Delta m < m_\tau$ (and exceeds 400~ns if
$\Delta m < 1.2$~GeV), (iii) if the lifetime is between 1 and 400~ns the ${\tilde \tau_1}$
may well decay inside the detector, whereas it is likely to escape the detector as a massive, 
slow-moving charged particle if it has a longer lifetime (which might be detectable
via measurements of its time-of-flight or anomalous heavy ionization). In the CMSSM
these properties of the ${\tilde \tau_1}$ are largely independent of other
parameters such as $\tan \beta$ and $A_0$, whereas a long-lived massive NLSP
would not be as favoured in the NUHM1.

The ATLAS $\ETslash$ searches used in this analysis are based on 5/fb of luminosity
at 7 and 8~TeV in the centre-of-mass. Both ATLAS and CMS have now accumulated
$> 20$/fb of luminosity at 8 TeV, and these samples will enable them to extend
significantly the reaches of such $\ETslash$ searches, reaching towards the tip of the
coannihilation strip. However, the results summarized in the previous paragraph imply
that complete coverage of the coannhiliation strip will require combining $\ETslash$ searches
with searches for massive charged particles. So far, the published results of such searches
use only 7-TeV data, so there is considerable scope for improving their sensitivity, and
our analysis suggests this would be an interesting priority for ATLAS and CMS.

However, it also follows from points (ii) and (iii) above that the standard searches for
$\ETslash$ and massive charged particles could usefully be complemented by
searches for decays into one or more soft charged particles inside the detector, so
we advocate optimizing searches for such decays using the 7- and 8-TeV LHC data.
We present in the Appendix detailed calculations of the most important ${\tilde \tau_1}$
decay modes. Simulating searches for such events would require interfacing {\tt PYTHIA}
with a ${\tilde \tau_1}$ decay code such as that described in the Appendix.
Since the ATLAS and CMS searches for long-lived particles use modified tracking and 
object reconstruction algorithms not included in {\tt Delphes}, we have not attempted to 
simulate the cases where such long-lived particles decay inside the detector with a lifetime
between $\sim 1$ and $\sim 100$~ns. We limit ourselves here to highlighting the importance
of such experimental searches for driving a silver stake through the coannihilation strip
of the CMSSM.

Beyond the specific CMSSM scenario that motivated this analysis, we think that this study
exemplifies the importance of an integrated approach to supersymmetry searches that
combines `traditional' $\ETslash$ searches with searches for more `exotic' signatures.

\section*{Acknowledgements}

We would like to thank B. Acharya, N. Desai and M. Takeuchi for helpful discussions. 
We are grateful to the other members of the MasterCode Collaboration
for their contributions to~\cite{MC8}, some of whose results have been used here.
This work was supported partly by the London
Centre for Terauniverse Studies (LCTS), using funding from the European
Research Council via the Advanced Investigator Grant 267352.
The work of K.A.O. and F.L. at the University of Minnesota was supported in part by DOE grant
DE-FG02-94ER-40823.

\section*{Appendix: Metastable Stau Decay Modes and Lifetime}

We first discuss the ${\tilde \tau_1} - \chi - \tau$ couplings, taking into account
the facts that the ${\tilde \tau_1}$ is a mixture of ${\tilde \tau_L}$ and ${\tilde \tau_R}$,
and that the lightest neutralino $\chi$ is a mixture of $\tilde B$, $\tilde W^3$ and ${\tilde H_{1,2}}$
components. It is convenient to discuss the effects of ${\tilde \tau_L} - {\tilde \tau_R}$ mixing
by referring to the couplings of the chiral supermultiplet partners of the two-component
left-handed spinors $\tau_L$ and $\tau^c_L$. In the MSSM, the interaction Lagrangian which contributes to the 
${\tilde \tau_1} - \chi - \tau$ couplings is, using the convention of~\cite{Haber:1984rc} and~\cite{Gunion:1984yn},  
\beq
{\cal L} = -\frac{1}{2} \frac{\partial^2 W}{\partial A_i \partial A_j} \psi_i \, \psi_j 
                             +  i \sqrt{2} g^{(\alpha)} A_i^* T^{(\alpha) a}_{i j}  \psi_j \, \lambda^{(\alpha) a} 
                             + {\rm h.c.} \, ,
\label{eq:two-com-lag-start}
\eeq
where $W$ is the superpotential, $A_i$ and $\psi_i$ denote the scalar and the fermionic components of the chiral MSSM superfields, respectively, $\lambda^{(\alpha) a}$ is the gaugino field, $g^{(\alpha)}$ is the gauge coupling constant, and $T^{(\alpha) a}$ is the (Hermitian) gauge group generator. The indices $i$ and $a$ label the chiral and gauge multiplets, respectively, and the index $(\alpha)$ labels the gauge group. In the context of the CMSSM, the generation mixings are not considered for sfermion fields, and the left-right mixing needs to be taken into account only for the third generation of sfermion fields. Therefore, the relevant superpotential is 
\beq
W = h_\tau {\tilde \tau_R^*} {\tilde \tau_L} H_1^1 \, ,
\label{eq:super-potential}
\eeq
where $h_\tau$ is the $\tau$ Yukawa coupling, $H_1^1$ is the upper component of one of the two Higgs doublets
\beq
H_1 \equiv \left( \begin{array}{c} H_1^1 \\ H_1^2 \end{array} \right) ,\;\;
H_2 \equiv \left( \begin{array}{c} H_2^1 \\ H_2^2 \end{array} \right) ,\;\;
\eeq
and we denote their vacuum expectation values by $\langle H_1^1 \rangle \equiv v_1 \,,\,
\langle H_2^2 \rangle \equiv v_2 \,,\, \langle H_1^2 \rangle = \langle H_2^1 \rangle = 0$. 
From eqs.~(\ref{eq:two-com-lag-start}) and (\ref{eq:super-potential}), we derive the ${\tilde \tau_1} - \chi - \tau$ Lagrangian in two-component notation, as
\beq
{\cal L}_{\rm int} = - h_\tau \left({\tilde \tau_R^*} {\tau_L} + {\tilde \tau_L} {\tau^c_L}\right) \psi_{H_1}^0
                             + \frac{i}{\sqrt{2}} \left( g^\prime y_L {\tilde \tau_L^*} {\tau_L} \lambda^\prime 
                             - g \, {\tilde \tau_L^*} {\tau_L} \lambda^3 
                             + g^\prime y_{\tau^c_L} {\tilde \tau_R} {\tau^c_L} \lambda^\prime \right)
                              + {\rm h.c.} \, ,
\label{eq:two-com-lag-done}
\eeq
where $\psi_{H_1}^0$ is the superpartner of the $H_1^1$ field, $g^\prime$ and $g$ are the ${\rm U(1)_y}$ hypercharge coupling and the ${\rm SU(2)_L}$ weak coupling, respectively, $\lambda^\prime$ and $\lambda^3$ are the gaugino fields corresponding to the ${\rm U(1)_y}$ generators, $\frac{1}{2}y_L = -\frac{1}{2}$ or $\frac{1}{2}y_{\tau^c_L} = 1$, and the third component of the ${\rm SU(2)_L}$ generator, $\frac{1}{2} \sigma^3$, respectively.  We note that since $y_{\tau^c_L}$ corresponds to an antiparticle, it has a positive sign. 

To transform eq.~(\ref{eq:two-com-lag-done}) into a four-component expression, we first define the four-component Dirac spinor
\beq
{\tau} \equiv  \begin{pmatrix} \tau_L \\ \bar{\tau}^c_L \end{pmatrix} , \;
\eeq
and the Majorana spinners
\beq
{\tilde B} \equiv \begin{pmatrix} -i \lambda^\prime \\ i \bar{\lambda}^\prime \end{pmatrix} ,\;
{\tilde W^3} \equiv \begin{pmatrix} -i \lambda^3 \\ i \bar{\lambda}^3 \end{pmatrix} , \;
{\tilde H_1} \equiv \begin{pmatrix} \psi_{H_1}^0 \\ \bar{\psi}_{H_1}^0 \end{pmatrix} . \;
\eeq
Then by using the formula
\beq
\bar{\Psi}_1 P_L \Psi_2 = \eta_1 \xi_2 \, ,
\eeq
where the $\Psi$'s are general four-component Dirac or Majorana spinors, defined as 
\beq
{\Psi} \equiv \begin{pmatrix} \xi \\ \bar{\eta} \end{pmatrix} \;
\text{or} \;\;
{\Psi} \equiv \begin{pmatrix} \xi \\ \bar{\xi} \end{pmatrix} ,\;
\eeq
we get
\bea
{\cal L}_{\rm int} = &-& \frac{g  m_\tau}{\sqrt{2} m_W \cos \beta} \left({\tilde \tau_R^*} \bar{\tilde H}_1 P_L \tau + {\tilde \tau_L} \bar{\tau} P_L {\tilde H_1} \right) \nonumber \\
                             &-& \frac{g}{\sqrt{2}} \left( - \tan \theta_W {\tilde \tau_L^*} \bar{\tilde B} P_L \tau - {\tilde \tau_L^*} \bar{\tilde W}^3 P_L \tau + 2 \tan \theta_W {\tilde \tau_R} \bar{\tau} P_L {\tilde B} \right) \nonumber \\
                              &+& {\rm h.c.} \, ,
\label{eq:four-com-lag-start}
\eea
where we have used $h_\tau = m_\tau / v_1$, $\frac{1}{2} g^2 (v_1^2 + v_2^2) = m_W^2$, $\tan \beta \equiv v_2 / v_1$, and $g^\prime = g\tan \theta_W$. 

Finally, we rewrite eq.~(\ref{eq:four-com-lag-start}) in terms of mass eigenstate fields. Using 
\bea
&&P_L \, {\tilde H}_1 = N_{i\,3}^* \, P_L \, {\tilde \chi}^0_i \,, \;\;
P_L \, {\tilde B} = N_{i\,1}^* \, P_L \, {\tilde \chi}^0_i \,, \;\;
P_L \, {\tilde W}^3 = N_{i\,2}^* \, P_L \, {\tilde \chi}^0_i \,, \;\; \nonumber \\
&&P_R \, {\tilde H}_1 = N_{i\,3} \, P_R \, {\tilde \chi}^0_i \,, \;\;
P_R \, {\tilde B} = N_{i\,1} \, P_R \, {\tilde \chi}^0_i \,, \;\;
P_R \, {\tilde W}^3 = N_{i\,2} \, P_R \, {\tilde \chi}^0_i \,, \;\;
\eea
and the relation between the mass eigenstates, 
$\stau_{1,2}$, and the interaction eigenstates, $\stau_{L,R}$,  
\bea
\label{eq:stau_rotation_matrix}
  \begin{pmatrix}
            \stau_L \\ \stau_R
        \end{pmatrix}  =
   \begin{pmatrix}
            U_{\stau  1L} & U_{\stau  2L} \\
             U_{\stau  1R} & U_{\stau 
             2R}
        \end{pmatrix}
     \begin{pmatrix}
            \stau_1 \\ \stau_2
        \end{pmatrix} \, ,
\eea
we get the ${\tilde \tau_1} - \chi - \tau$ Lagrangian
\bea
{\cal L}_{{\tilde \tau_1} - \chi - \tau} &=& {\tilde \tau_1}^* \bar{\tilde \chi}^0_i 
\left[
\left( - \dfrac{g  m_\tau}{\sqrt{2} m_W \cos \beta} U_{\stau  1R}^* N_{i\,3}^* 
+ \dfrac{g}{\sqrt{2}} \tan \theta_W U_{\stau  1L}^* N_{i\,1}^* 
+ \dfrac{g}{\sqrt{2}} U_{\stau  1L}^* N_{i\,2}^* 
\right) P_L  \right. \nonumber \\
&& \;\;\;\;\;\;\;\;\;\; \left.  +  \left( - \dfrac{g  m_\tau}{\sqrt{2} m_W \cos \beta} U_{\stau  1L}^* N_{i\,3} 
- \sqrt{2} g \tan \theta_W U_{\stau  1R}^* N_{i\,1} 
\right) P_R
\right]
\tau
+ {\rm h.c.} \nonumber \\
&\equiv& {\tilde \tau_1}^* \bar{\tilde \chi}^0_i \left(c_L P_L + c_R P_R \right) \tau + {\rm h.c.} \, ,
\label{eq:four-com-lag-done}
\eea
where we assume the $i$-th neutralino is the lightest one, namely $\chi$. 

We now discuss the $\tilde \tau_1$ lifetime and branching ratios, assuming that 
$\tilde \tau_1$ is the NLSP and the lightest neutralino $\chi$ is the LSP.
When the mass difference between $\tilde \tau_1$ and $\chi$, 
$\Delta m \equiv m_{\tilde \tau_1} - m_\chi$, is larger than $m_\tau$, 
the two-body decay, $\stau_1 \to \chi  \tau$, is kinematically allowed and it dominates the $\stau_1$ total decay rate.
The two-body decay rate is given explicitly by 
\bea
\Gamma_{\rm 2-body} & = & \frac{1}{16 \pi m_{\stau_1}^3} \left(m_{\stau_1}^4 + m_\chi^4 + m_\tau^4 - 2 m_{\stau_1}^2 m_\chi^2 
- 2 m_{\stau_1}^2 m_\tau^2 - 2 m_\chi^2 m_\tau^2 \right)^{1 \over 2} \nonumber \\
 && \times \left[ \left(|c_L|^2 + |c_R|^2 \right) \left((\Delta m)^2 + 2 (\Delta m) m_\chi - m_\tau^2 \right) - 2 \left(c_L c_R^* + c_R c_L^* \right) m_\tau m_\chi \right] \, .
\eea

We are interested in scenarios with a metastable $\stau_1$ having a lifetime in the range
$\sim 1$ ns to $\sim 400$ ns or more. When it is kinematically accessible,
the two-body decay mode gives a lifetime many orders of magnitude smaller than $1$ ns,
 as one can see from, for example, the left panel of Fig.~\ref{fig:ourlifetime}, 
 where we take the same parameters as in Fig.~2 of~\cite{Jittoh}, namely 
 a pure bino-like neutralino with $m_\chi = 300$ GeV, $U_{\stau  1L} = 1/2$ and 
 $U_{\stau  1R} = \sqrt{3}/2$. Therefore we neglect the contributions from all other decay modes when $\Delta m > m_\tau$. 

\begin{figure}
\vskip 0.5in
\vspace*{-0.75in}
\begin{minipage}{6.7in}
\centering
\epsfig{file=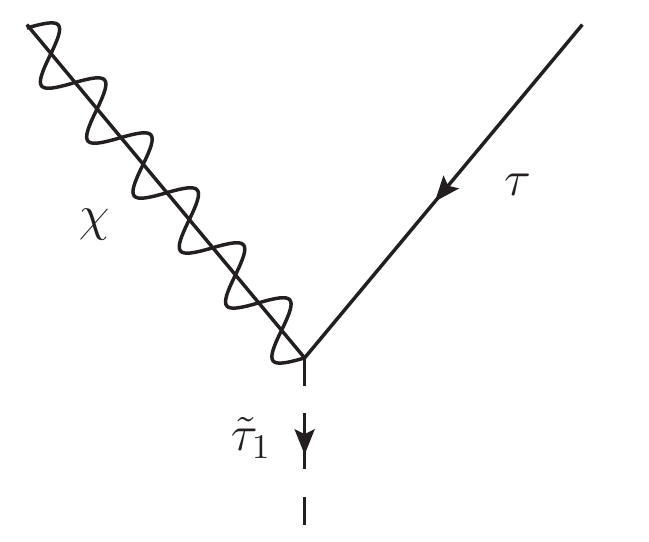,height=1.5in}
\epsfig{file=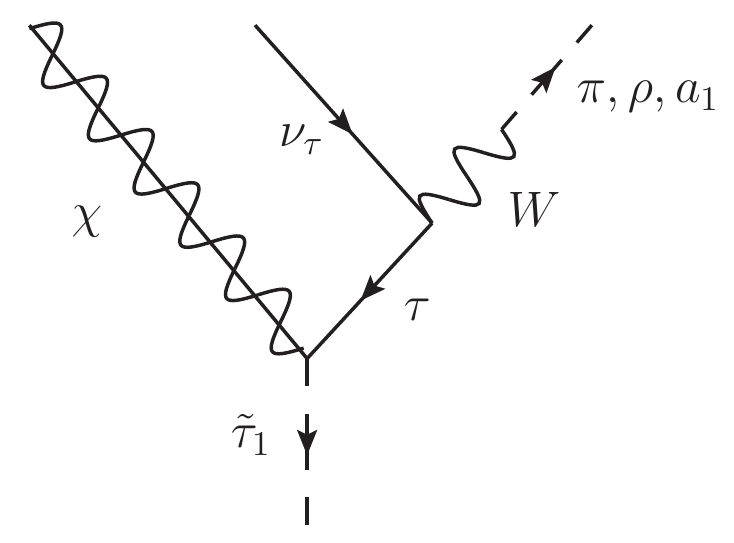,height=1.5in}
\epsfig{file=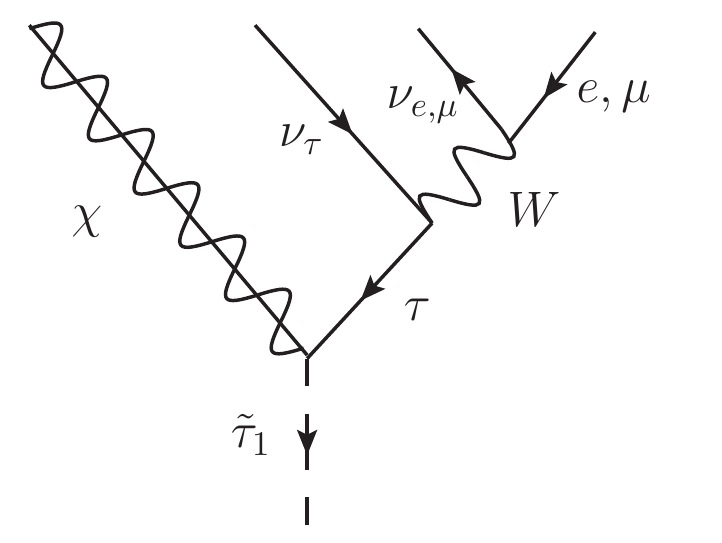,height=1.5in}
\hfill
\end{minipage}
\caption{
{\it Feynman diagrams for two-, three- and four-body $\stau_1$ decay modes. 
These diagrams are created using JaxoDraw~\protect\cite{Binosi:2003yf}. 
}}
\label{fig:feynman-diagrams}
\end{figure}

It was emphasized in~\cite{Jittoh} that the $\stau_1$ would be metastable if
$\Delta m < m_\tau$, so that the two-body decay is not kinematically allowed. 
We consider here the following dominant three- and four-body decay modes, namely, 
${\tilde \tau_1} \to \chi  \nu_\tau  a_1(1260)$, ${\tilde \tau_1} \to \chi  \nu_\tau  \rho(770)$, 
${\tilde \tau_1} \to \chi  \nu_\tau  \pi$, ${\tilde \tau_1} \to \chi  \nu_\tau  \nu_\mu  \mu$ and 
${\tilde \tau_1} \to \chi  \nu_\tau  \nu_e  e$ (the `bars' for antiparticles are suppressed). 
These modes close in sequence as $\Delta m$ decreases. 
The three-body decay mode ${\tilde \tau_1} \to \chi  \nu_\tau  \pi$ and the four-body decay modes
were considered in~\cite{Jittoh}, but not the decay modes involving $\rho$ and $a_1$ mesons. 
Just as $\Gamma(\tau \to \nu_\tau  \rho(770))$ and $\Gamma(\tau \to \nu_\tau  a_1(1260))$
are larger than $\Gamma(\tau \to \nu_\tau  \pi)$,
so also ${\tilde \tau_1} \to \chi  \nu_\tau  a_1(1260)$ and 
${\tilde \tau_1} \to \chi  \nu_\tau  \rho(770)$ may be important compared to
${\tilde \tau_1} \to \chi  \nu_\tau  \pi$ if these modes are kinematically allowed,
so we include these modes in our calculations.
The relevant Feynman diagrams are shown in Fig.~\ref{fig:feynman-diagrams}. We have
ignored the diagrams in which some supersymmetric particle, e.g., chargino, is in the propagator, 
because the contributions from these diagrams are suppressed by the large propagator masses 
compared to the diagrams with only Standard Model particles in the propagators.   

For ${\tilde \tau_1} \to \chi  \nu_\tau  \pi$, the coupling between the $\pi^\pm$ and the $W^\mp$ boson is given by chiral perturbation theory (see, for example,\cite{Scherer:2005ri}), as
\beq
{\cal L}_{W-\pi} = - \frac{g f_\pi V_{ud}}{2} \, W_\mu^+ \, \partial^\mu \pi^- + {\rm h. c.} \, ,
\eeq
where $f_\pi \approx 92.4~$MeV is the pion decay constant and $V_{ud}$ is the quark CKM matrix element.  
Apart from the coupling given in eq.~(\ref{eq:four-com-lag-done}), 
the only other coupling involved in this mode is the W-lepton-neutrino coupling given by the Standard Model, 
\beq
{\cal L}_{W-l-\nu} = - \frac{g}{\sqrt 2} W_\mu^+ \bar{\nu} \, \gamma^\mu P_L l^- + {\rm h. c.} \, .
\eeq
The decay rate for this process is 
\bea
\Gamma_{{\tilde \tau_1} \to \chi  \nu_\tau  \pi} &=&  
\frac{G_F^2 f_\pi^2 V_{ud}^2 ((\Delta m)^2 - m_\pi^2)}{128 \pi^3 m_{\stau_1}^3} 
 \int_0^1 dx \left[\left((\Delta m)^2 - q_f^2 \right) \left((\Delta m) + 2 m_\chi \right)^2 - q_f^2 \right]^{1 \over 2} \nonumber \\
&& \times \frac{1}{\left(q_f^2 - m_\tau^2\right)^2 + (m_\tau \Gamma_\tau)^2} \frac{\left(q_f^2 - m_\pi^2 \right)^2}{q_f^2} \nonumber \\
&& \times  \left[\left(|c_L|^2 q_f^2 + |c_R|^2 m_\tau^2 \right) \left( (\Delta m)^2 + 2 m_\chi \Delta m - q_f^2 \right) 
 - 2 \left(c_L c_R^* + c_R c_L^* \right) m_\chi m_\tau q_f^2 \right] , \nonumber \\
 \label{3b-to-pion}
\eea
where $G_F$ is the Fermi constant. In doing the phase space integration, we have
employed the same substitution as in~\cite{Jittoh}, namely
\beq
q_f^2 \equiv (\Delta m)^2 - \left((\Delta m)^2 - m_f^2 \right) x \, ,
\eeq
where the index $f$ denotes the massive particle in the final state other than the $\chi$, 
namely, $f  = \pi$ here, and it denotes $a_1$, $\rho$, $\mu$ and $e$, respectively, 
in the other three-body and four-body decay rates.  

For ${\tilde \tau_1} \to \chi  \nu_\tau  a_1(1260)$ and ${\tilde \tau_1} \to \chi  \nu_\tau  \rho(770)$, 
we use the $W^\pm-a_1^\mp(1260)$ and $W^\pm-\rho^\mp(770)$ couplings suggested 
by the idea of meson dominance~\cite{Lichard:1997ya}, 
\beq
{\cal L} = - \frac{g V_{ud}}{2 g_\rho} W_\mu^+  \left( m_\rho^2 w_\rho \rho^{- \mu} - m_{a_1}^2 w_{a_1} a_1^{- \mu} \right) + {\rm h.c.} \, ,
\eeq
where $g_\rho$ is the $\rho \pi \pi$ coupling constant, 
and $w_\rho$ and $w_{a_1}$ are phenomenological parameters. 
There is one difference between our treatment of $w_{\rho}$ and the way done in~\cite{Lichard:1997ya}: 
we take the $w_{\rho}$ as a phenomenological parameter to be fixed by the $\tau$ decay data, 
while in~\cite{Lichard:1997ya} $w_{\rho} = 1$ exactly. This is equivalent to saying
that we introduce one more phenomenological parameter as compared to~\cite{Lichard:1997ya}, 
as we get the value of $w_{\rho}$ by equating the experimental decay rate data for 
$\tau \to \pi^-  \pi^0  \nu_\tau$ given in~\cite{pdg-2012} ($\Gamma_{14}$) to the tree-level 
two-body decay rate of $\tau \to \nu_\tau \rho(770)$,
\beq
\Gamma_{\tau \to \nu_\tau \rho(770)} = \frac{G_F^2 V_{ud}^2 w_\rho^2 m_\rho^2}{8 \pi m_\tau^3  g_\rho^2} \frac{m_W^4}{\left( m_\rho^2 - m_W^2 \right)^2} \left(m_\tau^2-m_\rho^2 \right)^2 \left(m_\tau^2 + 2 m_\rho^2 \right) \, . 
\label{tau_to_rho_nu}
\eeq
Similarly, we get the value of $w_{a_1}$ by equating the sum of the decay rates of 
$\tau \to \pi^- 2\pi^0 \nu_\tau$ and $\tau \to \pi^- \pi^+ \pi^- \nu_\tau$ given in~\cite{pdg-2012} 
($\Gamma_{20}$ and $\Gamma_{62}$) to the tree-level two-body decay rate of 
${\tau \to \nu_\tau a_1(1260)}$, which is obtained by the substitutions 
$m_\rho \to m_{a_1}$ and $w_\rho \to w_{a_1}$ in eq.~(\ref{tau_to_rho_nu}).

The decay rate expression for ${\tilde \tau_1} \to \chi \nu_\tau \rho(770)$ is
\bea
\Gamma_{{\tilde \tau_1} \to \chi \nu_\tau \rho(770)} &=&  
\frac{G_F^2 V_{ud}^2 w_\rho^2 m_\rho^2  m_W^4 ((\Delta m)^2 - m_\rho^2)}{128 \pi^3 m_{\stau_1}^3 g_\rho^2 \left(m_\rho^2 - m_W^2 \right)^2} 
 \int_0^1 dx \left[\left((\Delta m)^2 - q_f^2 \right) \left((\Delta m) + 2 m_\chi \right)^2 - q_f^2 \right]^{1 \over 2} \nonumber \\
&& \times \frac{1}{\left(q_f^2 - m_\tau^2\right)^2 + (m_\tau \Gamma_\tau)^2} \frac{\left(q_f^2 - m_\rho^2 \right)^2 \left(q_f^2 + 2 m_\rho^2 \right)}{q_f^4} \nonumber \\
&& \times  \left[\left(|c_L|^2 q_f^2 + |c_R|^2 m_\tau^2 \right) \left( (\Delta m)^2 + 2 m_\chi \Delta m - q_f^2 \right) 
 - 2 \left(c_L c_R^* + c_R c_L^* \right) m_\chi m_\tau q_f^2 \right] . \nonumber \\
\eea
By substituting $m_\rho \to m_{a_1}$ and $w_\rho \to w_{a_1}$ in the above equation, we get the decay rate for ${\tilde \tau_1} \to \chi \nu_\tau  a_1(1260)$.

In calculating the four-body decay rates ${\tilde \tau_1} \to \chi  \nu_\tau  \nu_\mu  \mu$ and 
${\tilde \tau_1} \to \chi  \nu_\tau  \nu_e  e$, we take the leading-order result in the $1/m_W$ 
expansion, since we are interested in scenarios where $\Delta m \ll m_W$. The result is 
\bea
\Gamma_{\rm 4-body} &=& \frac{G_F^2 \left((\Delta m)^2 - m_l^2 \right)}{96 (2 \pi)^5 m_{\stau_1}^3}
\int_0^1 dx \left[\left((\Delta m)^2 - q_f^2 \right) \left((\Delta m) + 2 m_\chi \right)^2 - q_f^2 \right]^{1 \over 2}  \frac{1}{q_f^4} \nonumber \\
&& \times \frac{1}{\left(q_f^2 - m_\tau^2\right)^2 + (m_\tau \Gamma_\tau)^2} 
 \left[12 m_l^4 q_f^4 \log \left(\frac{q_f^2}{m_l^2} \right) +\left( q_f^4 - m_l^4 \right) \left( q_f^4 - 8 m_l^2 q_f^2 + m_l^4 \right) \right] \nonumber \\
&& \times \left[ \left(|c_L|^2 q_f^2 + |c_R|^2 m_\tau^2 \right) \left( (\Delta m)^2 + 2 m_\chi \Delta m - q_f^2 \right) - 2 \left( c_L c_R^* + c_R c_L^* \right) m_\chi m_\tau q_f^2  \right], \nonumber \\
\eea
where $m_l = m_e$ and $m_l = m_\mu$ for $e$ and $\mu$ final states, respectively. 

The branching ratios for all decay modes are 
shown in Fig.~\ref{fig:BRs}, where the same parameters as for Fig.~2 of~\cite{Jittoh} are used. 

Apart from the inclusion of the decay modes ${\tilde \tau_1} \to \chi  \nu_\tau  a_1(1260)$ and 
${\tilde \tau_1} \to \chi  \nu_\tau  \rho(770)$, we found a couple of differences between our results 
and those given in~\cite{Jittoh}, in which a bino-like neutralino is considered. 
In terms of the notation there, we get a different expression for 
$g_R$ from eq. (3) of~\cite{Jittoh}, and we note that our three-body decay rate expression 
$\Gamma_{{\tilde \tau_1} \to \chi  \nu_\tau  \pi}$ given in eq.~(\ref{3b-to-pion}) is different from 
eq. (A3) of~\cite{Jittoh}~ \footnote{We 
thank the authors of~\cite{Jittoh} for helpful correspondence.}.

\begin{figure}
\vskip 0.5in
\vspace*{-0.75in}
\begin{minipage}{8in}
\epsfig{file=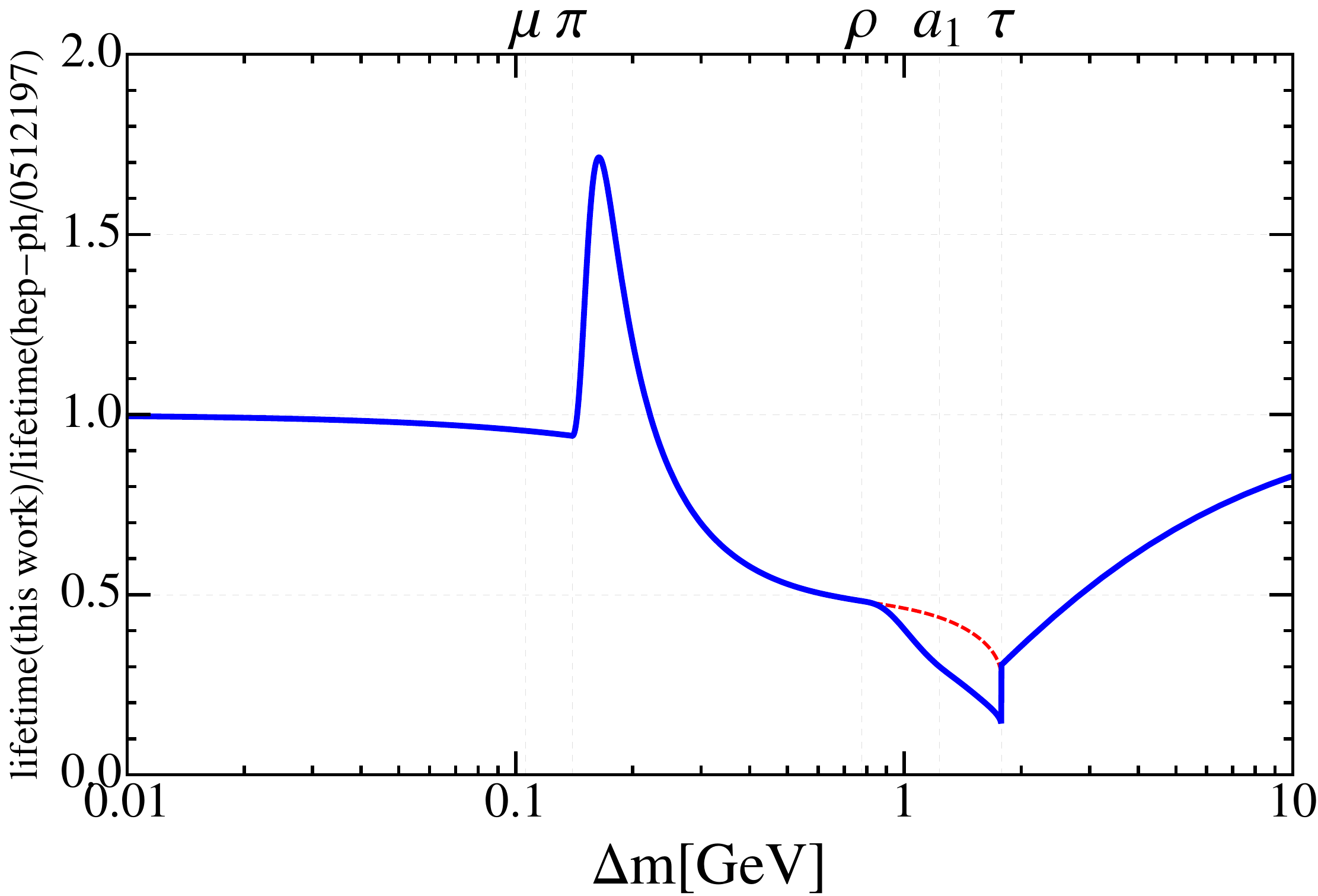,height=2.1in}
\epsfig{file=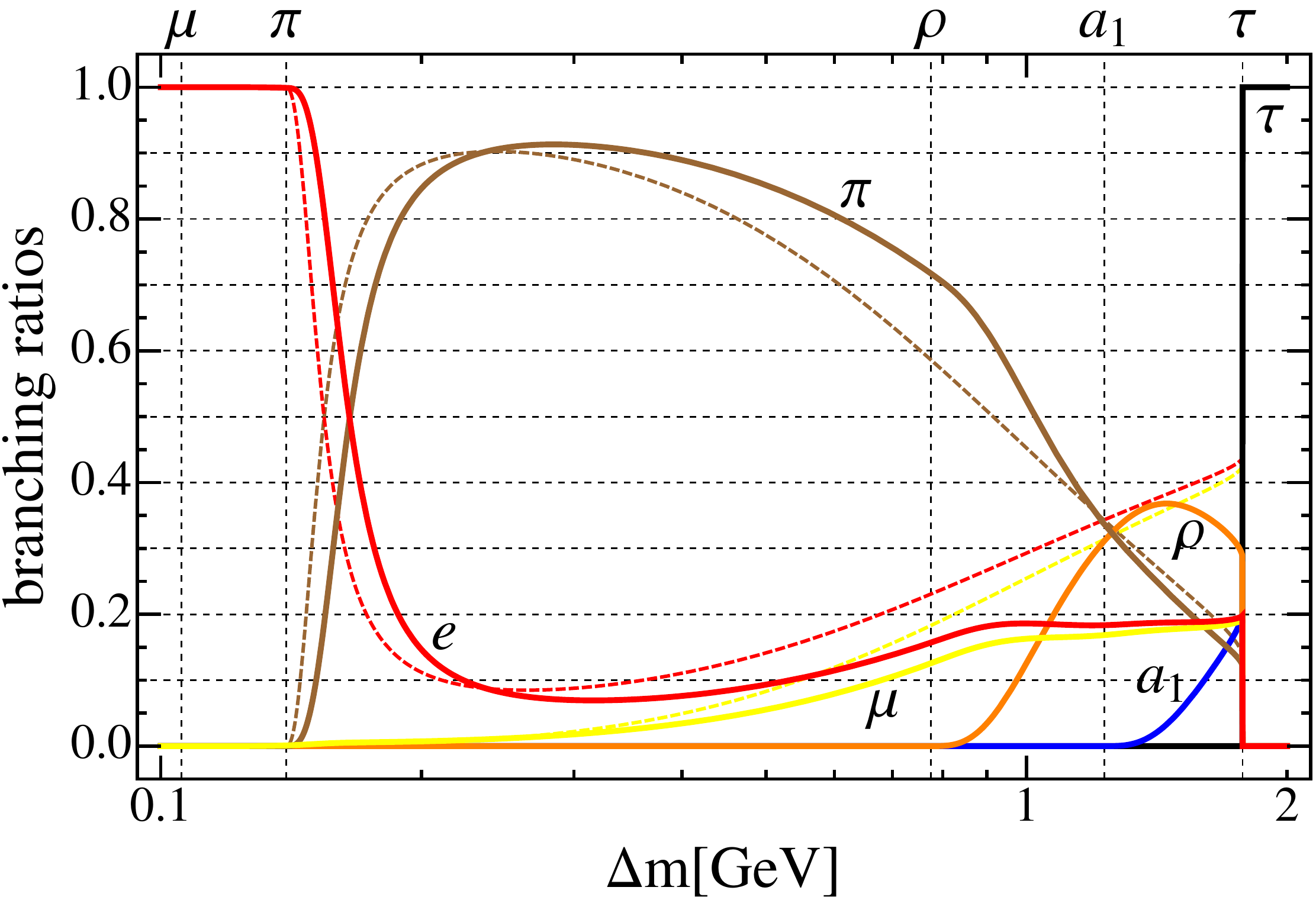,height=2.1in}
\hfill
\end{minipage}
\caption{
{\it Left panel: the ratio between the ${\tilde \tau_1}$ lifetime calculated in this work and
the result of~\protect\cite{Jittoh}, as a function of
$\Delta m \equiv m_{\tilde \tau_1} - m_\chi$, using the same parameters values as Fig.~2 of ~\protect\cite{Jittoh}. 
The red dashed line is the result if we would drop the contributions from the 
${\tilde \tau_1} \to \chi  \nu_\tau  a_1(1260)$ and ${\tilde \tau_1} \to \chi  \nu_\tau  \rho(770)$ modes.
Right panel: a comparison between
the corresponding calculations of the dominant ${\tilde \tau_1}$ branching ratios. 
The solid and dashed lines are the results calculated in this work and~\protect\cite{Jittoh}, respectively. 
The black, blue, orange, brown, yellow, and red lines are for the final states with $\tau$, $a_1(1260)$, 
$\rho(770)$, $\pi$, $\mu$, and $e$, respectively. 
}} 
\label{fig:compare}
\end{figure}

Our results for the ${\tilde \tau_1}$ lifetime and decay branching ratios
are compared with those of~\cite{Jittoh} in Fig.~\ref{fig:compare}.
The effect of the difference in $g_R$ compared with~\cite{Jittoh} is limited to the interference
term $\propto m_\tau$, which is important mainly when $\Delta m$ is small and ${\cal O}(m_\tau)$,
as seen in Fig.~\ref{fig:compare}. The difference in the expression
for $\Gamma_{{\tilde \tau_1} \to \chi  \nu_\tau  \pi}$ is responsible for the difference 
visible in Fig.~\ref{fig:compare} when $\Delta m \sim 200$~MeV. 
The contributions from the three-body decays ${\tilde \tau_1} \to \chi  \nu_\tau  a_1(1260)$ 
and ${\tilde \tau_1} \to \chi  \nu_\tau  \rho(770)$ are
important for $\Delta m \gappeq 1$~GeV, with a combined branching fraction $\sim 50$\%
when $\Delta m \sim 1.75$~GeV, and still $> 30$\% when $\Delta m \sim 1.25$~GeV.  This
feature is also visible in Fig.~\ref{fig:compare}. 

\begin{thebibliography}{99}

\bibitem{funnel}
M.~Drees and M.~M.~Nojiri,
Phys.\ Rev.\ D {\bf 47} (1993) 376 [arXiv:hep-ph/9207234];
H.~Baer and M.~Brhlik,
Phys.\ Rev.\ D {\bf 53} (1996) 597 [arXiv:hep-ph/9508321];
  Phys.\ Rev.\  D {\bf 57} (1998) 567
  [arXiv:hep-ph/9706509];
  H.~Baer, M.~Brhlik, M.~A.~Diaz, J.~Ferrandis, P.~Mercadante, P.~Quintana and X.~Tata,
    Phys.\ Rev.\  D {\bf 63} (2001) 015007
  [arXiv:hep-ph/0005027].
  
 \bibitem{cmssm}
 G.~L.~Kane, C.~F.~Kolda, L.~Roszkowski and J.~D.~Wells,
  Phys.\ Rev.\  D {\bf 49} (1994) 6173
  [arXiv:hep-ph/9312272];
  J.~R.~Ellis, T.~Falk, K.~A.~Olive and M.~Schmitt,
Phys.\ Lett.\ B {\bf 388} (1996) 97
[arXiv:hep-ph/9607292];
Phys.\ Lett.\ B {\bf 413} (1997) 355
[arXiv:hep-ph/9705444];
J.~R.~Ellis, T.~Falk, G.~Ganis, K.~A.~Olive and M.~Schmitt,
Phys.\ Rev.\ D {\bf 58} (1998) 095002
[arXiv:hep-ph/9801445];
V.~D.~Barger and C.~Kao,
Phys.\ Rev.\ D {\bf 57} (1998) 3131
[arXiv:hep-ph/9704403];
J.~R.~Ellis, T.~Falk, G.~Ganis and K.~A.~Olive,
Phys.\ Rev.\ D {\bf 62} (2000) 075010
[arXiv:hep-ph/0004169].

\bibitem{ehnos} H.~Goldberg,
                Phys.\ Rev.\ Lett.\ {\bf 50} (1983) 1419;
                J.~Ellis, J.~Hagelin, D.~Nanopoulos, K.~Olive and M.~Srednicki,
                Nucl.\ Phys.\ B {\bf 238} (1984) 453.
                
\bibitem{cmssmwmap}
J.~R.~Ellis, K.~A.~Olive, Y.~Santoso and V.~C.~Spanos,
Phys.\ Lett.\ B {\bf 565} (2003) 176
[arXiv:hep-ph/0303043];
H.~Baer and C.~Balazs,
  JCAP {\bf 0305}, 006 (2003)
  [arXiv:hep-ph/0303114];
A.~B.~Lahanas and D.~V.~Nanopoulos,
  Phys.\ Lett.\  B {\bf 568}, 55 (2003)
  [arXiv:hep-ph/0303130];
U.~Chattopadhyay, A.~Corsetti and P.~Nath,
  Phys.\ Rev.\  D {\bf 68}, 035005 (2003)
  [arXiv:hep-ph/0303201];
   C.~Munoz,
  Int.\ J.\ Mod.\ Phys.\  A {\bf 19}, 3093 (2004)
  [arXiv:hep-ph/0309346];
    R.~Arnowitt, B.~Dutta and B.~Hu,
arXiv:hep-ph/0310103;
   J.~Ellis and K.~A.~Olive,
  in {\it Particle dark matter}, ed. G.~Bertone (Cambridge University Press, 2010) pp142-163
  [arXiv:1001.3651 [astro-ph.CO]];
J.~Ellis and K.~A.~Olive,
  Eur.\ Phys.\ J.\ C {\bf 72}, 2005 (2012)
  [arXiv:1202.3262 [hep-ph]].


  \bibitem{wmap}
  E.~Komatsu {\it et al.}  [WMAP Collaboration],
  Astrophys.\ J.\ Suppl.\  {\bf 192} (2011) 18
  [arXiv:1001.4538 [astro-ph.CO]].

\bibitem{efo} J. Ellis, T. Falk, and K.A. Olive, Phys. Lett.  {\bf B444} (1998) 367
[arXiv:hep-ph/9810360];
J. Ellis, T. Falk, K.A. Olive, and M. Srednicki, {\it Astr. Part. Phys.}
{\bf 13} (2000) 181
[Erratum-ibid.\  {\bf 15} (2001) 413]
[arXiv:hep-ph/9905481].

\bibitem{efgosi}
A.~B.~Lahanas, D.~V.~Nanopoulos and V.~C.~Spanos,
  Mod.\ Phys.\ Lett.\  A {\bf 16} (2001) 1229
  [arXiv:hep-ph/0009065];
  A.~B.~Lahanas and V.~C.~Spanos,
  Eur.\ Phys.\ J.\ C {\bf 23} (2002) 185
  [arXiv:hep-ph/0106345];
J.~R.~Ellis, T.~Falk, G.~Ganis, K.~A.~Olive and M.~Srednicki,
Phys.\ Lett.\ B {\bf 510} (2001) 236
[arXiv:hep-ph/0102098].

\bibitem{fp}
  J.~L.~Feng, K.~T.~Matchev and T.~Moroi,
  Phys.\ Rev.\ Lett.\  {\bf 84}, 2322 (2000)
  [arXiv:hep-ph/9908309];
  Phys.\ Rev.\  D {\bf 61}, 075005 (2000)
  [arXiv:hep-ph/9909334]; 
  J.~L.~Feng, K.~T.~Matchev and F.~Wilczek,
  Phys.\ Lett.\  B {\bf 482}, 388 (2000)
  [arXiv:hep-ph/0004043].
  
 \bibitem{ATLAS7}
 G.~Aad {\it et al.}  [ATLAS Collaboration],
  arXiv:1208.0949 [hep-ex].
  
\bibitem{CMS7}
S.~Chatrchyan {\it et al.}  [CMS Collaboration],
  JHEP {\bf 1210}, 018 (2012)
  [arXiv:1207.1798 [hep-ex]];
  Phys.\ Rev.\ Lett.\  {\bf 109}, 171803 (2012)
  [arXiv:1207.1898 [hep-ex]].
  
\bibitem{ATLAS8}
ATLAS Collaboration,
{\tt http://cdsweb.cern.ch/record/1472710}.

\bibitem{CMS8}
CMS Collaboration,
{\tt http://cdsweb.cern.ch/record/1460095/files/SUS-12-016-pas.pdf}.

\bibitem{ATLASH}
G.~Aad {\it et al.}  [ATLAS Collaboration],
  Phys.\ Lett.\ B {\bf 716}, 1 (2012)
  [arXiv:1207.7214 [hep-ex]].
  
  \bibitem{CMSH}
   S.~Chatrchyan {\it et al.}  [CMS Collaboration],
  Phys.\ Lett.\ B {\bf 716}, 30 (2012)
  [arXiv:1207.7235 [hep-ex]].

   \bibitem{bmm}
G.~Aad {\it et al.}  [ATLAS Collaboration],
  Phys.\ Lett.\ B {\bf 713}, 387 (2012)
  [arXiv:1204.0735 [hep-ex]];
 T.~Aaltonen {\it et al.}  [CDF Collaboration],
  Phys.\ Rev.\ Lett.\  {\bf 107}, 239903 (2011)
  [Phys.\ Rev.\ Lett.\  {\bf 107}, 191801 (2011)]
  [arXiv:1107.2304 [hep-ex]];
  updated results presented at Aspen in Feb. 2012 by M.~Rescigno, 
  {\tt https://indico.cern.ch/getFile.py/} {\tt access?contribId=28\&sessionId=7\&resId} {\tt =1\&materialId=slides\&confId=143360}.;
S.~Chatrchyan {\it et al.}  [CMS Collaboration],
  Phys.\ Rev.\ Lett.\  {\bf 107}, 191802 (2011)
  [arXiv:1107.5834 [hep-ex]];
R.~Aaij {\it et al.}  [LHCb Collaboration],
  Phys.\ Lett.\  B {\bf 699} (2011) 330
  [arXiv:1103.2465 [hep-ex]];
arXiv:1203.4493 [hep-ex];
  For an official combination of the ATLAS, CMS and LHCb results, see:
ATLAS, CMS, and LHCb Collaborations, {\tt http://cdsweb.cern.ch/record/1452186/} {\tt files/LHCb-CONF-2012-017.pdf}.

  
\bibitem{LHCbnew} 
  R.~Aaij {\it et al.}  [LHCb Collaboration],
  arXiv:1211.2674 [hep-ex].
  
   \bibitem{XE100}
  E.~Aprile {\it et al.}  [XENON100 Collaboration],
  Phys.\ Rev.\ Lett.\  {\bf 107} (2011) 131302
  [arXiv:1104.2549 [astro-ph.CO]].

\bibitem{MC8}
O.~Buchmueller,  {\it et al.},
  arXiv:1207.7315 [hep-ph].
  
 
  \bibitem{Jittoh}
  T.~Jittoh, J.~Sato, T.~Shimomura and M.~Yamanaka,
  Phys.\ Rev.\ D {\bf 73}, 055009 (2006)
  [hep-ph/0512197].
  
 \bibitem{Kaneko}
  S.~Kaneko, J.~Sato, T.~Shimomura, O.~Vives and M.~Yamanaka,
  Phys.\ Rev.\ D {\bf 78} (2008) 116013
  [arXiv:0811.0703 [hep-ph]].
  
  \bibitem{CMSmcp}
  S.~Chatrchyan {\it et al.}  [CMS Collaboration],
  Phys.\ Lett.\ B {\bf 713} (2012) 408
  [arXiv:1205.0272 [hep-ex]].
  
  \bibitem{ATLASmcp}
   G.~Aad {\it et al.}  [ATLAS Collaboration],
  arXiv:1211.1597 [hep-ex].
  
   \bibitem{SSARD}  Information about this code is available from K.~A.~Olive: it contains important contributions 
from T.~Falk, A.~Ferstl, G.~Ganis, F.~Luo, A.~Mustafayev, J.~McDonald, K.~A.~Olive, P.~Sandick, Y.~Santoso and M.~Srednicki. 


\bibitem{PYTHIA}
T.~Sjostrand, S.~Mrenna and P.~Z.~Skands,
  Comput.\ Phys.\ Commun.\  {\bf 178} (2008) 852
  [arXiv:0710.3820 [hep-ph]].

 \bibitem{Delphes}
For a description of {\tt Delphes}, 
written by S.~Ovyn and X.~Rouby, see
{\tt http://www.fynu.ucl.ac.be/users/s.ovyn/} {\tt Delphes/index.html}.
  

\bibitem{PROSPINO}
  W.~Beenakker, R.~H\"opker, M.~Spira, P.~M.~Zerwas,
  Nucl.\ Phys.\  {\bf B492 } (1997)  51-103
  [hep-ph/9610490];
  W.~Beenakker, M.~Kr\"amer, T.~Plehn, M.~Spira, P.~M.~Zerwas,
  Nucl.\ Phys.\  {\bf B515 } (1998)  3-14
  [hep-ph/9710451];
  W.~Beenakker, M.~Klasen, M.~Kr\"amer, T.~Plehn, M.~Spira, P.~M.~Zerwas,
  Phys.\ Rev.\ Lett.\  {\bf 83 } (1999)  3780-3783
  [hep-ph/9906298].

\bibitem{MoEDAL}
MoEDAL Collaboration, {\tt http://moedal.web.cern.ch/}.

\bibitem{Haber:1984rc} 
  H.~E.~Haber and G.~L.~Kane,
  Phys.\ Rept.\  {\bf 117}, 75 (1985).
  
\bibitem{Gunion:1984yn} 
  J.~F.~Gunion and H.~E.~Haber,
  Nucl.\ Phys.\ B {\bf 272}, 1 (1986)
  [Erratum-ibid.\ B {\bf 402}, 567 (1993)].
  
\bibitem{Binosi:2003yf} 
  D.~Binosi and L.~Theussl,
  Comput.\ Phys.\ Commun.\  {\bf 161}, 76 (2004)
  [hep-ph/0309015].
  
\bibitem{Scherer:2005ri} 
  S.~Scherer and M.~R.~Schindler,
  hep-ph/0505265.
  
\bibitem{Lichard:1997ya} 
  P.~Lichard,
  Phys.\ Rev.\ D {\bf 55}, 5385 (1997)
  [hep-ph/9702345].
  
\bibitem{pdg-2012} 
  J.~Beringer {\it et al.}  [Particle Data Group Collaboration],
  Phys.\ Rev.\ D {\bf 86}, 010001 (2012).
  
\end{thebibliography}
\end{document}